\documentclass[journal,onecolumn,letterpaper,12pt,romanappendices]{IEEEtran}
\usepackage{color}
\usepackage{setspace}
\usepackage{array}
\usepackage{cite}
\usepackage{multirow}
\usepackage{stmaryrd}
\usepackage{setspace}
\usepackage{amssymb}
\usepackage[cmex10]{amsmath}
\usepackage{amsthm}
\usepackage[pdftex]{graphicx}
\usepackage{epstopdf}
\usepackage{epsfig}
\usepackage{verbatim}
\usepackage{bbding}
\usepackage{dsfont}

\newcommand{\ra}[1]{\ensuremath\raisebox{2pt}{$#1$}}
\newcommand{\ma}[1]{\ensuremath\mathcal{#1}}
\newcommand{\bs}[1]{\ensuremath\boldsymbol{#1}}
\newcommand{\ds}{\ensuremath\displaystyle\sum}
\newcommand{\df}[2]{\ensuremath\displaystyle\frac{#1}{#2}}
\newsavebox{\subfigure}
\newsavebox{\theorembox}
\newsavebox{\lemmabox}
\newsavebox{\corollarybox}
\newsavebox{\propositionbox}
\newsavebox{\examplebox}
\newsavebox{\conjecturebox}
\newsavebox{\algbox}
\newsavebox{\qbox}
\newsavebox{\problembox}
\newsavebox{\definitionbox}
\newsavebox{\assumptionbox}
\newsavebox{\hypothesisbox}
\newsavebox{\factbox}
\newsavebox{\remarkbox}
\savebox{\theorembox}{\noindent\bf Theorem}
\savebox{\lemmabox}{\noindent\bf Lemma}
\savebox{\corollarybox}{\noindent\bf Corollary}
\savebox{\propositionbox}{\noindent\bf Proposition}
\savebox{\examplebox}{\noindent\bf Example}
\savebox{\conjecturebox}{\noindent\bf Conjecture}
\savebox{\algbox}{\noindent\bf Algorithm}
\savebox{\qbox}{\noindent\bf Question}
\savebox{\definitionbox}{\noindent\bf Definition}
\savebox{\problembox}{\noindent\bf Problem}
\savebox{\assumptionbox}{\noindent\bf Assumption}
\savebox{\hypothesisbox}{\noindent\bf Hypothesis}
\savebox{\factbox}{\noindent\bf Fact}
\savebox{\remarkbox}{\noindent\bf Remark}
\newtheorem{theorem}{\usebox{\theorembox}}

\newtheorem{fact}{\usebox{\factbox}}

\begin{document}
\title{\huge Density Evolution Analysis of Node-Based Verification-Based Algorithms in Compressed Sensing}
\author{Yaser Eftekhari, Anoosheh Heidarzadeh, Amir H. Banihashemi, Ioannis Lambadaris\\
			Department of Systems and Computer Engineering, Carleton University, Ottawa, ON, Canada\\
			E-mails: \{eft-yas, anoosheh, Amir.Banihashemi, Ioannis\}@sce.carleton.ca}
\maketitle
\thispagestyle{empty}
%\newpage
%\linenumbers
%\linenumbersep 30pt\relax
%\linenumbersep 30pt
\begin{abstract}
In this paper, we present a new approach for the analysis of iterative node-based verification-based (NB-VB) recovery algorithms in the context of compressive sensing. These algorithms are particularly interesting due to their low complexity (linear in the signal dimension $n$). The asymptotic analysis predicts the fraction of unverified signal elements at each iteration $\ell$ in the asymptotic regime where $n \rightarrow \infty$. The analysis is similar in nature to the well-known density evolution technique commonly used to analyze iterative decoding algorithms. To perform the analysis, a message-passing interpretation of NB-VB algorithms is provided. This interpretation lacks the extrinsic nature of standard message-passing algorithms to which density evolution is usually applied. This requires a number of non-trivial modifications in the analysis. The analysis tracks the average performance of the recovery algorithms over the ensembles of input signals and sensing matrices as a function of $\ell$. Concentration results are devised to demonstrate that the performance of the recovery algorithms applied to any choice of the input signal over any realization of the sensing matrix follows the deterministic results of the analysis closely. Simulation results are also provided which demonstrate that the proposed asymptotic analysis matches the performance of recovery algorithms for large but finite values of $n$. Compared to the existing technique for the analysis of NB-VB algorithms, which is based on numerically solving a large system of coupled differential equations, the proposed method is much simpler and more accurate.
\end{abstract}
%%%%%%%%%%%%%%%%%%%%%%%%%%%%%%%%%%%%%%%%%%%%%%%%%%%%%%%%%%%%%%%%%%%%%%%%%%%%%%%%%%%%%%%%%%%%%%%%%%%%%%%%%%%%%
%%%%%%%%%%%%%%%%%%%%%%%%%%%%%%%%%%%%%%%%%%%%%%%%%%%%%%%%%%%%%%%%%%%%%%%%%%%%%%%%%%%%%%%%%%%%%%%%%%%%%%%%%%%%%
\section{Introduction}
\label{firstintro}
Compressed sensing was introduced with the idea to represent a signal $\bs{v}\in\mathbb{R}^n$ with $k$ nonzero elements with measurements $\bs{c}\in\mathbb{R}^m$, where $k<m\ll n$, and yet to be able to recover back the original signal $\bs{v}$ \cite{D06,CRTFeb06}. In the measuring process, also referred to as \textit{encoding}, signal elements are mapped to measurements through a linear transformation represented by the matrix multiplication $\bs{c} = \bs{G}\bs{v}$, where the matrix $\bs{G}\in \mathbb{R}^{m\times n}$ is referred to as the \emph{sensing matrix}. This linear mapping can also be characterized by a bipartite graph \cite{XH07}, referred to as the \textit{sensing graph}.

In the recovery process, also referred to as \textit{decoding}, based on the knowledge of the measurements and the sensing matrix, we estimate the original signal. Decoding process is successful if $\bs{v}$ is estimated correctly. Three performance measures namely, density ratio $\gamma \triangleq k/n$, compression ratio $r_c \triangleq m/n$, and oversampling ratio $r_o \triangleq m/k$ are used in order to measure and compare the performance of the recovery algorithms in the context of compressed sensing.\footnote{For successful decoding clearly we need $r_o \geq 1$. It is desirable to have this parameter as small as possible. Indeed, in \cite{WV09} authors proved that $r_o = 1$ is achievable in the asymptotic case ($n \rightarrow \infty$). This means that the highest density ratio that an algorithm can possibly handle is $\gamma^* = r_c$. Authors in \cite{B07} have shown that if the sensing matrix consists of i.i.d. Gaussian elements, then a decoder based on the $\ell_0$ norm can recover the original signal with $m=k+1$ measurements; i.e., $r_o \approx 1$. To find the solution based on the $\ell_0$ recovery, however, one has to perform an exhaustive search, which is computationally too complex \cite{T04}.}

The sensing matrix in compressive sensing can be either \textit{dense} or \textit{sparse}. A sensing matrix is considered dense if it has few, if not none, zero entries. Sparse matrices, on the other hand, have few nonzero entries in each row and column. One major difference between these two types of matrices is the encoding complexity associated with each class. For sparse matrices, the number of operations needed to calculate the measurements is considerably lower than the one needed for dense matrices.

Decoding algorithms can be classified based on the class of sensing matrix they use. The decoding algorithms in each class have certain properties in common, which follows. (For a comprehensive study on the topic we refer the interested readers to \cite{BGIKS08}.) Decoding algorithms associated with dense matrices have, generally, high complexity (between $\ma{O}(n^2)$ and $\ma{O}(n^3)$) compared to the lower complexity of algorithms utilizing sparse matrices (between $\ma{O}(n)$ and $\ma{O}(n^2)$). To have a better feeling about the complexity and running time of these two classes of algorithms, we have included (Section \ref{simulation}, Fig. \ref{Total_Exe_time}) the comparison between two standard recovery algorithms for dense matrices ($\ell_1$ minimization and weighted $\ell_1$ minimization) and one algorithm (SBB) for sparse matrices. As can be seen, the decoding algorithm on sparse matrices is faster by about two orders of magnitude. Decoding algorithms for dense matrices are mostly based on linear or convex programming \cite{D06,CRTFeb06,TG07, NV09}. The reason is that random dense matrices satisfy \textit{restricted isometry property} (RIP) with overwhelming probability \cite{MPTJ06,BDDW08}. The RIP was introduced by Cand\`{e}s and Tao in \cite{CT05} as the main restriction on the sensing matrix so that the recovery based on linear programming will be able to successfully recover the signal. Sparse matrices, on the other hand, do not satisfy RIP unless $m=\Omega(k^2)$ \cite{C08}.\footnote{Authors in \cite{BGIKS08} extended the definition of RIP and showed that sparse matrices satisfy a generalized RIP constraint. The generalized RIP suffices for the linear programming decoders to succeed in recovering sparse signals. However, the resulting bound on the reconstruction error is weaker compared to the case where the sensing matrix satisfies the original RIP condition.} In fact, many of the decoders based on sparse sensing matrices are iterative \cite{CM06, I08, GSTV06, GSTV07,SBB206, CSW10, XH07, ZP09, ZP08, ZP07, ZP07J, LMPDK08, APT10, BSB10}. Although more computationally complex, decoding algorithms on dense matrices tend to recover signals with larger number of nonzero elements (higher density ratio) compared to decoders on sparse matrices. Nevertheless, the high complexity of decoding algorithms on dense matrices hinders their application to high-dimensional signal recovery (signals with large $n$).

Focusing on recovery algorithms based on sparse graphs, we can further divide them into two major groups. In one group, we have algorithms that use group testing and similar techniques from estimation theory \cite{GSTV06, CM06, I08, GSTV07}. These are referred to as \textit{combinatorial algorithms}. In the other group, recovery algorithms work with the bipartite graph associated with the sensing matrix by passing messages over the edges of the graph \cite{ZP07, ZP07J, ZP08, LMPDK08, CSW10, BSB10, APT10, SBB206, ZP09, XH07}. These are referred to as \textit{message-passing} algorithms. Combinatorial algorithms, generally, assume that the decoder knows the size of the support set, $k$ \cite{CM06,GSTV06,I08,GSTV07}. These algorithms, have two main steps. In the first step, the algorithm outputs an estimate which has more nonzero values than the original signal. In the next step, knowing the parameter $k$, the estimate is pruned so that the output estimate has the same support size as the original signal. Combinatorial algorithms are, in general, more computationally complex than message-passing algorithms. For example, the algorithm introduced in \cite{CM06} has complexity $\ma{O}(k^2 \text{polylog}(n))$, which translates to $\ma{O}(n^2 \text{polylog}(n))$ in the regime where $k$ scales linearly with $n$. Message-passing algorithms, on the other hand, have computational complexity $\ma{O}(n)$.

In this work, we are interested in low-complexity recovery algorithms that exploit the sparsity of the sensing matrix. In particular, we are interested in message-passing recovery algorithms. In \cite{CSW10}, the authors propose a simple message-passing algorithm to reconstruct non-negative signals. This algorithm assumes lower and upper bounds for the signal elements. It then shrinks the difference between the two bounds through iterations. The important feature of the recovery algorithm introduced in \cite{CSW10} is its uniform guarantee on signal reconstruction. Another approach in message-passing algorithms is to assume a prior distribution for the signal elements and try to maximize the a-posteriori distribution of the elements based on the observed measurements. In \cite{BSB10}, the authors assume Gaussian mixture priors. The main problem associated with this approach is that the length of the messages passed over the edges of the graph grow exponentially fast with the number of iterations. In another work \cite{APT10}, the authors assume Jeffreys' priors \cite{R94} and aim at recovering the support set of the original signal using message-passing algorithms. Then, they apply well-known least-square algorithms, such as LSQR \cite{PS82}, to estimate the value of signal elements. Moreover, in \cite{APT10}, it is assumed that the size of the support set, $k$, is known. Algorithms discussed so far are either restrictive, in the sense that they assume some knowledge of the support set at the decoder, or have a high computational complexity that makes them impractical in applications with large $n$.

In this paper, we are interested in a sub-class of message-passing algorithms called \textit{Verification-Based} (VB) algorithms. These algorithms were originally introduced in the context of channel coding with non-binary alphabet \cite{LM05}. Thanks to the connection between compressive sensing and linear channel codes over real numbers noted in \cite{ZP08}, the authors in \cite{SBB206} and \cite{ZP09} used VB algorithms in the context of compressed sensing. This class of algorithms has certain properties that make it perhaps one of the most interesting classes of recovery algorithms in compressed sensing. The VB algorithms recover signal elements in iterations. When an element is recovered, its value is kept unchanged in future iterations. The algorithms in this class have decoding complexity $\ma{O}(n)$, which makes them suitable for applications involving recovery of signals with large $n$. Moreover, these algorithms operate on sparse sensing graphs, which translates to less computations in the encoding process. Another main advantage of VB algorithms is that they are not sensitive to the distribution of nonzero elements of the sensing matrix as well as the distribution of nonzero elements of the signal, if certain conditions are satisfied. We will elaborate on this topic further in section \ref{enc}. These properties make the VB algorithms a suitable choice for low-complexity recovery of sparse signals. The VB algorithms are, however, sensitive to the presence of noise in the measured data. In this work, our main focus is on the study of the noiseless case. This case is important because i) the noise-free analysis of recovery algorithms can serve as an upper bound for the performance of the noisy versions, and ii) noiseless compressive sensing has its own applications such as those in \cite{LMPDK08,LMP08}. For the sake of being thorough and to demonstrate the potential of VB algorithms in recovering signals from noisy measurements, we will also comment on using standard thresholding techniques to deal with noisy measurements in Section \ref{generalizations}. An in-depth analysis of the approach, however, is beyond the scope of this paper.

Another interesting feature of VB algorithms is that their performance can be analyzed in the asymptotic case ($n \rightarrow \infty$). Assume a probabilistic input model, in which a signal element is nonzero (and takes a value from a certain distribution) with probability $\alpha$ and is zero with probability $1- \alpha$. In the sequel, we refer to parameter $\alpha$ as the \emph{density factor}. Furthermore, let $\alpha^{(\ell)}$ denote the probability that a signal element is nonzero and unverified before iteration $\ell$ over the ensemble of all sensing graphs and inputs of interest. So, $\alpha^{(0)} = \alpha$. If $\lim_{\ell \rightarrow \infty} \alpha^{(\ell)} = 0$, then the algorithm is called successful for the initial density factor $\alpha$.\footnote{It is easy to prove that the probability of a zero-valued signal element being unverified at iteration $\ell$ is upper bounded by $d_c \frac{\alpha^{(\ell)}}{1-\alpha^{(\ell)}}$. Hence, when $\alpha^{(\ell)}$ tends to zero, this probability also tends to zero.} On the other hand, if there exists $\epsilon > 0$, such that $\lim_{\ell \rightarrow \infty} \alpha^{(\ell)} > \epsilon$, then the algorithm is said to fail for the initial density factor $\alpha$. Using the combinatorial arguments in \cite{LMSS01} and \cite{LMSSF01}, one can see that the algorithms complete the recovery once the probability $\alpha^{(\ell)} \rightarrow 0$.

Authors in \cite{LM05,ZP07J,ZP07,ZP09} have shown that for each VB recovery algorithm in the asymptotic regime as $n \rightarrow \infty$ and $\ell \rightarrow \infty$, a limiting value exists for $\alpha$, before which the recovery algorithm is successful and beyond which it is not. We refer to this limit as the \textit{success threshold}. The success threshold serves as an asymptotic measure of performance for VB algorithms. It can also be used to estimate the performance of these algorithms for finite but large values of $n$. To this end, researchers have analyzed VB algorithms in the asymptotic regime ($n \rightarrow \infty, \ell \rightarrow \infty$) in order to find the success threshold associated with each VB algorithm. There are two categories of VB algorithms: \emph{node-based (NB)} and \emph{message-based (MB)} \cite{ZP07J}. The two categories yield different success thresholds and are analyzed using different techniques. In general, NB algorithms have higher success thresholds and are harder to analyze. We elaborate on the differences between the two categories and their corresponding analytical tools in Section \ref{VB}. The focus of this work is on the analysis of NB algorithms.

Asymptotic analysis of VB algorithms can be found in \cite{ZP07J,ZP09,LM05}. Algorithms considered in \cite{LM05,ZP09} are of MB type, while the authors in \cite{ZP07J} considered the NB type recovery algorithms. Moreover, for their analysis, the authors of \cite{ZP09} made the assumption that $k/n \rightarrow 0 \text{ as } n \rightarrow \infty$. From a practical point of view, however, it is more desirable to analyze the cases in which the number of nonzero elements of the signal, $k$, scales linearly with its dimension $n$. In fact, in this paper, we show that VB algorithms are capable of recovering signals whose density ratio is a nonzero constant, i.e., $k/n$ remains constant, as $n$ is increased.

The analysis of NB-VB algorithms discussed in \cite{ZP07J} results in a system of coupled differential equations. Due to the high complexity of solving the resulting differential equations, the authors used numerical methods to approximate the asymptotic results. They further assumed that the algorithms resolve at most one signal element in each iteration. Therefore, the number of iterations needed for the numerical analysis equals $n$; the number of signal elements. The challenges associated with the analysis of \cite{ZP07J} are twofold: 1) as the analysis is only valid for $n \rightarrow \infty$, one has to choose very large $n$ for the numerical approximation, which directly translates to long running time and high computational complexity, and 2) since the numerical approach is used to approximately solve the differential equations, the approximation errors can potentially propagate through the iterations. This makes it hard to evaluate how close the success threshold reported by this analysis (even for large values of $n$) is to the real success threshold. In comparison, the analysis proposed in this paper is much faster (by about two orders of magnitude for the tested cases) and is more robust against numerical errors. The general approach for the analysis is also different and is based on basic probability theory.

The goal of this work is to develop a low-complexity framework for the asymptotic analysis (as $n \rightarrow \infty, \ell \rightarrow \infty$) of NB-VB algorithms over sparse random sensing graphs and extend it to include recovery algorithms of similar nature such as that of \cite{XH07}. In our analysis, we assume that the measurements are noiseless. The main analytical tool used in this paper is probability theory. We demonstrate that the recovery algorithms can be described by a first order time-varying Markov chain. We thus track the distribution of the states of this Markov chain through iterations in the analysis. The purpose is to find the transition probabilities between different states of the Markov chain as the iterations progress. The computational complexity of the proposed analysis thus increases linearly with the number of iterations. The calculation of transition probabilities includes simple mathematical operations, more specifically addition and multiplication, as opposed to solving complex systems of coupled differential equations, as is the case in \cite{ZP07J}. One should however note that for a sensing graph in which each signal element affects $d_v$ measurements and each measurement is a linear combination of $d_c$ signal elements, where $d_v$ and $d_c$ are fixed and positive integers, the number of states in the proposed analysis is $\ma{O}(d_v + d_c^2)$, which is in the same order as the number of differential equations in \cite{ZP07J} is. As part of our asymptotic analysis, we also prove concentration results which certify that the performance of a recovery algorithm for a random choice of the input signal and the sensing matrix is very close to what is predicted by the density evolution results at the limit of $n \rightarrow \infty$.

Using the proposed analysis, we can determine the distribution of the decoder states at any desired iteration. By tracking the distribution of the decoder states with iterations, we then find the success threshold of different NB-VB algorithms. Moreover, using our approach, we perform a comprehensive study and comparison of performance of different VB recovery algorithms over a variety of sparse graphs. Our simulations show that the behavior of VB algorithms, when applied to signals with large lengths (in the order of $10^5$), are in good agreement with the asymptotic analytical results.

The rest of the paper is organized as follows. In section \ref{back_knowledge}, we introduce the class of bipartite graphs and inputs signals of interest in this paper. We also provide a more detailed description of VB algorithms in this section. In section \ref{enc}, the decoding process in each VB algorithm is discussed further. Also in this section, we discuss the important notion of false verification and its probability in VB algorithms. A message-passing interpretation of the recovery algorithms is presented in section \ref{Decoding}. In this section, we also make a more detailed distinction between NB and MB recovery algorithms. The analysis framework will be introduced in section \ref{analysis}. We propose a simple modification of VB algorithms to deal with noisy measurements in section \ref{generalizations}. Simulation results will be presented in section \ref{simulation}. Appendices \ref{app_original_Genie}, \ref{app_original_LM}, and \ref{app_original_SBB} are devoted to the derivation of the transition probabilities. Appendix \ref{conc_bound} consists of some bounds needed for the concentration theorem, presented and proved in section \ref{analysis}.
%%%%%%%%%%%%%%%%%%%%%%%%%%%%%%%%%%%%%%%%%%%%%%%%%%%%%%%%%%%%%%%%%%%%%%%%%%%%%%%%%%%%%%%%%%%%%%%%%%%%%%%%%%%%%
%%%%%%%%%%%%%%%%%%%%%%%%%%%%%%%%%%%%%%%%%%%%%%%%%%%%%%%%%%%%%%%%%%%%%%%%%%%%%%%%%%%%%%%%%%%%%%%%%%%%%%%%%%%%%
\section{Background}
\label{back_knowledge}
%%%%%%%%%%%%%%%%%%%%%%%%%%%%%%%%%%%%%%%%%%%%%%%%%%%%%%%%%%%%%%%%%%%%%%%%%%%%%%%%%%%%%%%%%%%%%%%%%%%%%%%%%%%%%
\subsection{Ensembles of Sensing Graphs and Inputs}
\label{Defs}
A \textit{bipartite graph} (or \textit{bigraph}) $\ma{G}({V}\cup{C},{E})$ is defined as a graph whose set of vertices ${V}\cup{C}$ is divided into two disjoint sets ${V}$ and ${C}$, so that every edge in the set of edges ${E}$ connects a vertex in ${V}$ to one in ${C}$. Corresponding to each such graph, a \textit{biadjacency matrix} $\bs{A}(\ma{G})$ of size $|{C}|\times |{V}|$ is formed as follows: the entry $a_{ij}$ is $1$ if there exists an edge $e_{ij}\in E$ connecting the vertex $c_i\in C$ to the vertex $v_j\in V$; and is $0$, otherwise.

Let $d_v$ and $d_c$ be two positive integers. Consider a bigraph $\ma{G}(V\cup C, E)$ with $|V| = n$ and $|C| = m$, so that each vertex in $V$ ($C$) is incident to $d_v$ ($d_c$) vertices in $C$ ($V$). Clearly, $n d_v=m d_c$. We refer to this bigraph as an $(n,d_v,d_c)$-\textit{biregular graph}. The biadjacenecy matrix $\bs{A}(\ma{G})$ associated to an $(n,d_v,d_c)$-biregular graph has $d_c$ $1$'s in each row and $d_v$ $1$'s in each column.

Moreover, a \textit{bipartite weighted graph} (or \textit{weighted bigraph}) $\ma{G'}({V}\cup{C},W(E))$ is a generalization of the bigraph $\ma{G}({V}\cup{C},{E})$ in the sense that a weight $w_{ij}:= w(e_{ij})\in \mathbb{R} \backslash \{0\}$ is associated with each edge $e_{ij}\in E$. The biadjacency matrix $\bs{A}(\ma{G'})$ corresponding to the weighted bigraph $\ma{G'}$ is acquired from the biadjacency matrix $\bs{A}(\ma{G})$ of the underlying bigraph $\ma{G}$ by replacing nonzero $a_{ij}$ values in $\bs{A}(\ma{G})$ with $w_{ij}$. A \textit{regular bipartite weighted graph} (or \textit{weighted biregular graph}) is defined similarly.

For given parameters $d_v,d_c$ and $n$ ($m=n d_v/d_c$), let $\ma{G}^{n}(d_v,d_c)$ denote the ensemble of all $(n,d_v,d_c)$-biregular graphs. Let us assume an arbitrary, but fixed, labeling scheme for vertex sets $V$ and $C$ over the ensemble. Further, let $\bs{W}$ be a matrix of size $m \times n$ of weights $w$ drawn i.i.d. according to a distribution $f(w)$, and $\ma{W}_{f}^{m\times n}$ be the ensemble of all such matrices. Now for any biregular graph $\ma{G}({V}\cup{C},E) \in \ma{G}^{n}(d_v,d_c)$ and any weight matrix $\bs{W} \in \ma{W}_{f}^{m\times n}$, we form the corresponding $(n,d_v,d_c)$-weighted biregular graph $\ma{G'}({V}\cup{C},W(E))$ as follows. To every edge $e_{ij}\in E$, $1\leq i \leq m, 1 \leq j \leq n$, connecting $i$th vertex from $C$ and $j$th vertex from $V$, we assign the weight $w(e_{ij}) = w_{ij}$; i.e., the weight in row $i$ and column $j$ of the weight matrix $\bs{W}$. Thus, we construct the ensemble of all $(n,d_v,d_c)$-weighted biregular graphs, denoted by $\ma{G}_{f}^{n}(d_v,d_c)$, by freely combining elements in $\ma{G}^{n}(d_v,d_c)$ and $\ma{W}_{f}^{m\times n}$.

Thus far, we have described the ensemble of graphs that are of interest in this work. In what fallows, we describe the ensemble of inputs of interest. Let $\alpha\in [0,1]$ be a fixed real number and $\bs{v}$ be a vector of length $n$ with elements $\bs{v}_i$ drawn i.i.d. according to a probability distribution function defined as follows: the element is zero with probability $1-\alpha$, or follows a distribution $g$ with probability $\alpha$ (i.e., $\Pr[\bs{v}_i=v]=\alpha g(v)+(1-\alpha)\delta(v)$, where $\delta$ is the Dirac delta function). We denote the ensemble of all such vectors by $\ma{V}_{g}^{n}(\alpha)$.\footnote{It is worth noting that the expected fraction of nonzero elements in such a vector is $\alpha$. Using a Chernoff bound, it can be shown that the actual fraction of nonzero elements in a randomly chosen vector from this ensemble is tightly concentrated around its expected value ($\alpha$) with high probability.}

In compressive sensing, each measurement is a linear combination of the signal elements. With a slight abuse of notation, we use $c_i$ ($v_j$) for both the label and the value of the $i$th measurement (the $j$th signal element). We denote by $\bs{c}$ and $\bs{v}$, the column vectors of the measurements $c_i$'s ($1\leq i\leq m$), and the signal elements $v_j$'s ($1\leq j\leq n$), respectively. The underlying system of linear combinations can then be represented by the matrix multiplication $\bs{c} = \bs{G} \bs{v}$. In this paper, the sensing matrix $\bs{G}$ is the biadjacency matrix of a weighted bigraph $\ma{G}(V\cup C,W(E))$ drawn uniformly at random from the ensemble $\ma{G}_f^{n}(d_v,d_c)$. Henceforth, we refer to the graph $\ma{G}$ as the \textit{sensing graph}. Moreover, the signal vector $\bs{v}$ is drawn uniformly at random from the ensemble $\ma{V}_{g}^{n}(\alpha)$. The sets of signal elements and measurements are respectively mapped to the vertex sets $V$ and $C$ ($|V|=n$, $|C|=m$). The coefficient of the $j^\text{th}$ signal element (${v}_j\in V$) in the linear combination associated with the $i^\text{th}$ measurement ${c}_i\in C$, the entry $g_{ij}$ in $\bs{G}$, is the entry $w_{ij}$ of the biadjacency matrix $\bs{A}(\ma{G})$ of $\ma{G}$.

Following the terminology frequently used in the context of coding,\footnote{In the context of coding, a linear code can be represented by a bigraph, where the two sets of nodes represent the code symbols, and the linear constraints that the symbols have to satisfy \cite{M}.} we refer to the sets ${V}$ and $C$ as the \textit{variable nodes} and \textit{check nodes}, respectively. We will interchangeably use the terms variable nodes and signal elements as well as check nodes and measurements. The main focus of this paper is on the weighted biregular graphs. The results however, can be generalized to irregular graphs.
%%%%%%%%%%%%%%%%%%%%%%%%%%%%%%%%%%%%%%%%%%%%%%%%%%%%%%%%%%%%%%%%%%%%%%%%%%%%%%%%%%%%%%%%%%%%%%%%%%%%%%%%%%%%%
\subsection{Previous Work on VB Algorithms}
\label{VB}
Luby and Mitzenmacher \cite{LM05} proposed and analyzed two iterative algorithms over bigraphs for packet-based error correction in the context of channel coding. In these algorithms, a variable node can be in one of the two states: ``verified'' or ``unverified''. Under certain circumstances, a variable node is verified and a value is assigned to it. This node then contributes to the verification of other variable nodes. The decoding process continues until either the entire set of unverified variable nodes is verified, or the process makes no further progress while there are still some unverified variables. Due to the verification nature of the procedure, the two algorithms in \cite{LM05} are called \textit{verification-based} (VB) algorithms. If the assigned value to a variable node at a certain iteration is different from its true value, a \textit{false verification} has occurred. In section \ref{enc}, we discuss sufficient conditions for VB algorithms so that the probability of false verification is zero.

The verification process in VB algorithms can be seen as a message-passing procedure. In general, a variable node sends its current state (either verified or unverified) to its neighboring check nodes along with its value (if verified). A check node processes the received messages and subsequently sends some messages to its neighboring variable nodes. Each unverified variable node decides on its next state, either verified or unverified, based on the received messages from check nodes. The process of passing messages between variable nodes and check nodes continues until all variable nodes are verified, or no variable node changes its state.

In message-passing algorithms, a node can take two approaches in order to produce a message based on the set of received messages. In the first approach, the outgoing message is a function of \emph{all} received messages. In this case, all messages leaving a node at a certain iteration are the same. In the second approach, the message passed from node $a$ to node $b$ in the bigraph, is a function of all the received messages by node $a$ except the received message from node $b$. Therefore, the outgoing messages of a node at a certain iteration may be different, depending on the received messages. In the context of VB algorithms, the first approach is known as \emph{node-based (NB)}, while the second approach is called \emph{message-based (MB)} \cite{ZP07J,ZP09}.\footnote{In the context of iterative decoding algorithms, NB and MB approaches are known as non-extrinsic and extrinsic message-passing, respectively \cite{BG96}.} So, for a NB-VB algorithm, the state of a variable node is reported identically by all its outgoing messages, while in an MB-VB algorithm, different states may be reported by different outgoing messages from a variable node.

As noted in \cite{ZP07J}, the authors in \cite{LM05} defined the two VB algorithms using the NB representation but analyzed them using the MB representation. In \cite{ZP07J}, the authors proved that for one of the VB algorithms, the NB and MB versions perform the same, but for the other VB algorithm, the NB version outperforms the MB one. In compressed sensing, this implies that NB versions, in general, have higher success thresholds; i.e., can successfully recover signals with larger density ratios \cite{ZP08}.

A well-known method to analyze iterative message-passing algorithms in coding theory is density evolution \cite{RU01}. In density evolution, the distribution of messages is tracked with the iteration number. The evolution of the message distributions with iterations will then reveal important properties of the decoding algorithm such as {\em decoding threshold} and {\em convergence speed} \cite{RU01}. The derivation of the message distribution however, requires the independence among the incoming messages to a node. The analysis is thus only applicable to extrinsic message-passing algorithms (MB decoders). To extend density evolution to NB algorithms, Zhang and Pfister \cite{ZP07J} used a system of differential equations as the analytical tool. Applying their analysis to $(d_v,d_c)$ graphs, the number of differential equations is $\ma{O}(d_v^3+d_c^2)$. This rapidly becomes too complex to handle for large values of $d_v$ and $d_c$.\footnote{The number of differential equations for a (3,6) bigraph is about 30.} Numerical methods were used in \cite{ZP07J} to solve the system of differential equations and consequently evaluate the performance of the NB algorithms. It is important to note that the authors in \cite{ZP07J} analyzed a serial version of NB-VB algorithms, i.e., the version which allows only one variable node to be verified in each iteration. The total number of iterations in their calculations is thus equal to the size of the signal $n$. Another important issue regarding the analysis of \cite{ZP07J} is that while the validity of the analysis is proved for the asymptotic scenario of $n \rightarrow \infty$, the numerical results are highly dependent on the selected, large but still finite, value of $n$.
%%%%%%%%%%%%%%%%%%%%%%%%%%%%%%%%%%%%%%%%%%%%%%%%%%%%%%%%%%%%%%%%%%%%%%%%%%%%%%%%%%%%%%%%%%%%%%%%%%%%%%%%%%%%%
%%%%%%%%%%%%%%%%%%%%%%%%%%%%%%%%%%%%%%%%%%%%%%%%%%%%%%%%%%%%%%%%%%%%%%%%%%%%%%%%%%%%%%%%%%%%%%%%%%%%%%%%%%%%%
\section{VB Algorithms, Verification Rules and False Verification}
\label{enc}
%%%%%%%%%%%%%%%%%%%%%%%%%%%%%%%%%%%%%%%%%%%%%%%%%%%%%%%%%%%%%%%%%%%%%%%%%%%%%%%%%%%%%%%%%%%%%%%%%%%%%%%%%%%%%
\subsection{VB Algorithms and Verification Rules}
\label{VBenc}
In compressive sensing, the decoder receives the vector of measurements and aims at estimating the original signal based on the knowledge of measurements and the sensing graph. In this section, we discuss four VB decoding algorithms.

The first algorithm, here referred to as ``Genie'', is a benchmark VB algorithm in which the support set of the signal is known to the decoder.\footnote{The Genie algorithm is similar to the peeling algorithm over the BEC proposed in \cite{LMSS01}.} We use the Genie algorithm and its analysis to motivate and explain the analytical framework. The success threshold associated with this algorithm serves as an upper bound for the performance of other VB algorithms.

In other recovery algorithms, the decoder has no information about the support set. The next two decoders considered in this paper, are the two main VB decoding algorithms in the context of CS. The first algorithm is referred to as LM and is the first algorithm discussed in \cite{ZP09}; LM1. The second main VB algorithm is the algorithm introduced in \cite{SBB206}, which is the same as the second algorithm discussed in \cite{ZP09}; LM2. We refer to this algorithm as SBB.

By the description given in Section \ref{VB}, the algorithm in \cite{XH07}, here referred to as XH, also falls into the category of VB algorithms, and can be also analyzed using the proposed framework. The details of the analysis for this algorithm however, is not included in this paper. We just report some numerical results on the success threshold and the convergence speed of this algorithm in Section \ref{simulation}.

In what follows, we give the general description of the aforementioned VB algorithms, as found in the literature (\cite{ZP07, ZP07J, ZP09, XH07}). We then use this description to discuss the issue of false verification. In Section \ref{Decoding}, we present the equivalent message-passing description of the VB algorithms.

In the VB algorithms, when a variable node is verified at an iteration, its verified value is subtracted from the value of its neighboring check nodes. The variable node, then, is removed from the sensing bigraph along with all its adjacent edges. Hence, all the neighboring check nodes of the verified variable node face a reduction in their degree. In the next iteration, some variable nodes may be verified based on the degree and/or the value of their neighboring check nodes. The rules based on which the variable nodes are verified at each iteration are called \emph{verification rules} and are as follows:

\begin{itemize}
	\item Zero Check Node (ZCN): If a check node has a zero value, all its neighboring variable nodes are verified with a zero value.
	\item Degree One Check Node (D1CN): If a check node has degree 1 in a graph, its unique neighboring variable node is verified with the value of the check node.
	\item Equal Check Nodes (ECN): Suppose we have $N$ check nodes with the same nonzero value, then 1) all variable nodes neighboring a subset of these $N$ check nodes (not all of them) are verified with the value zero; 2) if there exists a unique variable node neighboring all $N$ check nodes, then it is verified with the common value of the check nodes.
\end{itemize}

Verification rules ZCN and ECN are responsible for verifying variable nodes not in the support set. Since, the Genie algorithm has the complete knowledge of the support set, it has no need to apply these two rules. Hence, D1CN is the only rule used by the Genie. Other VB algorithms, each uses a combination of verification rules in order to verify and resolve unverified variable nodes. Assuming zero probability for false verification, the order in which the rules are applied does not affect the overall performance of the algorithm; it will only change the order in which variable nodes are verified. Verification rules adopted by different algorithms are summarized in Table \ref{verification_rules}.

\begin{table}[!h]
	\caption{Verification rules adopted in each VB algorithm}
	\centering	
%	{\footnotesize{
	\begin{tabular}{|l|c|c|c|}
		\hline
		 & ZCN & D1CN & ECN\\
		\hline
		\hline
		 Genie & Not Needed & \Checkmark & Not Needed\\
		\hline
		 LM & \Checkmark & \Checkmark & \XSolidBrush\\
		\hline
		 SBB & \Checkmark & \Checkmark & \Checkmark\\
		\hline
		 XH & \Checkmark & \XSolidBrush & \Checkmark\\
		\hline
	\end{tabular}
%	}}
	\label{verification_rules}
\end{table}

Based on Table \ref{verification_rules}, SBB applies the union of all rules to verify variable nodes. Therefore, this algorithm is expected to have the highest success threshold amongst the practical VB algorithms discussed here. This is verified in Section \ref{simulation}.

The ECN rule as stated above can not be easily captured in the analysis. Based on our extensive simulations, we conjecture that this recovery rule can be modified to read as follows (without affecting the asymptotic behavior of the recovery algorithms):

Modified Equal Check Nodes (MECN): Suppose we have $N$ check nodes with the same nonzero value. Then if there exists a unique variable node neighbor to all such check nodes, it is verified with the common value of the check nodes. It is only in this case that any other variable node connected to such check nodes is verified as zero.

%%%%%%%%%%%%%%%%%%%%%%%%%%%%%%%%%%%%%%%%%%%%%%%%%%%%%%%%%%%%%%%%%%%%%%%%%%%%%%%%%%%%%%%%%%%%%%%%%%%%%%%%%%%%%
\subsection{False Verification}
\label{FVSection}
Let $\ma{K}$ denote the set of nonzero variable nodes in the signal; the support set. Also, let $\ma{M}(c)$ denote the set of variable nodes neighbor to a check node $c$. Now, consider the following facts:

\begin{itemize}
\item[($1$)] Let $\mathcal{C}$ be an arbitrary subset of check nodes. If all the check nodes in $\ma{C}$  are neighbor to the same subset of nodes in $\mathcal{K}$, then all these check nodes have the same value.
\item[($2$)] Any check node with no neighbor in $\ma{K}$ has a zero value.
\end{itemize}

Verification rules ZCN and ECN in VB algorithms are designed based on the following assumptions:

\begin{itemize}
	\item[($1'$)] Let $\ma{C}'$ be any arbitrary subset of check nodes with the same value. Then all these check nodes are neighbor to the same subset of $\ma{K}$.
	\item[($2'$)] For any zero valued check node, none of its neighboring variable nodes belong to the set $\ma{K}$.
\end{itemize}

It is worth noting that the assumptions ($1'$) and ($2'$) are the converses of the facts (1) and (2), respectively. To any choice of distributions $f$ and $g$ for nonzero weights of the sensing graph and nonzero signal elements, respectively, corresponds a certain probability that the converses fail to hold. Those distributions which make the converses hold true with probability $1$ (almost surely), are of interest in this paper. In the following theorem, we give an example of distributions that make the statements ($1'$) and ($2'$) hold true almost surely.

\begin{theorem}
\label{unique}
Let $\bs{c}_i$ and $\bs{c}_j$ be two distinct check nodes and $\ma{V}_i$ and $\ma{V}_j$ be their corresponding set of neighboring variable nodes in $\ma{K}$; i.e., $\ma{V}_i=\ma{M}(\bs{c}_i)\cap \ma{K}$ and $\ma{V}_j=\ma{M}(\bs{c}_j)\cap \ma{K}$. Suppose that at least one of the distributions $f$ or $g$ described before is continuous. Then the statements ($1'$) and ($2'$), described above, are correct with probability one for $\bs{c}_i$ and $\bs{c}_j$.
\end{theorem}

To prove the theorem, we need the following fact.

\begin{fact}
\label{Uniqueness}
Let $x_i$ and $x_j$ be two independent samples drawn from a continuous distribution. It follows that:
\[
\Pr\left(x_i = x_j\right) = 0.
\]
Stated differently, no two independent samples of a continuous distribution will have the same value, almost surely. Moreover, for any constant $c$, we have
\[
\Pr\left(x_i = c\right) = 0.
\]
\end{fact}

\begin{IEEEproof}[Proof of Theorem \ref{unique}]
Let $\bs{G}$$=[w_{ij}]$ be the sensing matrix. The value of a check node $\bs{c}_i$ is $\sum_{j:\bs{v}_j\in\ma{M}(\bs{c}_i)}{w_{ij}\bs{v}_j}$. Therefore, if at least one of the following conditions is satisfied, then the value of a check node follows a continuous distribution:
\begin{enumerate}
	\item the nonzero elements of the sensing matrix (weights associated to the edges of the sensing graph), follow a continuous distribution $f$.
	\item the value of a nonzero variable node (a variable node in the support set $\ma{K}$) follows a continuous distribution $g$.
\end{enumerate}

Therefore, according to Fact~\ref{Uniqueness}, the subset of check nodes with the same value can not be independent almost surely. Their dependence implies that they are neighbor to the same set of nonzero variable nodes. This proves the statement ($1'$). Similarly, if a check node is neighbor to at least one nonzero variable node, its value follows a continuous distribution, which according to Fact~\ref{Uniqueness} is zero with probability zero. This proves the statement ($2'$).
\end{IEEEproof}

So, the continuity of $f$ or $g$ is a sufficient condition to have the probability of false verification equal to zero. In the rest of the paper, we assume that the statements ($1'$) and ($2'$) are correct with probability one and consequently, the probability of false verification for a variable node in any iteration of the VB algorithms is zero. Using the union bound, one can see that the probability of false verification in any iteration and also in the whole recovery algorithm is zero.
%%%%%%%%%%%%%%%%%%%%%%%%%%%%%%%%%%%%%%%%%%%%%%%%%%%%%%%%%%%%%%%%%%%%%%%%%%%%%%%%%%%%%%%%%%%%%%%%%%%%%%%%%%%%%
%%%%%%%%%%%%%%%%%%%%%%%%%%%%%%%%%%%%%%%%%%%%%%%%%%%%%%%%%%%%%%%%%%%%%%%%%%%%%%%%%%%%%%%%%%%%%%%%%%%%%%%%%%%%%
\section{VB Recovery Algorithms as Message-Passing Algorithms}
\label{Decoding}
%%%%%%%%%%%%%%%%%%%%%%%%%%%%%%%%%%%%%%%%%%%%%%%%%%%%%%%%%%%%%%%%%%%%%%%%%%%%%%%%%%%%%%%%%%%%%%%%%%%%%%%%%%%%%
\subsection{Definitions and Setup}
There are a number of VB decoding algorithms that can be formulated as \emph{node-based message-passing} (NB-MP) algorithms. These are the algorithms that are of interest to us in this paper. Each algorithm works in iterations through exchanging messages between the check nodes and the variable nodes along the edges in the graph. Any message sent from a variable node to its neighboring check nodes belongs to an alphabet set $\mathcal{M}:\{0,1\}\times \mathbb{R}$. The first coordinate of such a message is a status flag, sometimes referred to as ``recovery flag'', taking binary values. The flag indicates the verification status of the variable node. If this flag is $0$, then the variable node is not verified. If, on the other hand, the flag is $1$, then the variable node has been verified. In this case, the second coordinate, which is a real number, is interpreted as the verified value of the variable node.

Similarly, any message sent from a check node to all its neighboring variable nodes belongs to an alphabet set $\mathcal{O}: \mathbb{Z}^{{}^{+}}\times \mathbb{R}$. The first coordinate of such a message indicates the number of unverified variable nodes neighbor to the check node. The first coordinate is in fact the \emph{degree} of the check node in the subgraph induced by the unverified variable nodes. The second coordinate indicates the current value of the check node, i.e., the result of the linear combination of the unverified neighboring variable nodes.

The edges, in NB-MP algorithms, do not simply forward messages from check nodes to variable nodes and vice versa. Instead, based on the traveling direction of the message, edges multiply or divide the second coordinate of the message by their associated weight. More specifically, if the message is sent from a variable node to a check node, its second coordinate is multiplied by the weight. The second coordinate of the message is divided by the weight, if the message is sent from a check node to a variable node. So, although messages generated by a node (either variable node or check node) are sent identically over all adjacent edges, the fact that the edges may have different weights will result in different messages being received at the destination nodes. All such messages are independent if the weights associated with the corresponding edges are independent.

Any iteration $\ell\geq 1$ in NB-VB algorithms, consists of two rounds, each with two half-rounds. In each round, every check node processes all received messages from the previous round together with its associated measurement and sends out a message from the alphabet $\mathcal{O}$ to all its neighboring variable nodes (first half-round). In the second half-round, each (unverified) variable node decides on its next state by processing all its received messages. Whichever the decision is, the variable node sends back a message, from the alphabet $\mathcal{M}$, to all its neighboring check nodes. So, a round starts with check nodes processing the received messages from neighboring variable nodes, proceeds with the transmission of messages from check nodes to variable nodes, continues by variable nodes processing the received messages from neighboring check nodes, and ends with the transmission of messages from variable nodes to check nodes. The two rounds in each iteration follow the same general structure. They only differ in the processing carried out in the variable nodes.

Let $\Phi_v^{(1,\ell)}: \ma{O}^{d_v} \rightarrow \ma{M}$ and $\Phi_v^{(2,\ell)}: \ma{O}^{d_v} \rightarrow \ma{M}$, $\ell \in \mathbb{N}$, represent the mappings used at any \emph{unverified} variable node to map the incoming messages to the outgoing message in the first and the second round of iteration $\ell$, respectively. Obviously, due to the verification-based nature of the algorithms, when a variable node becomes verified at an iteration, its outgoing message remains unchanged, irrespective of its incoming messages. In contrast to the variable nodes, the mapping function used in check nodes is identical for both the first and the second round of each iteration. Every check node $c_i, i \in [m]$ has an associated received measurement $\bs{c}_i$, a random variable taking values in $\mathbb{R}$. So, we use the notation $\Phi_c^{(\ell)}: \mathbb{R} \times \ma{M}^{d_c} \rightarrow \ma{O}$, $\ell \in \mathbb{N}$, to denote the mapping function used in all check nodes at iteration $\ell$. For the sake of completeness, let $\Phi_v^{(0)} = \Phi_v^{(2,0)}: \ma{O}^{d_v} \rightarrow \ma{M}$ and $\Phi_c^{(0)}: \mathbb{R} \rightarrow \ma{O}$ represent the mappings used, respectively in all variable nodes and check nodes at iteration $0$. This iteration consists of only one round. For the VB algorithms under consideration, the mapping functions in the variable nodes and check nodes are not a function of the iteration number. Therefore, we omit the superscript $(\ell)$ henceforth.

In what follows, we describe VB algorithms of Section \ref{enc} as message-passing algorithms with the general structure explained above.\footnote{It is worth mentioning that the message-passing description of the NB-VB algorithms, presented in Section \ref{RecAlg}, is only valid for the cases in which the nonzero weights of the sensing graph are drawn from an uncountable or countably infinite alphabet set. If the elements of the sensing matrix are drawn from a finite alphabet set, such as binary $0$ and $1$, the outgoing messages from a check node should also include the list of all unverified variable nodes neighbor to the check node. The mapping function in the variable nodes should also change in order to use the extra information in the incoming messages.}
%%%%%%%%%%%%%%%%%%%%%%%%%%%%%%%%%%%%%%%%%%%%%%%%%%%%%%%%%%%%%%%%%%%%%%%%%%%%%%%%%%%%%%%%%%%%%%%%%%%%%%%%%%%%%
\subsection{Message-Passing Description of Recovery Algorithms}
\label{RecAlg}
To describe the four VB recovery algorithms using the message-passing approach, we need to define the mappings $\Phi_v^{(1)}$, $\Phi_v^{(2)}$ and $\Phi_c$. Mapping $\Phi_v^{(1)}$ embeds the verification rules D1CN and ECN, while the mapping $\Phi_v^{(2)}$ embeds the ZCN rule. To make the description of mappings $\Phi_v$ and $\Phi_c$ simpler, we introduce some notations to represent the incoming messages to variable and check nodes from the alphabet sets $\ma{O}$ and $\ma{M}$, respectively. A message $\bs{o} \in \ma{O}$, incoming to a variable node, is an ordered pair of elements $(d,\xi)$, where $d \in \mathbb{Z}^{^{+}}, \xi \in \mathbb{R}$. A message $\bs{m} \in \ma{M}$, incoming to a check node, is an ordered pair of elements $(s,\omega)$, where $s \in \{0,1\}, \omega \in \mathbb{R}$. Moreover, we assume that there is an arbitrary numbering for edges adjacent to a node (either variable node or check node). So, we use the notations $\bs{o}_{i}, i \in [d_v]$ and $\bs{m}_{j}, j \in [d_c]$, to denote the incoming messages to variable nodes and check nodes, respectively.

At iteration zero, all variable nodes are unverified and there is no received message at the check nodes. At this stage, all check nodes send their corresponding measurements along with their degree ($d_c$) to their neighboring variable nodes. For the following iterations $\ell \geq 1$, the mapping function at any check node $c_i$ is as follows:
\[
\Phi_c(\bs{c}_i, \bs{m}_1, \cdots, \bs{m}_{d_c}) = (d_c - \ds_{i=1}^{d_c}{s_i},\bs{c}_i - \ds_{i=1}^{d_c}{s_i \omega_i}),
\]
where, $\bs{c}_i$ is the measurement associated with the check node $c_i$, and $\bs{m}_i = (s_i,\omega_i)$ is the message received along the $i$th edge. The mapping functions $\Phi_v^{(1)}$, $\Phi_v^{(2)}$ are algorithm dependent and are discussed for each VB algorithm separately next.

The decoder stops at an iteration $\ell$, $\ell \geq 1$, if the algorithm makes no further progress, i.e., the set of verified variable nodes are the same for the two consecutive iterations $\ell-1$ and $\ell$. Equivalently, the algorithm stops if the messages sent from variable nodes to check nodes, and also from check nodes to variable nodes, are the same for two consecutive iterations $\ell$ and $\ell-1$. At this point, if the decoder is able to verify all the variable nodes, then the decoding is called successful. Otherwise, the decoder will declare a failure.

\textbf{Genie}\\
In this algorithm, the decoder has the knowledge of the support set $\ma{K}$. So, the verification rules ZCN and ECN are not needed for this algorithm. Hence, each iteration in this algorithm consists of only one round, in which one verification rule (D1CN) is applied to all variable nodes. For variable nodes not in the support set, the outgoing message in all iterations is fixed and equals $\bs{m} = (1,0)$.

For any variable node in the support set, the mapping $\Phi_v(\bs{o}_1, \cdots, \bs{o}_{d_v})$ is defined based on the following rules.
\begin{itemize}
	\item If among all received messages (from the neighboring check node), there exists only one message, say $\bs{o}_{i}, i \in [d_c]$, such that $\bs{o}_{i} = (1,\xi_i), \xi_i\in \mathbb{R}$, then $\Phi_v(\bs{o}_1, \cdots, \bs{o}_{d_v}) = (1,\xi_i)$. In this case, the variable node is verified with the value $\xi_i$.
	\item If multiple messages exist in the form $(1,\xi)$ (any $\xi \in \mathbb{R}$), then choose one at random, say $\bs{o}_{i} = (1,\xi_i), \xi_i\in \mathbb{R}$, and set $\Phi_v(\bs{o}_1, \cdots, \bs{o}_{d_v}) = (1,\xi_i)$. In this case, the variable node is verified with the value $\xi_i$.
	\item If none of the above happens, then $\Phi_v(\bs{o}_1, \cdots, \bs{o}_{d_v}) = (0,0)$. In this case, the variable node is still unverified.
\end{itemize}

\textbf{LM}\\
For any unverified variable node in this algorithm, the mappings $\Phi^{(1)}_v$ and $\Phi^{(2)}_v$ are defined according to the following rules.
\begin{itemize}
	\item $\Phi^{(1)}_v(\bs{o}_1, \cdots, \bs{o}_{d_v})$:
	\begin{itemize}
		\item If among all received messages (from the neighboring check nodes), there exists only one message, say $\bs{o}_{i}, i \in [d_c]$, such that $\bs{o}_{i} = (1,\xi_i), \xi_i\in \mathbb{R}$, then $\Phi^{(1)}_v(\bs{o}_1, \cdots, \bs{o}_{d_v}) = (1,\xi_i)$. In this case, the variable node is verified with the value $\xi_i$.
		\item If multiple messages exist in the form $(1,\xi)$ (any $\xi \in \mathbb{R}$), then choose one at random, say $\bs{o}_{i} = (1,\xi_i), \xi_i\in \mathbb{R}$, and set $\Phi^{(1)}_v(\bs{o}_1, \cdots, \bs{o}_{d_v}) = (1,\xi_i)$. In this case, the variable node is verified with the value $\xi_i$.
		\item If none of the above happens, then $\Phi^{(1)}_v(\bs{o}_1, \cdots, \bs{o}_{d_v}) = (0,0)$. In this case, the variable node is still unverified.
	\end{itemize}
	\item $\Phi^{(2)}_v(\bs{o}_1, \cdots, \bs{o}_{d_v})$:
	\begin{itemize}
		\item If there exists at least one message $\bs{o}_i$ such that $\bs{o}_i = (d_i,0)$ (for any $d_i \in \mathbb{Z}^{^{+}}$), then \\ $\Phi^{(2)}_v(\bs{o}_1, \cdots, \bs{o}_{d_v}) = (1,0)$. In this case, the variable node is verified with the value equal to $0$.
		\item If no incoming message exists in the form $(d_i,0)$ (for any $d_i \in \mathbb{Z}^{^{+}}$), then \\
		$\Phi^{(2)}_v(\bs{o}_1, \cdots, \bs{o}_{d_v}) = (0,0)$. In this case, the variable node is still unverified.
	\end{itemize}
\end{itemize}

\textbf{SBB}\footnote{The original recovery algorithm introduced in \cite{SBB206} has a computational complexity of $O(m\cdot\log m)$ (which translates to $O(n\cdot\log n)$ when using biregular graphs of fixed degrees). It is easy to prove that the message-passing description provided here does not change the recovery capability of the algorithm but results in the reduction of decoding complexity from $O(n\cdot\log n)$ to $O(n)$.}\\
For any unverified variable node in this algorithm, the mappings $\Phi^{(1)}_v$ and $\Phi^{(2)}_v$ are defined by the following rules.
\begin{itemize}
	\item $\Phi^{(1)}_v(\bs{o}_1, \cdots, \bs{o}_{d_v})$:
	\begin{itemize}
		\item If among all received messages (from the neighboring check nodes), there exists only one message, say $\bs{o}_{i}, i \in [d_c]$, such that $\bs{o}_{i} = (1,\xi_i), \xi_i\in \mathbb{R}$, then $\Phi^{(1)}_v(\bs{o}_1, \cdots, \bs{o}_{d_v}) = (1,\xi_i)$. In this case, the variable node is verified with the value $\xi_i$.
		\item If there exist $N$ messages ($2 \leq N \leq d_v$) $\bs{o}_{i_1}=(d_{i_1},\xi_{i_1})$, $\bs{o}_{i_2}=(d_{i_2},\xi_{i_2})$, $\cdots$, $\bs{o}_{i_N}=(d_{i_N},\xi_{i_N})$, such that $\xi_{i_1} = \xi_{i_2} = \cdots = \xi_{i_N}$, then $\Phi^{(1)}_v(\bs{o}_1, \cdots, \bs{o}_{d_v}) = (1,\xi_{i_1})$. In this case, the variable node is verified with the common value $\xi_{i_1}$.\footnote{We note that the message received by the variable node equals the message sent from the check node divided by the weight of the connecting edge. Therefore, receiving two messages with the same value would imply that, almost surely, the unverified variable node under consideration is the unique nonzero variable node neighbor to the $N$ check nodes. Other unverified variable nodes neighbor to these $N$ check nodes do not belong to the support set and should be verified with a value equal to zero. This, however, happens in the next round.}
		\item If a variable node is verified to different values according to verification rules above, then choose one at random and generate the outgoing message accordingly.
		\item If none of the above happens, then $\Phi^{(1)}_v(\bs{o}_1, \cdots, \bs{o}_{d_v}) = (0,0)$. In this case, the variable node is still unverified.
	\end{itemize}
	\item $\Phi^{(2)}_v(\bs{o}_1, \cdots, \bs{o}_{d_v})$:
	\begin{itemize}
		\item If there exists at least one message $\bs{o}_i$ such that $\bs{o}_i = (d_i,0)$ (for any $d_i \in \mathbb{Z}^{^{+}}$), then \\ $\Phi^{(2)}_v(\bs{o}_1, \cdots, \bs{o}_{d_v}) = (1,0)$. In this case, the variable node is verified with the value equal to $0$.
		\item If no incoming message exists in the form $(d_i,0)$ (for any $d_i \in \mathbb{Z}^{^{+}}$), then \\
		$\Phi^{(2)}_v(\bs{o}_1, \cdots, \bs{o}_{d_v}) = (0,0)$. In this case, the variable node is still unverified.
	\end{itemize}
\end{itemize}

\textbf{XH}\\
For any unverified variable node in this algorithm, the mappings $\Phi^{(1)}_v$ and $\Phi^{(2)}_v$ are defined according to the following rules.
\begin{itemize}
	\item $\Phi^{(1)}_v(\bs{o}_1, \cdots, \bs{o}_{d_v})$:
	\begin{itemize}
		\item If there exist $M$ messages ($\lceil d_v/2 \rceil \leq M \leq d_v$) $\bs{o}_{i_1}=(d_{i_1},\xi_{i_1})$, $\bs{o}_{i_2}=(d_{i_2},\xi_{i_2})$, $\cdots$, $\bs{o}_{i_M}=(d_{i_M},\xi_{i_M})$, such that $\xi_{i_1} = \xi_{i_2} = \cdots = \xi_{i_M}$, then $\Phi^{(1)}_v(\bs{o}_1, \cdots, \bs{o}_{d_v}) = (1,\xi_{i_1})$. In this case, the variable node is verified with the common value $\xi_{i_1}$.
		\item If a variable node is verified to different values according to the verification rule above, i.e., if two groups of messages both at least of size $d_v/2$ satisfy the above condition, then choose one at random and generate the outgoing message accordingly.
		\item If none of the above happens, then $\Phi^{(1)}_v(\bs{o}_1, \cdots, \bs{o}_{d_v}) = (0,0)$. In this case, the variable node is still unverified.
	\end{itemize}
	\item $\Phi^{(2)}_v(\bs{o}_1, \cdots, \bs{o}_{d_v})$:
	\begin{itemize}
		\item If there exists at least one message $\bs{o}_i$ such that $\bs{o}_i = (d_i,0)$ (for any $d_i \in \mathbb{Z}^{^{+}}$), then \\ $\Phi^{(2)}_v(\bs{o}_1, \cdots, \bs{o}_{d_v}) = (1,0)$. In this case, the variable node is verified with the value equal to $0$.
		\item If no incoming message exists in the form $(d_i,0)$ (for any $d_i \in \mathbb{Z}^{^{+}}$), then \\
		$\Phi^{(2)}_v(\bs{o}_1, \cdots, \bs{o}_{d_v}) = (0,0)$. In this case, the variable node is still unverified.
	\end{itemize}
\end{itemize}
%%%%%%%%%%%%%%%%%%%%%%%%%%%%%%%%%%%%%%%%%%%%%%%%%%%%%%%%%%%%%%%%%%%%%%%%%%%%%%%%%%%%%%%%%%%%%%%%%%%%%%%%%%%%%
\subsection{A Short Note on False Verification}
In the above description of recovery algorithms, there may be cases where a variable node can be verified to different values by different rules. Using the same assumption made in Section \ref{FVSection}, it is easy to see that the probability of this event is equal to zero. In such cases, we have thus assumed that the variable node is verified by one of the rules selected randomly. Clearly, the probability of false verification as a result of such selection is zero.
%%%%%%%%%%%%%%%%%%%%%%%%%%%%%%%%%%%%%%%%%%%%%%%%%%%%%%%%%%%%%%%%%%%%%%%%%%%%%%%%%%%%%%%%%%%%%%%%%%%%%%%%%%%%%
%%%%%%%%%%%%%%%%%%%%%%%%%%%%%%%%%%%%%%%%%%%%%%%%%%%%%%%%%%%%%%%%%%%%%%%%%%%%%%%%%%%%%%%%%%%%%%%%%%%%%%%%%%%%%
\section{Asymptotic Analysis Framework}
\label{analysis}
%%%%%%%%%%%%%%%%%%%%%%%%%%%%%%%%%%%%%%%%%%%%%%%%%%%%%%%%%%%%%%%%%%%%%%%%%%%%%%%%%%%%%%%%%%%%%%%%%%%%%%%%%%%%%
In this section, we first show that (i) the performance of a realization of the sensing graph, with a certain selection of the edge weights for the recovery of a realization of the input signal concentrates around the average performance of the ensemble (where the average is taken over all the elements in the ensemble $\mathcal{G}^{n}_f(d_v,d_c)\times \mathcal{V}^{n}_g(\alpha)$, for given probability distribution functions $f,g$, and given constant parameters $d_v,d_c$ and $\alpha$), as $n$ tends to infinity, and (ii) the average performance of the ensemble, as $n$ goes to infinity, converges to the performance of the cycle-free case defined as follows.\footnote{Our method of proof is very similar to that of \cite{RU01}, though due to the differences in the nature of the problems (channel coding vs. compressed sensing) and the difference in the update rules at the graph nodes, some arguments are revised and some new components are added to the proof.}

Let $\mathcal{N}_v^{2\ell}$ be the neighborhood of node $v$ of depth $2\ell$, i.e., the subgraph consisting of the variable node $v$ and all those nodes that are connected to $v$ with any path of length less than or equal to $2\ell$. We say that we are working under the cycle-free assumption when for a fixed $\ell$, and for every $v$, $\mathcal{N}_v^{2\ell}$ is tree-like. Similar to \cite{RU01}, the concentration result presented here is valid for any iteration number $\ell$. We thus often omit the superscript $\ell$ in the notations.

Following the concentration results, we present the asymptotic analysis of the Genie algorithm. The analysis provides us with the average ensemble performance for the asymptotic case of $n \rightarrow \infty$. We then generalize the concepts used in the analysis of Genie and analyze LM and SBB algorithms. The analysis of XH is similar and is omitted to prevent redundancy. However, we shall report the thresholds of this decoder for different biregular graphs in section \ref{simulation}.
%%%%%%%%%%%%%%%%%%%%%%%%%%%%%%%%%%%%%%%%%%%%%%%%%%%%%%%%%%%%%%%%%%%%%%%%%%%%%%%%%%%%%%%%%%%%%%%%%%%%%%%%%%%%%
\subsection{Concentration Results and Convergence to Cycle-Free Case}
Consider a weighted graph selected at random from $\mathcal{G}_f^n(d_v,d_c)$. Such a graph can be represented by its corresponding unweighted graph $G$ (from the ensemble $\mathcal{G}^n(d_v,d_c)$) along with its vector of edge weights $\bs{w}$ (the vector of elements, with an arbitrary fixed order, in an $m\times n$ matrix $\bs{W}$ from the ensemble $\ma{W}_{f}^{m\times n}$). Also consider an input signal vector $\bs{v}$ chosen randomly from $\mathcal{V}_g^n(\alpha)$. Suppose that a VB algorithm is applied to the measurement vector $\bs{c}=\bs{G} \bs{v}$ to recover $\bs{v}$ iteratively, where $\bs{G}$ is the biadjacency matrix of the chosen weighted graph. For this scenario, let $\beta^{(\ell)}$ ($:=\beta^{(\ell)}(G,\bs{w},\bs{v})$) be the fraction of unverified nonzero variable nodes at the beginning of iteration $\ell$, i.e., the fraction of variable to check node messages passed along the edges of the chosen weighted graph with unverified status (sent by nonzero variable nodes); further, let $\mathbf{E}[\beta^{(\ell)}]$ denote the expected value of $\beta^{(\ell)}$, where the expectation is taken over the ensembles $\mathcal{G}_f^n(d_v,d_c)$ and $\mathcal{V}_g^n(\alpha)$. Now, consider the corresponding cycle-free case, and let $\alpha^{(\ell)}$ be the expected number of messages with unverified status passed along an edge emanating from a nonzero variable node with a tree-like neighborhood of depth at least $2\ell$ at the $\ell$th iteration. Here, again, the expectation is taken over the ensembles $\ma{W}_{f}^{m\times n}$ and $\mathcal{V}_g^n(\alpha)$.

In the subsequent subsection, we will show how $\alpha^{(\ell)}$ can be calculated. It should be clear that $\alpha^{(\ell)}$, being defined as the ``average'' over the ensemble of weights and input vectors, is the same as the ``probability'' that a message from a non-zero variable node with a tree-like neighborhood of depth at least $2\ell$, at the $\ell$th iteration, carries an unverified status. In this section, we use the interpretation of $\alpha^{(\ell)}$ as an average. The interpretation of $\alpha^{(\ell)}$ as a probability will be used in the analysis section. In the following, we will show that over all realizations, with high probability, $\beta^{(\ell)}$ does not deviate much from $\mathbf{E}[\beta^{(\ell)}]$, and $\mathbf{E}[\beta^{(\ell)}]$, itself, is not far from $\alpha^{(\ell)}$, as $n$ tends to infinity.

\begin{theorem}\label{Concentration}Over the probability space of all weighted graphs $\mathcal{G}^{n}_f(d_v,d_c)$, and all signal inputs $\mathcal{V}^{n}_g(\alpha)$, for a fixed $\ell$, letting $\beta^{(\ell)}$ and $\alpha^{(\ell)}$ be defined as above, there exist positive constants $\mu(d_v,d_c,\ell)$ and $\gamma(d_v,d_c,\ell)$, such that (i) for any $\epsilon>0$, \begin{equation}\label{EqDot}\Pr\left[\left|\beta^{(\ell)}-\mathbf{E}[\beta^{(\ell)}]\right|>{\epsilon}/{2}\right]\leq 2e^{-\epsilon^2 n/\mu},\end{equation} and (ii) for any $\epsilon>0$, and $n>2{\gamma}/{\epsilon}$, \begin{equation}\label{EqStar}\left|\mathbf{E}[\beta^{(\ell)}]-\alpha^{(\ell)}\right|<{\epsilon}/{2}.\end{equation}\end{theorem}

Note that combining~\eqref{EqDot} and~\eqref{EqStar}, the following holds: for any $\epsilon>0$, and $n>2\gamma/\epsilon$, \[\Pr\left[\left|\beta^{(\ell)}-\alpha^{(\ell)}\right|>{\epsilon}\right]\leq 2e^{-\epsilon^2 n/\mu}.\] Hereafter, for ease of notation, we drop the superscript $\ell$, as we are studying the parameters of interest at a fixed iteration $\ell$. 

\begin{proof}We start by proving~\eqref{EqDot}. Let the triple $T:=(G,\bs{w},\bs{v})$ represent one particular realization $G$, $\bs{w}$, and $\bs{v}$ of the ensembles of graphs, edge weights and input vectors, respectively. For any $i$, $0\leq i\leq (2d_v+1)n:=\xi$, the $i$th element of $T$, with an arbitrary order which is fixed during the discussion, is referred to as the $i$th coordinate of $T$. Consider two arbitrary realizations $T'$ and $T''$. Let the symbol ``$\equiv_i$,'' for all $0\leq i\leq \xi$, be a binary relation as follows: $T'\equiv_i T''$ implies $T'\equiv_{i-1} T''$, where $T'\equiv_i T''$ if the first $i$ coordinates of the triples $T'$ and $T''$ are the same.\footnote{An example: If, for $i=n d_v$, we have $T'\equiv_i T''$, then the two realizations $T'$ and $T''$ have the same graph, but may have different edge weights and different input vectors.} Suppose that we first expose the $n d_v$ edges of the graph one at a time. In the next $n d_v$ steps, we proceed with exposing the $n d_v$ edge weights one by one, and finally, we expose the $n$ variable nodes' input values one at a time. Then, $T'\equiv_i T''$ if and only if the information that we reveal in the first $i$ steps of exposure for both triples is the same.

We construct a martingale sequence $\beta_0,\beta_1,...,\beta_{\xi}$, by defining \[\beta_i(T)= \mathbf{E}[\beta(T')|T'\equiv_i T].\] ($\beta_0$ is a constant, i.e., the expected value of $\beta(T')$ over all graphs $G'$, all edge weights $\bs{w}'$, and all input vectors $\bs{v}'$; $\beta_{\xi}$ is $\beta(T)$ itself.) By the application of Azuma's inequality (see \cite[Chapter~7]{AS:2008}), one can give an upper bound on \[\Pr\left[\left|\beta(T)-\mathbf{E}\left[\beta(T')\right]\right|>\epsilon/2\right]=\Pr\left[\left|\beta_{\xi}-\beta_0\right|>\epsilon/2\right],\] so long as for all $0\leq i<\xi$, $\left|\beta_{i+1}-\beta_{i}\right|\leq \Delta_i$, for some bounded $\Delta_i$, as $n$ tends to infinity.

In the following, we find $\Delta_i$, for all $0\leq i<\xi$, for the Genie algorithm, and by following similar steps, one can find $\Delta_i$ for the other VB algorithms. We will explain the details of the proof for Genie, and the proofs for the other algorithms will not be presented as the method of the proofs is quite similar.

First, consider the steps where we expose the edges of the graph, i.e., for all $0 \leq i < n d_v$. We want to upper bound \[\left|\beta_{i+1}(T)-\beta_i(T)\right|.\] Let $\mathcal{G}(G,i)$ be the set of graphs in the ensemble $\mathcal{G}^n(d_v,d_c)$, such that their first $i$ edges are the same as the edges in $G$, and let $\mathcal{G}_j(G,i)$ be the subset of $\mathcal{G}(G,i)$ whose $(i+1)$th edge from the perspective of variable nodes is the $j$th edge from the perspective of check nodes. It should be clear that $\mathcal{G}(G,i)=\cup_{j\in J_i}\mathcal{G}_j(G,i)$, where $J_i$ is the set of indices of those edges from the perspective of check nodes that have not been exposed before revealing the $(i+1)$th edge from the perspective of variable nodes. Thus, by definition, 
\begin{eqnarray}
\beta_i(T) &=& \mathbf{E}\left[\beta(T')|G'\in\mathcal{G}(G,i)\right] \nonumber \\ &=&\sum_{j\in J_i}\mathbf{E}\left[\beta(T')|G'\in\mathcal{G}_j(G,i)\right] \nonumber \\ & & \cdot\Pr\left[G'\in \mathcal{G}_j(G,i)|G'\in\mathcal{G}(G,i)\right], \label{EQ1} 
\end{eqnarray} 
and 
\[\beta_{i+1}(T)=\mathbf{E}\left[\beta(T')|G'\in\mathcal{G}_j(G,i)\right],\] 
for some $j\in J_i$. Since $\sum_{j\in J_i}{\Pr[G'\in \mathcal{G}_j(G,i)|G'\in \mathcal{G}(G,i)]}=1$, by using \eqref{EQ1}, one can show that 
\begin{eqnarray}
\beta_i(T) &\geq & \min_{j\in J_i}\mathbf{E}\left[\beta(T')|G'\in\mathcal{G}_j(G,i)\right] \nonumber \\ &=& \mathbf{E}\left[\beta(T')|G'\in\mathcal{G}_k(G,i)\right], \label{EQ2} 
\end{eqnarray} 
for some $k\in J_i$. Therefore, 
\begin{eqnarray}\lefteqn{\left|\beta_{i+1}(T)-\beta_i(T)\right|} \nonumber \\ &\leq& \max_{j\in J_i}\left|\mathbf{E}\left[\beta(T')|G'\in\mathcal{G}_j(G,i)\right]-\beta_i(T)\right| \label{EQ3} \\ &\leq & \max_{j,k\in J_i}\left|\mathbf{E}\left[\beta(T')|G'\in\mathcal{G}_j(G,i)\right] - \mathbf{E}\left[\beta(T')|G'\in\mathcal{G}_k(G,i)\right]\right|. \label{EQ4} 
\end{eqnarray} 
Then, for all $1\leq j,k \leq m d_c$, we need to bound the right hand side of the above inequality.

Let $\phi_{j,k}:\mathcal{G}_j(G,i)\rightarrow\mathcal{G}_k(G,i)$ be a map such that for any given graph $H\in\mathcal{G}_j(G,i)$, the graph $H':=\phi_{j,k}(H)$ is the same as $H$, except in one pair of edges, i.e., if the $j$th and the $k$th edges in $H$ from the perspective of check nodes are the $(i+1)$th and the $i'$th edges from the perspective of variable nodes, respectively, then, these two edges will be the $i'$th and the $(i+1)$th edges in $H'$ from the perspective of variable nodes, respectively. By construction, $\phi_{j,k}$ is a bijection, and it preserves probabilities. Thus, $\mathbf{E}[\beta(G',\bs{w}',\bs{v}')|G'\in\mathcal{G}_k(G,i)]=\mathbf{E}[\beta(\phi_{j,k}(G'),\bs{w}',\bs{v}')|G'\in\mathcal{G}_j(G,i)]$. Further, since the graphs $H$ and $H'$ defined as above are different only in one pair of edges, we will show later that for each VB algorithm of interest, regardless of the choices of $\bs{w}$ and $\bs{v}$, the difference between $\beta(H,\bs{w},\bs{v})$ and $\beta(H',\bs{w},\bs{v})$ is bounded from above. To be precise, in the case of the Genie algorithm, for all edge weights $\bs{w}$, and for all inputs $\bs{v}$, we can write
\begin{equation}
\label{GenieBound}\left|\beta(H,\bs{w},\bs{v})-\beta(\phi_{j,k}(H),\bs{w},\bs{v})\right|\leq {d_v^{\ell}}/{n}.
\end{equation} 
The proof of inequality~\eqref{GenieBound}, and similar results for the cases of LM and SBB algorithms are given in Appendix~\ref{AppD3}.\footnote{Arguments similar to what was used in upper bounding the difference between $\beta(T)$ and $\beta(T')$, when $G$ and $G'$ are different in one and only one pair of edges, will be used to give an upper bound on $\beta(T)$ and $\beta(T')$, when $(G,\bs{w})=(G',\bs{w}')$, but $\bs{v}$ and $\bs{v}'$ are different for one and only one variable node. Such a bound will be used later on in the proof.} By~\eqref{GenieBound}, any pair $\beta(H,\bs{w},\bs{v})$ and $\beta(\phi_{j,k}(H),\bs{w},\bs{v})$ has a bounded difference and thus, for any $j,k\in J_i$, \begin{eqnarray}\lefteqn{\left|\mathbf{E}[\beta(T')|G'\in\mathcal{G}_j(G,i)] - \mathbf{E}[\beta(T')|G'\in\mathcal{G}_k(G,i)]\right|} \nonumber \\ & & = \left|\mathbf{E}[\beta'|G'\in\mathcal{G}_j(G,i)] - \mathbf{E}[\beta'_{\phi_{j,k}}|G'\in\mathcal{G}_j(G,i)]\right| \nonumber \\ & & =\left|\mathbf{E}[\beta'-\beta'_{\phi_{j,k}}|G'\in\mathcal{G}_j(G,i)]\right| \nonumber \\ & & \leq \mathbf{E}[|\beta'-\beta'_{\phi_{j,k}}||G'\in\mathcal{G}_j(G,i)] \nonumber \\ & & \leq \max_{G',\bs{w}',\bs{v}'}|\beta(G',\bs{w}',\bs{v}')-\beta(\phi_{j,k}(G'),\bs{w}',\bs{v}')| \nonumber \\ & & \leq {d_v^{\ell}}/{n}, \label{EQ5} \end{eqnarray} where $\beta':=\beta(G',\bs{w}',\bs{v}')$, and $\beta'_{\phi_{j,k}}:=\beta(\phi_{j,k}(G'),\bs{w}',\bs{v}')$. Thus, combining~\eqref{EQ4} and~\eqref{EQ5}, for all $0 \leq i < n d_v$, one can take $\Delta_i=d_v^{\ell}/n$.

% either $(G,\bs{v})=(G',\bs{v}')$, but $\bs{w}$ and $\bs{w}'$ are different for one and only one edge, or 

In the following, we shall find $\Delta_i$ for all $nd_v \leq i< 2 n d_v$, or all $2 n d_v\leq i<\xi$, i.e., when we reveal the edge weights or the input values one at a time, respectively.

Since all the edge weights are nonzero and drawn independently from a continuous alphabet, one can see that for all $nd_v\leq i< 2nd_v$, and for all realizations $T$, $\beta_i(T)=\beta_{i+1}(T)$. We can, therefore, take $\Delta_i=0$.

The method of upper bounding the difference between $\beta_i(T)$ and $\beta_{i+1}(T)$, for all $2 n d_v\leq i < \xi$, is different than what was used earlier. By definition, $\beta_i(T)$ is the expected value of the random variables $\beta(T')$, for all realizations $T'$, such that $T'\equiv_i T$. For a given $T$, consider the value of $\beta_i(T)$. In the corresponding expectation, the value of $(i+1)$th coordinate of $T$ is free (not fixed). It should be clear that $\beta_i(G,W,R)$ cannot be less than the expected value of $\beta(T')$, where $T'\equiv_i T$, and the $(i+1)$th coordinate of $T'$ is zero. Now, consider the value of $\beta_{i+1}(T)$. This quantity cannot be larger than the expected value of $\beta(T'')$, where $T''\equiv_i T$, and the $(i+1)$th coordinate of $T''$ has a nonzero value. For a given $T$, for any realization $T'$ in the ensemble over which we take the average in order to calculate $\beta_{i}(T)$, there are counterpart realizations $T''$ in the ensemble over which we take the average with regards to $\beta_{i+1}(T)$ with the following properties: (i) $(G'',\bs{w}'')=(G',\bs{w}')$, and $\bs{v}''$ and $\bs{v}'$ are different only in the value of the $(i+1)$th coordinate of the underlying realizations $T'$ or $T''$, and (ii) the number of edges emanating from nonzero variable nodes which carry unverified status messages in $T'$ is not larger than that in $T''$. Let $T(i+1)$, $T'(i+1)$ and $T''(i+1)$ represent the $(i+1)$th coordinate of the realizations $T$, $T'$ and $T''$, respectively. From the above argument, one can conclude that (i) if $T(i+1)=0$, then \begin{equation}\label{NEQ1}\beta_{i+1}(T)\leq \beta_{i}(T),\end{equation} and (ii) if $T(i+1)\neq 0$, then \begin{equation}\label{NEQ2}\beta_{i}(T)\leq \beta_{i+1}(T).\end{equation} Furthermore, by definition, it should be clear that (i) if $T(i+1)=0$, \begin{equation}\label{NEQ3}\beta_{i}(T)\leq \mathbf{E}[\beta(T'')|T''\equiv_i T, T''(i+1)\neq 0], \end{equation} and \begin{equation}\label{NEQ4}\beta_{i+1}(T)= \mathbf{E}[\beta(T')|T'\equiv_i T, T'(i+1)=0], \end{equation} and (ii) if $T(i+1)\neq 0$, \begin{equation}\label{NEQ5}\beta_{i}(T)\geq \mathbf{E}[\beta(T')|T'\equiv_i T, T'(i+1)=0], \end{equation} and \begin{equation}\label{NEQ6}\beta_{i+1}(T)\leq \mathbf{E}[\beta(T'')|T''\equiv_i T, T''(i+1)\neq 0].\end{equation} Thus, combining~\eqref{NEQ1}, \eqref{NEQ3} and~\eqref{NEQ4} in the case of $T(i+1)=0$, or combining \eqref{NEQ2}, \eqref{NEQ5} and~\eqref{NEQ6} in the case of $T(i+1)\neq 0$, one can see that 
\begin{eqnarray}
\lefteqn{|\beta_i(T)-\beta_{i+1}(T)|} \nonumber \\ & & \leq |\mathbf{E}[\beta(T')|T'\equiv_i T, T'(i+1)= 0] - \mathbf{E}[\beta(T'')|T''\equiv_i T, T''(i+1)\neq 0]| \nonumber \\ & & \leq \max_{T',T''}\left|\beta(T')-\beta(T'')\right|, \label{EQ6}\end{eqnarray} where the maximization is over all the realizations $T'$ and $T''$, such that $T'\equiv_i T, T''\equiv_i T$, and $T'$ and $T''$ are the same except in their $(i+1)$th coordinate which is zero or nonzero in $T'$ or $T''$, respectively. By using the bounds for Genie algorithm in Appendix~\ref{AppD3}, for all possible realizations $T'$ and $T''$ which satisfy the above conditions, one can write 
\begin{equation}\label{EQ7}\left|\beta(T')-\beta(T'')\right|\leq d_v^{\ell}/n.
\end{equation} 
By combining~\eqref{EQ6} and~\eqref{EQ7}, for all $2 n d_v\leq i< \xi$, we can have $\Delta_i=d_v^{\ell}/n$.

Now, applying Azuma's inequality, i.e., for any $\lambda>0$, \[\Pr[|\beta-\mathbf{E}[\beta]|>\lambda]\leq 2e^{-\frac{\lambda^2}{2\sum_{0\leq i<\xi}\Delta_i^2}},\] followed by setting $\lambda=\epsilon/2$, for any $\epsilon>0$, we obtain \[\Pr[|\beta-\mathbf{E}[\beta]|>\epsilon/2]\leq 2e^{-\epsilon^2 n/\mu},\] where $\mu=8(d_v+1)d_v^{2\ell}$. 

Similarly, for each of the LM and SBB algorithms, by replacing the value of $\Delta_i$, for each $0\leq i<\xi$, according to the bounds given in Appendix~\ref{AppD3}, one can prove inequality~\eqref{EqDot}, for some constant $\mu$ which depends on the recovery algorithm and the parameters $\ell,d_v,$ and $d_c$.

To complete the proof of Theorem~\ref{Concentration}, we now prove inequality~\eqref{EqStar}. Let $\mathbf{E}[\beta_i]$, for all $1\leq i\leq nd_v$, be the expected number of variable to check node messages with unverified status passed along the $i$th edge, denoted by $e_i$, connected to a nonzero variable node $v(e_i)$. Note that the expectation is over all graphs, all signal inputs and all edge weights. Then, by linearity of expectation, from the definition, it follows that \begin{eqnarray*}\mathbf{E}[\beta] &=& \sum_{i\in[n d_v]}\mathbf{E}[\beta_i]/n d_v \\ &=& \mathbf{E}[\beta_1],\end{eqnarray*} where we have used the fact that due to the regularity of the graph and having i.i.d.~edge weights and input vector's elements, $\mathbf{E}[\beta_i]=\mathbf{E}[\beta_1]$, for all $i\in[n d_v]$. Clearly, \begin{eqnarray*}\lefteqn{\mathbf{E}[\beta_1]= \mathbf{E}[\beta_1|\mathcal{N}_{v(e_1)}^{2\ell} \text{is tree-like}]\cdot\Pr[\mathcal{N}_{v(e_1)}^{2\ell} \text{is tree-like}]} \\ & & +\mathbf{E}[\beta_1|\mathcal{N}_{v(e_1)}^{2\ell} \text{is not tree-like}]\cdot\Pr[\mathcal{N}_{v(e_1)}^{2\ell} \text{is not tree-like}].\end{eqnarray*} In Appendix~\ref{AppD1}, the probability that, for a variable node $v$, $\mathcal{N}_v^{2\ell}$ is not tree-like is upper bounded by $\gamma/n$, where $\gamma$ is a constant with respect to $n$, i.e., \[\Pr[\mathcal{N}_v^{2\ell}\text{is not tree-like}]\leq \frac{\gamma}{n}.\] Further, by definition, $\mathbf{E}[\beta_1|\mathcal{N}_{v(e_1)}^{2\ell} \text{is tree-like}]=\alpha^{(\ell)}$, and obviously, $|\mathbf{E}[\beta_1|\mathcal{N}_{v(e_1)}^{2\ell} \text{is not tree-like}]|\leq 1$. Therefore \[\alpha^{(\ell)}(1-\gamma/n)\leq \mathbf{E}[\beta]\leq \alpha^{(\ell)}+\gamma/n,\] which results in \[|\mathbf{E}[\beta]-\alpha^{(\ell)}|\leq \gamma/n.\] Lastly, for any $\epsilon>0$, by taking $n>2\gamma/\epsilon$, we get \[|\mathbf{E}[\beta]-\alpha^{(\ell)}|< \epsilon/2,\] which completes the proof.\end{proof}

%%%%%%%%%%%%%%%%%%%%%%%%%%%%%%%%%%%%%%%%%%%%%%%%%%%%%%%%%%%%%%%%%%%%%%%%%%%%%%%%%%%%%%%%%%%%%%%%%%%%%%%%%%%%%
\subsection{Analysis of the Genie}
\label{simplified}
In the Genie algorithm, the support set is known. Therefore, the set of all variable nodes can be divided into two disjoint sets: verified $\ma{R}$ and unverified $\ma{K}$. At iteration zero, variable nodes in the support set are unverified, and the zero-valued variable nodes (variable nodes not in the support set) belong to the verified set (with a verified value equal to zero). In future iterations, during the verification process, variable nodes are removed from the set $\ma{K}$ and added to the set $\ma{R}$. We use the notation $\ma{K}^{(\ell)}$ and $\ma{R}^{(\ell)}$ to denote the set of unverified and verified variable nodes at (the beginning of) iteration $\ell$, respectively. We also use the superscript $\ell$ to indicate the iteration number for all the other sets in this section in the same way.

Each iteration of the Genie algorithm consists of only one round (two half-rounds). At a generic iteration $\ell$, in the first half-round, check nodes process the received messages from variable nodes sent at iteration $\ell-1$ and generate outgoing messages to be delivered to variable nodes. As we shall see, check nodes are grouped based on their degree also reflected in their outgoing messages. A check node with degree $j$ before half-round 1 may have degree $i \leq j$ after half-round 1. The set of check nodes with degree $i$ \emph{after} the (processing in the) first half-round of iteration $\ell$, is represented with $\ma{N}^{(\ell,1)}_i$. In the second half-round, the variable nodes process the incoming messages and generate outgoing messages accordingly. Variable nodes, are also grouped based on the number of neighboring check nodes of degree $1$. The group of unverified variable nodes with $i$ neighboring check nodes of degree $1$ \emph{after} the (processing in the) second half-round of iteration $\ell$ is represented with $\ma{K}^{(\ell,2)}_i$. Note that the grouping of check nodes remains unchanged during the second half-round of the same iteration. In a similar way, the grouping of variable nodes remains unchanged during the first half-round of the next iteration.

In the first half-round of iteration zero, every check node sends its corresponding measurement value along with its degree, $d_c$. In the second half-round, variable nodes in $\ma{R}^{(0)}$ return a verified message with a value equal to $0$, while variable nodes in $\ma{K}^{(0)}$ return a message with the status bit indicating their unverified status. At iteration $0$, the set $\ma{K}^{(0,2)}_{0}$ includes all unverified variable nodes $\ma{K}^{(0)}$. Therefore, at this stage, no additional variable node can be verified because all incoming messages to unverified variable nodes have $d_c$ in their first coordinate.

In the first half-round of iteration 1, received messages from variable nodes are processed at check nodes and outgoing messages to variable nodes are generated. Each such message has the following two properties: 1) the second coordinate of the message is the same as the second coordinate of the message sent over the same edge in the same half-round of iteration 0, and 2) the first coordinate of the message is at most the same as the first coordinate of the message sent over the same edge in the same half-round of iteration 0. The second coordinates are the same because no variable node from the support set was verified at iteration 0. The first coordinates may be different because the variable nodes not in the support set ($\ma{R}^{(0)}$) have been revealed thus reducing the number of unverified variable nodes connected to the check nodes. We use the notation $\ma{N}^{(1)}_{i \downarrow j}$ to refer to the set of check nodes $\ma{N}^{(0,1)}_{i}$ that are categorized in $\ma{N}^{(1,1)}_{j}$. The arrow points downward to emphasize that $j \leq i$. Note that for iteration $1$, $i=d_c$.

In the second half-round of iteration 1 and after receiving the messages from check nodes, variable nodes in $\ma{K}^{(0,2)}_{0}$ should be regrouped in $\ma{K}^{(1,2)}_{j}, 0\leq j\leq d_v$. We denote by $\ma{K}^{(1)}_{i \uparrow j}$ the set of variable nodes in $\ma{K}^{(0,2)}_i$ joining the set $\ma{K}^{(1,2)}_j$. In this case, $j \geq i$, hence the use of the arrow pointing up. Based on the verification rule for the Genie, at any iteration $\ell$ if an unverified variable node is neighbor to at least one check node in the set $\ma{N}^{(\ell,1)}_1$, it will be verified. So, variable nodes in the set $\bigcup_{j=1}^{d_v}\ma{K}^{(1,2)}_j$ are verified by the end of iteration 1. Therefore, the new sets $\ma{R}^{(2)}$ and $\ma{K}^{(2)}$ to be used at iteration 2 are calculated as follows.
\[
\ma{R}^{(2)} =  \ma{R}^{(1)} \cup \left\{ \ma{K}^{(1,2)}_1, \ma{K}^{(1,2)}_2, \cdots, \ma{K}^{(1,2)}_{d_v} \right\}, \hspace{10pt} \ma{K}^{(2)} = \ma{K}^{(1,2)}_0.
\]
The process of sending, receiving, and processing messages between check nodes and variable nodes as well as the verification process continues in next iterations in the same fashion as we discussed above. In summary, in a generic iteration $\ell$, we have the following relationships:
\begin{align*}
\ma{N}^{(\ell-1,1)}_{i} &= \ds_{j=0}^{i} \ma{N}^{(\ell)}_{i \downarrow j}, \hspace{10pt} i = 0,1, \cdots, d_c, \hspace{20pt} \ma{N}^{(\ell)}_{1 \downarrow 1} = 0,\\
\ma{N}^{(\ell,1)}_{j} &= \ds_{i=j}^{d_c} \ma{N}^{(\ell)}_{i \downarrow j}, \hspace{10pt} j = 0,1, \cdots, d_c.
\end{align*}
\[
\ma{R}^{(\ell+1)} =  \ma{R}^{(\ell)} \cup \left\{ \ma{K}^{(\ell,2)}_1, \ma{K}^{(\ell,2)}_2, \cdots, \ma{K}^{(\ell,2)}_{d_v} \right\}, \hspace{10pt} \ma{K}^{(\ell+1)} = \ma{K}^{(\ell,2)}_0.
\]
\[
\ma{K}^{(\ell-1,2)}_{0} = \ds_{j=0}^{d_v} \ma{K}^{(\ell)}_{0 \uparrow j}, \hspace{20pt} \ma{K}^{(\ell,2)}_{j} = \ma{K}^{(\ell)}_{0 \uparrow j}, \hspace{10pt} j = 0,1, \cdots, d_v.
\]
By tracking the set $\ma{K}^{(\ell)}$ with iterations, we can decide on the success or failure of the algorithm. If the size of the set $\ma{K}^{(\ell)}$ shrinks to zero as $\ell \rightarrow \infty$, then the algorithm is successful. On the other hand, if there exists an $\epsilon > 0$ such that $|\ma{K}^{(\ell)}| \geq \epsilon$, $\forall \ell \geq 1$, then the algorithm is not successful. The success or failure of an algorithm depends on the parameters of the graph ($d_v$ and $d_c$) as well as the initial size of the support set $|\ma{K}^{(0)}|$.

To be able to use the concentration results discussed before and analyze the Genie in the asymptotic case, we track the probability $\alpha^{(\ell)}$ that a variable node belongs to the set $\ma{K}^{(\ell)}$. Hence, we focus on a tree-like graph with random weights and a random input signal (random support set and random nonzero values). Let $\ra{p}^{(\ell,1)}_{\ma{N}_i}, 0\leq i\leq d_c$, denote the probability that a check node belongs to the set $\ma{N}^{(\ell,1)}_i$. Furthermore, let $\ra{p}^{(\ell,2)}_{\ma{K}_j}, 0\leq j \leq d_v$, denote the probability that an unverified (nonzero) variable node belongs to the set $\ma{K}^{(\ell,2)}_j$. In what follows, we show the step-by-step procedure to find the update rules for probabilities $\ra{p}^{(\ell,1)}_{\ma{N}_i}$, $\ra{p}^{(\ell,2)}_{\ma{K}_j}$, and $\alpha^{(\ell+1)}$, in terms of probabilities $\alpha^{(\ell-1)}$, $\alpha^{(\ell)}$, $\ra{p}^{(\ell-1,1)}_{\ma{N}_i}$, and $\ra{p}^{(\ell-1,2)}_{\ma{K}_j}$. The formulas are calculated using probabilistic/counting arguments. The details can be found in Appendix \ref{app_original_Genie}.
\begin{enumerate}
	\item Find the set of probabilities $\ra{p}^{(\ell,1)}_{\ma{N}_i} ,i=0,\cdots,d_c$ from:
\[
\ra{p}^{(\ell,1)}_{\ma{N}_i} = \sum_{j=i}^{d_c}{\ra{p}^{(\ell-1,1)}_{\ma{N}_{j}} \ra{p}^{(\ell)}_{\ma{N}_{j \downarrow i}}}, \hspace{20pt} 0\leq i\leq d_c,
\]
where,

\[
\ra{p}^{(\ell)}_{\ma{N}_{j \downarrow i}} = {j\choose{j-i}} \left(A^{(\ell)}\right)^{j-i}\left(1-A^{(\ell)}\right)^i,\hspace{10pt} 2\leq j\leq d_c, \hspace{10pt} 0\leq i\leq j,
\]
\[
\ra{p}^{(\ell)}_{\ma{N}_{1 \downarrow 0}} = 1,
\hspace{10pt}
\ra{p}^{(\ell)}_{\ma{N}_{1 \downarrow 1}} = 0,
\hspace{10pt}
\ra{p}^{(\ell)}_{\ma{N}_{0 \downarrow 0}} = 1,
\]
\[
A^{(\ell)} = 1 - \left(1 - p^{(\ell)} \right)^{d_v - 1}, \hspace{10pt} \ra{p}^{(\ell)} = \df{\ra{p}^{(\ell-1,1)}_{\ma{N}_1}}{\alpha^{(\ell-1)} d_c}.
\]

	\item Find the set of probabilities $\ra{p}^{(\ell,2)}_{\ma{K}_j}$ from:
\[
\ra{p}^{(\ell,2)}_{\ma{K}_j} = {d_v\choose j}\left(\ra{p}^{(\ell+1)}\right)^{j}\left(1-\ra{p}^{(\ell+1)}\right)^{d_v-j},\hspace{20pt} 0\leq j\leq d_v.
\]
	\item Based on the set of probabilities $\ra{p}^{(\ell,2)}_{\ma{K}_i}$, find the probability $\alpha^{(\ell+1)}$ from:
\[
\alpha^{(\ell+1)} = \alpha^{(\ell)}\left(1-\sum_{i=1}^{d_v}{\ra{p}^{(\ell,2)}_{\ma{K}_i}}\right) = \alpha^{(\ell)} \ra{p}^{(\ell,2)}_{\ma{K}_0}.
\]
\end{enumerate}

The initial set of probabilities $\ra{p}^{(1,1)}_{\ma{N}_i}$ for Genie are as follows: 
\[
\ra{p}^{(1,1)}_{\ma{N}_i} = {d_c \choose{i}} \left(\alpha^{(0)}\right)^i \left(1 - \alpha^{(0)}\right)^{d_c-i},\hspace{20pt} 0\leq i\leq d_c.
\]
The set of probabilities $\ra{p}^{(1,2)}_{\ma{K}_j}$ are calculated following step (2) by noticing that $\alpha^{(1)} = \alpha^{(0)}$.
%%%%%%%%%%%%%%%%%%%%%%%%%%%%%%%%%%%%%%%%%%%%%%%%%%%%%%%%%%%%%%%%%%%%%%%%%%%%%%%%%%%%%%%%%%%%%%%%%%%%%%%%%%%%%
\subsection{Notation and Setup for the Analysis of LM and SBB}
\label{notation}
In LM and SBB, at the beginning of any iteration $\ell$, the set of all variable nodes can be divided into three disjoint sets: $\ma{K}^{(\ell)}$, $\ma{R}^{(\ell)}$, and $\Delta^{(\ell)}$. The set $\ma{K}^{(\ell)}$ consists of all unverified nonzero variable nodes, while the set $\Delta^{(\ell)}$ consists of all unverified zero-valued variable nodes. The set $\ma{R}^{(\ell)}$ includes all variable nodes recovered up to iteration $\ell$. Clearly, the decoder can not make the distinction between variable nodes in the sets $\ma{K}^{(\ell)}$ and $\Delta^{(\ell)}$. The distinction between the two sets, however, is very useful in the analysis.

Furthermore, at any iteration $\ell$, we partition the set of all check nodes into subsets $\ma{N}^{(\ell)}_{i,j}$. The index $i$ indicates the number of neighboring variable nodes in the set $\ma{K}^{(\ell)}$ while the index $j$ indicates the number of neighboring variable nodes in the set $\Delta^{(\ell)}$. Note that: 1) the degree of each check node in the subgraph induced by unverified variable nodes at iteration $\ell$ (reflected in the outgoing message of the check node), is $i+j$, and 2) the second coordinate of messages received by a variable node in the support set from a subset of check nodes all in the sets $\ma{N}^{(\ell)}_{1,j}, 0\leq j\leq d_c-1$, is the same.

In algorithms LM and SBB, each iteration consists of two rounds, each with two half-rounds. The configuration of the sets at the end of each half-round (1 or 2), each round (R1 or R2), and each iteration ($\ell$), is specified using the following 4 superscripts: $(\ell,R1,1), (\ell,R1,2), (\ell,R2,1),$ and $(\ell,R2,2)$. In the first half-rounds (any round and any iteration), messages are passed from check nodes to variable nodes, while in the second half-rounds, messages are passed from variable nodes to check nodes. Also, with the definition of mapping functions $\Phi^{(1,\ell)}_v$ and $\Phi^{(2,\ell)}_v$, the set of verified variable nodes in the first and the second rounds belong to the sets $\ma{K}^{(\ell)}$ and $\Delta^{(\ell)}$, respectively. We have summarized in Table \ref{T:changes} the sets that are affected in each half-round (HR) of each round (R) at any iteration.

\begin{table}
\caption{Sets that change in each half-round of each round at any iteration}
\label{T:changes}
\begin{center}
\renewcommand{\arraystretch}{1.5}
\begin{tabular}{|c|c|c|c|}
	\hline
	\multicolumn{2}{|c|}{R1} & \multicolumn{2}{c|}{R2} \\
	\hline
	HR1 & HR2 & HR1 & HR2 \\
	\hline
	$\ma{N}_{k,i} \rightarrow \ma{N}_{k,j}$ &	$\ma{K}_{i} \rightarrow \ma{K}_{j}$ & $\ma{N}_{i,k} \rightarrow \ma{N}_{j,k}$ & $\Delta_{i} \rightarrow \Delta_{j}$ \\
	\hline
\end{tabular}
\end{center}
\end{table}

In the Genie algorithm, the set $\ma{K}^{(\ell)}_{i}$ represents the set of unverified variable nodes in the support set that have $i$ neighboring check nodes of degree $1$. The definition of this set in the LM and SBB algorithms is different, as explained in the following. The set $\ma{K}^{(\ell)}_{i}$ in the LM algorithm represents the set of unverified variable nodes in the support set with $i$ neighboring check nodes in the set $\ma{N}^{(\ell)}_{1,0}$. To define this set in the SBB algorithm, let $\ma{N}^{(\ell)}_i := \bigcup_{j=0}^{d_c-i} \ma{N}^{(\ell,R1,1)}_{i,j}$. With this notation, the set $\ma{K}^{(\ell)}_{i}$ in the SBB algorithm is defined as the set of unverified variable nodes in the support set with $i$ neighboring check nodes in the set $\ma{N}^{(\ell)}_1$.

In order to track the evolution of the unverified support set, it is imperative to characterize the set of variable nodes in the support set that are recovered in each iteration. In Theorems \ref{LMModel} and \ref{SBBModel} below, we characterize the verification of unverified nonzero variable nodes in the set $\ma{K}^{(\ell)}$ in each iteration $\ell$ for the two algorithms LM and SBB. The theorems are proved in Appendices \ref{app_original_LM} and \ref{app_original_SBB}.

\begin{theorem}
\label{LMModel}
In the first round of any iteration $\ell$ in the LM algorithm, a nonzero variable node $v\in \ma{K}^{(\ell)}$ is verified if and only if it belongs to the set $\bigcup_{i=1}^{d_v}\ma{K}^{(\ell,R1,2)}_i$.
\end{theorem}

\begin{theorem}
\label{SBBModel}
In the first round of any iteration $\ell$ in the SBB algorithm, a nonzero variable node $v\in \ma{K}^{(\ell)}$ is verified if and only if it belongs to the set $\bigcup_{i=2}^{d_v}\ma{K}^{(\ell,R1,2)}_i \cup \hat{\ma{K}}^{(\ell,R1,2)}_1$, where the set $\hat{\ma{K}}^{(\ell,R1,2)}_1$ consists of all variable nodes in the set $\ma{K}^{(\ell,R1,2)}_1$ connected to the set $\ma{N}^{(\ell,R1,1)}_{1,0}$.
\end{theorem}

In the LM and SBB algorithms, unverified variable nodes with zero values are verified in R2. Note that a check node is zero-valued if it belongs to the set $\ma{N}^{(\ell,R2,1)}_{0,j}, 0\leq j\leq d_c$. Therefore, for the verification of zero-valued variable nodes in the second round of iteration $\ell$, we divide the set of variable nodes in $\Delta^{(\ell)}$ into subsets $\Delta^{(\ell)}_i, 0\leq i\leq d_v$ with the following definition: a variable node in the set $\Delta^{(\ell)}_i$ has $i$ neighboring check nodes in the set $\left\{\bigcup_{j=1}^{d_c} \ma{N}^{(\ell,R2,1)}_{0,j} \backslash \bigcup_{j=1}^{d_c} \ma{N}^{(\ell,R1,1)}_{0,j} \right\}$, i.e., the set of all check nodes which became zero-valued after HR1 of R2. In Theorem \ref{LMSBBModel} below, we characterize the verification of unverified zero-valued variable nodes in the set $\Delta^{(\ell)}$ at HR2-R2 in each iteration $\ell$ of LM and SBB algorithms.

\begin{theorem}
\label{LMSBBModel}
In the second half-round of the second round of any iteration $\ell$ in the LM and SBB algorithms a zero-valued variable node $v\in \Delta^{(\ell)}$ is verified if and only if it belongs to the set $\bigcup_{i=1}^{d_v}\Delta^{(\ell)}_i$.
\end{theorem}

We denote by $\ma{N}^{(\ell,R1)}_{k,i \downarrow j}$ the set of check nodes that are moved from $\ma{N}^{(\ell-1,R2,1)}_{k,i}$ to $\ma{N}^{(\ell,R1,1)}_{k,j}$ in HR1-R1 of iteration $\ell$. Similarly, the set of check nodes that are moved from $\ma{N}^{(\ell,R1,1)}_{i,k}$ to $\ma{N}^{(\ell,R2,1)}_{j,k}$ in HR1-R2 of iteration $\ell$ are denoted by $\ma{N}^{(\ell,R2)}_{i \downarrow j,k}$. Since variable nodes in $\ma{K}$ and $\Delta$ are verified through iterations, we always have $j \leq i$ and hence the use of notation $i \downarrow j$. Moreover, we denote the set of variable nodes that are moved from $\ma{K}^{(\ell-1,R1,2)}_{i}$ to $\ma{K}^{(\ell,R1,2)}_{j}$ and from $\Delta^{(\ell-1,R2,2)}_{i}$ to $\Delta^{(\ell,R2,2)}_{j}$ in HR2-R1 and HR2-R2 of iteration $\ell$ by $\ma{K}^{(\ell,R1)}_{i \uparrow j}$ and $\Delta^{(\ell,R2)}_{i \uparrow j}$, respectively. The notation $i \uparrow j$ implies that $j \geq i$.

The sets that fully describe the state of the decoder at the beginning of iteration $\ell$ are: $\ma{K}^{(\ell)}$, $\ma{R}^{(\ell)}$, $\Delta^{(\ell)}$, $\ma{N}^{(\ell-1,R2,1)}_{i,j}$, $\ma{K}^{(\ell-1,R1,2)}_{i}$, and $\Delta^{(\ell-1,R2,2)}_{i}$. For the analysis, we track the probability that a node (variable node or check node) belongs to a certain set at each half-round, round, or iteration. We use the notation $\alpha^{(\ell)}$ to denote the probability that a variable node belongs to the set $\ma{K}^{(\ell)}$. For the rest of the sets, we use the standard notation of probabilities that was applied in the analysis of the Genie algorithm. For instance, we denote the probability that a check node belongs to the set $\ma{N}^{(\ell,R1,1)}_{i,j}$ by $\ra{p}^{(\ell,R1,1)}_{\ma{N}_{i,j}}$.

In the analysis, the goal is to find the recursive equations that relate these probabilities for consecutive iterations. As we shall see, the analysis of the decoding process for LM and SBB results in a system of coupled recursive update rules. Moreover, we show that the update rules at iteration $\ell$ are functions of probabilities at iteration $\ell-1$. Hence, the complexity of the analysis scales linearly with the number of iterations. In the following two sections, we present the update rules for the LM and SBB algorithms. The derivation of formulas are discussed in detail in Appendices \ref{app_original_LM} and \ref{app_original_SBB}.
%%%%%%%%%%%%%%%%%%%%%%%%%%%%%%%%%%%%%%%%%%%%%%%%%%%%%%%%%%%%%%%%%%%%%%%%%%%%%%%%%%%%%%%%%%%%%%%%%%%%%%%%%%%%%
\subsection{Update Rules for the LM Algorithm}
\label{originalLM}
In one iteration $\ell\geq 1$ of the analysis of the LM algorithm, the following update rules are applied:
\begin{enumerate}
\item Find the set of probabilities $\ra{p}^{(\ell,R1,1)}_{\ma{N}_{i,j}}$, $1\leq i\leq d_c$, $0\leq j\leq d_c - i$, from:
\[
\ra{p}^{(\ell,R1,1)}_{\ma{N}_{i,j}} = \ds_{k=j}^{d_c-i} \ra{p}^{(\ell-1,R2,1)}_{\ma{N}_{i,k}} \ra{p}^{(\ell,R1)}_{\ma{N}_{i,k \downarrow j}},
\]
where,
\begin{equation}
\ra{p}^{(\ell,R1)}_{\ma{N}_{i,k \downarrow j}} = {k \choose j} \left( A^{(\ell)} \right)^j \left( 1 - A^{(\ell)} \right)^{k-j},\hspace{20pt} A^{(\ell)} = \df{\ra{p}^{(\ell)}_\Delta}{1 - D^{(\ell-1)}},
\label{eq:A}
\end{equation}

\begin{equation}
D^{(\ell-1)} = \ds_{j=1}^{d_c-1} j\df{\ra{p}^{(\ell-1,R2,1)}_{\ma{N}_{0,j}}}{\ds_{i=0}^{d_c} \ds_{j=1}^{d_c-1} j \ra{p}^{(\ell-1,R2,1)}_{\ma{N}_{i,j}}},
\label{eq:D}
\end{equation}
and $\ra{p}^{(\ell)}_\Delta$ is given in \eqref{eq:p_delta}.

\item Find the set of probabilities $\ra{p}^{(\ell,R1,2)}_{\ma{K}_i}$, $0\leq i\leq d_v$, from:
\[
\ra{p}^{(\ell,R1,2)}_{\ma{K}_i} = {d_v\choose i}\left(B^{(\ell)} \right)^i \left(1 - B^{(\ell)} \right)^{d_v-i},
\]
where,
\[
B^{(\ell)} = \df{\ra{p}^{(\ell,R1,1)}_{\ma{N}_{1,0}}}{\alpha^{(\ell)} d_c}.
\]

\item Find the updated probability that a variable node is in the support set and is not verified from:
\[
\alpha^{(\ell+1)} = \alpha^{(\ell)} \left(1 - \ds_{i=1}^{d_v} \ra{p}^{(\ell,R1,2)}_{\ma{K}_i} \right).
\]

\item Find the set of probabilities $\ra{p}^{(\ell,R2,1)}_{\ma{N}_{j,k}}$, $0\leq j\leq d_c$, $0\leq k\leq d_c - i$, from:
\[
\ra{p}^{(\ell,R2,1)}_{\ma{N}_{j,k}} = \ds_{i=j}^{d_c} \ra{p}^{(\ell,R1,1)}_{\ma{N}_{i,k}} \ra{p}^{(\ell,R2)}_{\ma{N}_{i\downarrow j,k}},
\]
where,
\[
\ra{p}^{(\ell,R2)}_{\ma{N}_{i\downarrow j,k}} = {i\choose{j}} \left( C^{(\ell)} \right)^j \left(1 - C^{(\ell)} \right)^{i-j},\hspace{20pt} C^{(\ell)} = \df{\ra{p}^{(\ell,R1,2)}_{\ma{K}_{0}}}{1 - B^{(\ell)}}.
\]
Note that:
\[
\ra{p}^{(\ell,R2)}_{\ma{N}_{1\downarrow 0,0}} = 1, \hspace{20pt} \ra{p}^{(\ell,R2)}_{\ma{N}_{1\downarrow 1,0}} = 0.
\]

\item Find the set of probabilities $\ra{p}^{(\ell,R2,2)}_{\Delta_i}$, $0\leq i\leq d_v$, from:
\[
\ra{p}^{(\ell,R2,2)}_{\Delta_i} = {d_v \choose i} \left( D^{(\ell)} \right)^i \left( 1 - D^{(\ell)} \right)^{d_v - i},
\]
where $D^{(\ell)}$ is calculated according to \eqref{eq:D}.

\item And lastly, find the probability that a variable node is zero-valued and unverified at iteration $\ell$, from:
\begin{equation}
\ra{p}^{(\ell+1)}_\Delta = \ra{p}^{(\ell)}_\Delta \ra{p}^{(\ell,R2,aHR2)}_{\Delta_0}.
\label{eq:p_delta}
\end{equation}
\end{enumerate}
We have the following initial probabilities, by letting $\alpha := \alpha^{(0)}$ denote the initial density factor. 
\[
\ra{p}^{(1)}_\Delta = (1-\alpha)\left(1 - \left(1 - \alpha \right)^{d_c - 1} \right)^{d_v}, \hspace{10pt} D^{(0)} = \left( 1 - \alpha \right)^{d_c - 1} \hspace{10pt }\Rightarrow \hspace{10pt} A^{(1)} = (1-\alpha)\left(1 - \left(1 - \alpha \right)^{d_c - 1} \right)^{d_v-1}.
\]
Since all check nodes have degree $d_c$ in the subgraph induced by the unverified variable nodes at iteration zero, we have:
\[
\ra{p}^{(0,R2,1)}_{\ma{N}_{i,d_c-i}} = {d_c \choose i} \left(\alpha \right)^i \left( 1- \alpha \right)^{d_c - i}, \hspace{20pt} i=0, \cdots, d_c,
\]
and thus
\[
\ra{p}^{(1,R1,1)}_{\ma{N}_{i,j}} = \ra{p}^{(0,R2,1)}_{\ma{N}_{i,d_c-i}} \ra{p}^{(1,R1)}_{\ma{N}_{i,d_c-i \downarrow j}},\hspace{20pt}i=1,\cdots,d_v,\hspace{20pt}j=0,\cdots,d_c-i,
\]
where,
\[
\ra{p}^{(1,R1)}_{\ma{N}_{i,d_c-i \downarrow j}} = {d_c-i \choose j} \left( A^{(1)} \right)^j \left(1 - A^{(1)} \right)^{d_c - i-j},\hspace{20pt}i=1,\cdots,d_v,\hspace{20pt}j=0,\cdots,d_c-i.
\]
Since no element of the support set is verified at iteration zero, we also have:
\[
\alpha^{(1)} = \alpha^{(0)} = \alpha.
\]
%%%%%%%%%%%%%%%%%%%%%%%%%%%%%%%%%%%%%%%%%%%%%%%%%%%%%%%%%%%%%%%%%%%%%%%%%%%%%%%%%%%%%%%%%%%%%%%%%%%%%%%%%%%%%
\subsection{Update Rules for the SBB Algorithm}
\label{originalSBB}
In one generic iteration $\ell$, $\ell \geq 2$, of the SBB algorithm, the following update rules are applied in the analysis:
\begin{enumerate}
	\item Find the set of probabilities $\ra{p}^{(\ell,R1,1,+)}_{\ma{N}_{1,j}}$, $\ra{p}^{(\ell,R1,1,C)}_{\ma{N}_{1,j}}$, and $\ra{p}^{(\ell,R1,1)}_{\ma{N}_{i,j}}$, $2\leq i\leq d_c$, $0\leq j\leq d_c - i$, from:
\begin{align*}
\ra{p}^{(\ell,R1,1,+)}_{\ma{N}_{1,j}} &= \ds_{k=j}^{d_c-1} \ra{p}^{(\ell-1,R2,1,+)}_{\ma{N}_{1,k}} \ra{p}^{(\ell,R1)}_{\ma{N}_{1,k \downarrow j}},\\
\ra{p}^{(\ell,R1,1,C)}_{\ma{N}_{1,j}} &= \ds_{k=j}^{d_c-1} \ra{p}^{(\ell-1,R2,1,C)}_{\ma{N}_{1,k}} \ra{p}^{(\ell,R1)}_{\ma{N}_{1,k \downarrow j}},\\
\ra{p}^{(\ell,R1,1)}_{\ma{N}_{i,j}} &= \ds_{k=j}^{d_c-i} \ra{p}^{(\ell-1,R2,1)}_{\ma{N}_{i,k}} \ra{p}^{(\ell,R1)}_{\ma{N}_{i,k \downarrow j}},
\end{align*}
where, $\ra{p}^{(\ell,R1)}_{\ma{N}_{i,k \downarrow j}}$, and parameters $A^{(\ell)}$ and $D^{(\ell-1)}$ therein, are obtained from equations \eqref{eq:A} and \eqref{eq:D}.

	\item Find the set of probabilities $\ra{p}^{(\ell,R1,2)}_{\ma{K}_0}$, $\ra{p}^{(\ell,R1,2,+)}_{\ma{K}_1}$, $\ra{p}^{(\ell,R1,2,C)}_{\ma{K}_1}$, and $\ra{p}^{(\ell,R1,2)}_{\ma{K}_j}$, $1\leq j\leq d_v$, from:
\begin{align*}
\ra{p}^{(\ell,R1,2)}_{\ma{K}_0} &= \df{1}{N^{(\ell,R1)}} \ra{p}^{(\ell-1,R1,2)}_{\ma{K}_0} \ra{p}^{(\ell,R1)}_{\ma{K}_{0 \uparrow 0}},\\
\ra{p}^{(\ell,R1,2,+)}_{\ma{K}_1} &= \df{1}{N^{(\ell,R1)}} \ra{p}^{(\ell-1,R1,2)}_{\ma{K}_0} \ra{p}^{(\ell,R1)}_{\ma{K}_{0 \uparrow 1}} \left(1 - f^{(\ell,R1,+)} \right),\\
\ra{p}^{(\ell,R1,2,C)}_{\ma{K}_1} &= \df{1}{N^{(\ell,R1)}} \ra{p}^{(\ell-1,R1,2)}_{\ma{K}_1} \ra{p}^{(\ell,R1)}_{\ma{K}_{1 \uparrow 1}} \left(1 - f^{(\ell,R1,C)} \right),\\
\ra{p}^{(\ell,R1,2)}_{\ma{K}_1} &= \ra{p}^{(\ell,R1,2,+)}_{\ma{K}_1} + \ra{p}^{(\ell,R1,2,C)}_{\ma{K}_1},\\
\ra{p}^{(\ell,R1,2)}_{\ma{K}_j} &= 0, \hspace{20pt} j=2,\cdots,d_v,
\end{align*}
where,
\[
\ra{p}^{(\ell,R1)}_{\ma{K}_{j \uparrow i}} = {d_v - j \choose i-j} \left( B^{(\ell)} \right)^{i-j} \left( 1 - B^{(\ell)} \right)^{d_v - i}, \hspace{20pt} B^{(\ell)} = \df{\ds_{j=0}^{d_c-1} \ra{p}^{(\ell,R1,1,+)}_{\ma{N}_{1,j}}}{\ds_{j=0}^{d_c-1} \ra{p}^{(\ell,R1,1,+)}_{\ma{N}_{1,j}} + \ds_{i=2}^{d_c} \ds_{j=0}^{d_c - i} i \ra{p}^{(\ell,R1,1)}_{\ma{N}_{i,j}}},
\]
and,
\[
f^{(\ell,R1,+)} = \df{\ra{p}^{(\ell,R1,1,+)}_{\ma{N}_{1,0}}}{\ds_{j=0}^{d_c - 1} \ra{p}^{(\ell,R1,1,+)}_{\ma{N}_{1,j}}}, \hspace{20pt}
f^{(\ell,R1,C)} = \df{\ra{p}^{(\ell,R1,1,C)}_{\ma{N}_{1,0}}}{\ds_{j=0}^{d_c - 1} \ra{p}^{(\ell,R1,1,C)}_{\ma{N}_{1,j}}},
\]
and,
\[
N^{(\ell,R1)} = \ra{p}^{(\ell-1,R1,2)}_{\ma{K}_0} \ra{p}^{(\ell,R1)}_{\ma{K}_{0 \uparrow 0}} + 
\ra{p}^{(\ell-1,R1,2)}_{\ma{K}_0} \ra{p}^{(\ell,R1)}_{\ma{K}_{0 \uparrow 1}} \left(1 - f^{(\ell,R1,+)} \right) + 
\ra{p}^{(\ell-1,R1,2)}_{\ma{K}_1} \ra{p}^{(\ell,R1)}_{\ma{K}_{1 \uparrow 1}} \left(1 - f^{(\ell,R1,C)} \right).
\]

	\item The probability that a variable node belongs to the support set and remains unverified after iteration $\ell$, $\alpha^{(\ell+1)}$ is calculated as follows:
\[
\alpha^{(\ell+1)} = \alpha^{(\ell)} N^{(\ell,R1)}.
\]

	\item Find the set of probabilities $\ra{p}^{(\ell,R2,1)}_{\ma{N}_{0,j}}$, $\ra{p}^{(\ell,R2,1,+)}_{\ma{N}_{1,j}}$, $\ra{p}^{(\ell,R2,1,C)}_{\ma{N}_{1,j}}$, and $\ra{p}^{(\ell,R2,1)}_{\ma{N}_{k,j}}$, $2\leq k\leq d_c$, $0\leq j\leq d_c-k$, from:
\begin{align*}
\ra{p}^{(\ell,R2,1)}_{\ma{N}_{0,j}} &= \ra{p}^{(\ell,R1,1)}_{\ma{N}_{0,j}} + \ra{p}^{(\ell,R2,C,O)}_{\ma{N}_{1,j}} + \ra{p}^{(\ell,R2,+,O)}_{\ma{N}_{1,j}} + \ds_{i=2}^{d_c} \ra{p}^{(\ell,R1,1)}_{\ma{N}_{i,j}} \ra{p}^{(\ell,R2)}_{\ma{N}_{i\downarrow 0,j}},\\
\ra{p}^{(\ell,R2,1,+)}_{\ma{N}_{1,j}} &= \ds_{i=2}^{d_c} \ra{p}^{(\ell,R1,1)}_{\ma{N}_{i,j}} \ra{p}^{(\ell,R2)}_{\ma{N}_{i\downarrow 2,j}},\\
\ra{p}^{(\ell,R2,1,C)}_{\ma{N}_{1,j}} &= \ra{p}^{(\ell,R2,C,F)}_{\ma{N}_{1,j}} + \ra{p}^{(\ell,R2,+,F)}_{\ma{N}_{1,j}},\\
\ra{p}^{(\ell,R2,1)}_{\ma{N}_{k,j}} &= \ds_{i=k}^{d_c} \ra{p}^{(\ell,R1,1)}_{\ma{N}_{i,j}} \ra{p}^{(\ell,R2)}_{\ma{N}_{i\downarrow k,j}}, \hspace{20pt} k=2,\cdots,d_c,
\end{align*}
where,
\begin{align*}
\ra{p}^{(\ell,R2,+,F)}_{\ma{N}_{1,i}} &= \ra{p}^{(\ell,R1,1,+)}_{\ma{N}_{1,i}} \df{\ra{p}^{(\ell-1,R1,2)}_{\ma{K}_0} \ra{p}^{(\ell,R1)}_{\ma{K}_{0 \uparrow 1}}}{\ds_{j=1}^{d_v} j \ra{p}^{(\ell-1,R1,2)}_{\ma{K}_0} \ra{p}^{(\ell,R1)}_{\ma{K}_{0 \uparrow j}} + \ds_{j=2}^{d_v} (j-1) \ra{p}^{(\ell-1,R1,2)}_{\ma{K}_1} \ra{p}^{(\ell,R1)}_{\ma{K}_{1 \uparrow j}}},\\
\ra{p}^{(\ell,R2,+,O)}_{\ma{N}_{1,i}} &= \ra{p}^{(\ell,R1,1,+)}_{\ma{N}_{1,i}} - \ra{p}^{(\ell,R2,+,F)}_{\ma{N}_{1,i}},\\
\ra{p}^{(\ell,R2,C,F)}_{\ma{N}_{1,i}} &= \ra{p}^{(\ell,R1,1,C)}_{\ma{N}_{1,i}} \df{\ra{p}^{(\ell,R1)}_{\ma{K}_{1 \uparrow 1}}}{\ds_{j=1}^{d_v} \ra{p}^{(\ell,R1)}_{\ma{K}_{1 \uparrow j}}},\\
\ra{p}^{(\ell,R2,C,O)}_{\ma{N}_{1,i}} &= \ra{p}^{(\ell,R1,1,C)}_{\ma{N}_{1,i}} - \ra{p}^{(\ell,R2,C,F)}_{\ma{N}_{1,i}},\\
\ra{p}^{(\ell,R2)}_{\ma{N}_{i\downarrow k,j}} &= {i\choose k} \left( C^{(\ell)} \right)^{k}\left( 1 - C^{(\ell)} \right)^{i-k}, \hspace{20pt} \ra{p}^{(\ell,R1)} = \df{\ds_{j=0}^{d_c-1} \ra{p}^{(\ell,R1,1,+)}_{\ma{N}_{1,j}} + \ds_{j=0}^{d_c-1} \ra{p}^{(\ell,R1,1,C)}_{\ma{N}_{1,j}}}{\alpha^{(\ell)} d_c},\\
C^{(\ell)} &= \df{\ra{p}^{(\ell-1,R1,2)}_{\ma{K}_0} \ra{p}^{(\ell,R1)}_{\ma{K}_{0 \uparrow 0}} }{1 - \ra{p}^{(\ell,R1)}}\\
&+ \df{\ra{p}^{(\ell-1,R1,2)}_{\ma{K}_0} \ra{p}^{(\ell,R1)}_{\ma{K}_{0 \uparrow 1}} \left(\df{d_v-1}{d_v}\right)}{1 - \ra{p}^{(\ell,R1)}} \left(1 - f^{(\ell,R1,+)} \right)\\
&+ \df{\ra{p}^{(\ell-1,R1,2)}_{\ma{K}_1} \ra{p}^{(\ell,R1)}_{\ma{K}_{1 \uparrow 1}} \left(\df{d_v-1}{d_v}\right)}{1 - \ra{p}^{(\ell,R1)}} \left(1 - f^{(\ell,R1,C)} \right).
\end{align*}
	\item Find the set of probabilities  $\ra{p}^{(\ell,R2,2)}_{\Delta_i}$, $0\leq i\leq d_v$, from:
\[
\ra{p}^{(\ell,R2,2)}_{\Delta_i} = {d_v \choose i} \left( D^{(\ell)} \right)^i \left( 1 - D^{(\ell)} \right)^{d_v - i},
\]
where $D^{(\ell)}$ is calculated according to \eqref{eq:D}.

	\item Find the probability that a variable node is zero-valued and remains unverified in iteration $\ell+1$ as follows:
\[
\ra{p}^{(\ell+1)}_\Delta = \ra{p}^{(\ell)}_\Delta \ra{p}^{(\ell,R2,2)}_{\Delta_0}.
\]
\end{enumerate}
The initial conditions in the SBB algorithm are for iterations 0 and 1, and are given below. The update rules for iteration $1$ serve as the initial conditions for iteration $2$.

\begin{itemize}
	\item HR1-R1:
\[
\ra{p}^{(1,R1,1)}_{\ma{N}_{i,j}} = {d_c\choose i} \left(\alpha^{(0)} \right)^i \left(1 - \alpha^{(0)} \right)^{d_c-i} \ra{p}^{(1,R1)}_{\ma{N}_{i,d_c-i \downarrow j}},\hspace{20pt}i=1,\cdots,d_c,\hspace{20pt}j=0,\cdots,d_c-i,
\]
where,
\[
\ra{p}^{(1,R1)}_{\ma{N}_{i,d_c-i \downarrow j}} = {d_c-i \choose j} \left( A^{(1)} \right)^j \left( 1 - A^{(1)} \right)^{d_c - i-j},\hspace{20pt} A^{(1)} = (1-\alpha^{(0)})\left(1 - \left(1 - \alpha^{(0)} \right)^{d_c - 1} \right)^{d_v-1}.
\]

	\item HR2-R1 (Regrouping variable nodes after the recovery):
\begin{align*}
\ra{p}^{(1,R1,2)}_{\ma{K}_0} &= \df{1}{N^{(1,R1)}} \left(1 - B^{(1)}\right)^{d_v}.\\
\ra{p}^{(1,R1,2)}_{\ma{K}_1} &= \df{d_v}{N^{(1,R1)}} \left(1 - f^{(1,R1)}\right) B^{(1)} \left(1 - B^{(1)}\right)^{d_v-1}.\\
\ra{p}^{(1,R1,2)}_{\ma{K}_i} &= 0, \hspace{20pt} 2\leq i\leq d_v,
\end{align*}

where,
\[
B^{(1)} = \ds_{j=0}^{d_c-1} \df{\ra{p}^{(1,R1,1)}_{\ma{N}_{1,j}}}{\alpha^{(0)} d_c},
\hspace{3cm}
f^{(1,R1)} = \df{\ra{p}^{(1,R1,1)}_{\ma{N}_{1,0}}}{\ds_{j=0}^{d_c-1} \ra{p}^{(1,R1,1)}_{\ma{N}_{1,j}}},
\]
and,
\[
N^{(1,R1)} = \left(1 - B^{(1)}\right)^{d_v} + d_v \left(1 - f^{(1,R1)}\right) B^{(1)} \left(1 - B^{(1)}\right)^{d_v-1}.
\]

\item HR2-R1 (Recovering variable nodes based on ECN):
\[
\alpha^{(2)} = \alpha^{(0)} N^{(1,R1)} \left( \ra{p}^{(1,R1,2)}_{\ma{K}_0} + \ra{p}^{(1,R1,2)}_{\ma{K}_1} \right),
\]

\item HR1-R2:
\begin{align*}
\ra{p}^{(1,R2,1)}_{\ma{N}_{k,j}} &= \ds_{i=k}^{d_c} \ra{p}^{(1,R1,1)}_{\ma{N}_{i,j}} \ra{p}^{(1,R2)}_{\ma{N}_{i\downarrow k,j}}, \hspace{20pt}k=2,\cdots,d_c, \hspace{20pt} j=0,\cdots,d_c-i,\\
\ra{p}^{(1,R2,1,+)}_{\ma{N}_{1,j}} &= \ds_{i=2}^{d_c} \ra{p}^{(1,R1,1)}_{\ma{N}_{i,j}} \ra{p}^{(1,R2)}_{\ma{N}_{i\downarrow 1,j}},\\
\ra{p}^{(1,R2,1,C)}_{\ma{N}_{1,j}} &= \ra{p}^{(1,R1,1)}_{\ma{N}_{1,j}} \ra{p}^{(1,R2)}_{\ma{N}_{1\downarrow 1,j}},
\end{align*}
where,
\begin{align*}
\ra{p}^{(1,R2)}_{\ma{N}_{0\downarrow 0,j}} &= 1, \hspace{20pt} 0\leq j \leq d_c,\\
\ra{p}^{(1,R2)}_{\ma{N}_{1\downarrow 0,0}} &= 1, \hspace{20pt} \ra{p}^{(1,R2)}_{\ma{N}_{1\downarrow 1,0}} = 0,\\
\ra{p}^{(1,R2)}_{\ma{N}_{1\downarrow 0,j}} &= 1 - \left(1 - B^{(1)}\right)^{d_v-1}, \hspace{20pt} \ra{p}^{(1,R2)}_{\ma{N}_{1\downarrow 1,j}} = \left(1 - B^{(1)}\right)^{d_v-1}, \hspace{20pt} 0\leq j\leq d_c-1,\\
\ra{p}^{(1,R2)}_{\ma{N}_{i\downarrow k,j}} &= {i\choose k} \left( C^{(1)} \right)^k \left( 1 - C^{(1)} \right)^{i-k}, \hspace{20pt} 2\leq i\leq d_c, \hspace{20pt} 0\leq k\leq i, \hspace{20pt} 0\leq j\leq d_c-i,\\
C^{(1)} &= \left(1 - B^{(1)}\right)^{d_v-1} + \left(d_v-1\right)B^{(1)} \left(1 - B^{(1)}\right)^{d_v-2} \left(1 - f^{(1,R1)} \right).\\
\end{align*}

\item HR2-R2:
\[
\ra{p}^{(2)}_\Delta = (1-\alpha^{(0)})\left(1 - \left(1 - \alpha^{(0)} \right)^{d_c - 1} \right)^{d_v} \left( 1 - D^{(1)} \right)^{d_v},
\]
where, $D^{(1)}$ is obtained by \eqref{eq:D}.

\end{itemize}
%%%%%%%%%%%%%%%%%%%%%%%%%%%%%%%%%%%%%%%%%%%%%%%%%%%%%%%%%%%%%%%%%%%%%%%%%%%%%%%%%%%%%%%%%%%%%%%%%%%%%%%%%%%%%
%%%%%%%%%%%%%%%%%%%%%%%%%%%%%%%%%%%%%%%%%%%%%%%%%%%%%%%%%%%%%%%%%%%%%%%%%%%%%%%%%%%%%%%%%%%%%%%%%%%%%%%%%%%%%
\section{Noisy Measurements}
\label{generalizations}
We adopt the following model for the case where the measurements are noisy \cite{CRTAug06}:
\[
\bs{c} = \bs{G}\bs{v} + \bs{n}.
\]
In this new model, $\bs{v}$ and $\bs{G}$ are the original signal and the sensing matrix, respectively. The new term, $\bs{n}$, represents the noise vector added to the noiseless measurements $\bs{G}\bs{v}$, resulting in the noisy measurement vector $\bs{c}$. Elements of the noise vector are assumed to be i.i.d. Gaussian random variables with mean $0$ and variance $\sigma^2$. The addition of noise to the measurements results in the following two probabilities to be zero: 1) the probability of having a zero measurement, and 2) the probability of having two equal measurements. This will disable the ZCN and ECN rules in recovering the signal elements. Without the ZCN and ECN rules, zero-valued variable nodes are not verified, and consequently, no check node will have a reduced degree in the subgraph induced by unverified variable nodes. Therefore, the D1CN rule will also be ineffective.

In the context of message-passing algorithms, there are generally two approaches to deal with noisy measurements. In the first approach, the original formulation of the problem is changed so that the noise is taken into consideration \cite{BSB10,APT10}. In the other approach, the algorithms are changed in order to cope with the presence of noise in the measurements \cite{XH07}. Although the first approach may result in lower reconstruction noise, it requires unbounded message size and is susceptible to approximation errors. The authors in \cite{XH07} instead equipped their VB algorithm with some thresholding techniques and proved that if the original signal is sparse enough, they are able to recover the location and the sign of the non-zero signal elements successfully. In what follows, we propose a similar thresholding technique to deal with the noisy measurements.

Thresholding is a common technique in detection theory to deal with noisy measurements \cite{T01}. We apply this technique to VB algorithms by defining two thresholds $\epsilon_1$ and $\epsilon_2$. We use $\epsilon_1$ to convert small noisy measurements to zero; i.e., any measurement $c$, such that $|c| < \epsilon_1$, is set to zero. We use $\epsilon_2$ as the acceptable tolerance for the equality of two noisy measurements; i.e., we consider two measurements $c_1$ and $c_2$ equal if $|c_1 - c_2| < \epsilon_2$. In this case, we assign $c_1$ and $c_2$ a new common value equal to $(c_1 + c_2)/2$. While the scope of this work is not to optimize thresholds $\epsilon_1$ and $\epsilon_2$, our goal is to demonstrate the potential of thresholding in desensitizing the VB algorithms to the measurement noise. We explain this through an example and by comparing the performance of the SBB algorithm equipped with thresholding and two methods based on $\ell_1$ minimization in the case where the measurements are noisy.

Consider a signal of length $n=1000$. We let the size of the support set, $k$, to increase from $10$ to $150$ in steps of $10$. For each such support size $k$, we pick $k$ out of the $n$ elements randomly, and assign to each element an even integer uniformly distributed in the range $[-1000,1000]$, independent of the value of the other nonzero elements. In the case of the SBB algorithm, the signal is measured through a $(3,6)$ unweighted bigraph. In the case of $\ell_1$-based algorithms, we use sensing matrices consisting of orthonormal columns with Standard Gaussian elements \cite{CRTAug06}. In all cases, the number of measurements, $m$, is fixed at $500$. Each measurement is independently contaminated with a Gaussian noise of mean $0$ and variance equal to $\sigma^2 = 0.25$.

For the SBB, we set both thresholds $\epsilon_1$ and $\epsilon_2$ equal to 1.99. Since the value of nonzero signal elements are drawn from a finite alphabet and since the graph is unweighted, false alarm may occur with nonzero probability in the recovery process. In our simulations, we consider a recovery algorithm ``successful'' if it can fully recover the support set.

The first $\ell_1$-based recovery algorithm is the $\ell_1$ regularization method introduced in \cite{CRTAug06}. For the second algorithm, we empower the $\ell_1$ regularization algorithm with the knowledge of the size of the support set. The algorithm thus keeps the $k$ components that are the largest in magnitude and converts the rest to zero. To simulate the two $\ell_1$-based decoders, we use the L1MAGIC package available in \cite{L1}.

\begin{figure}[!ht]
\vspace{1cm}
\centering
\includegraphics[height=250 pt]{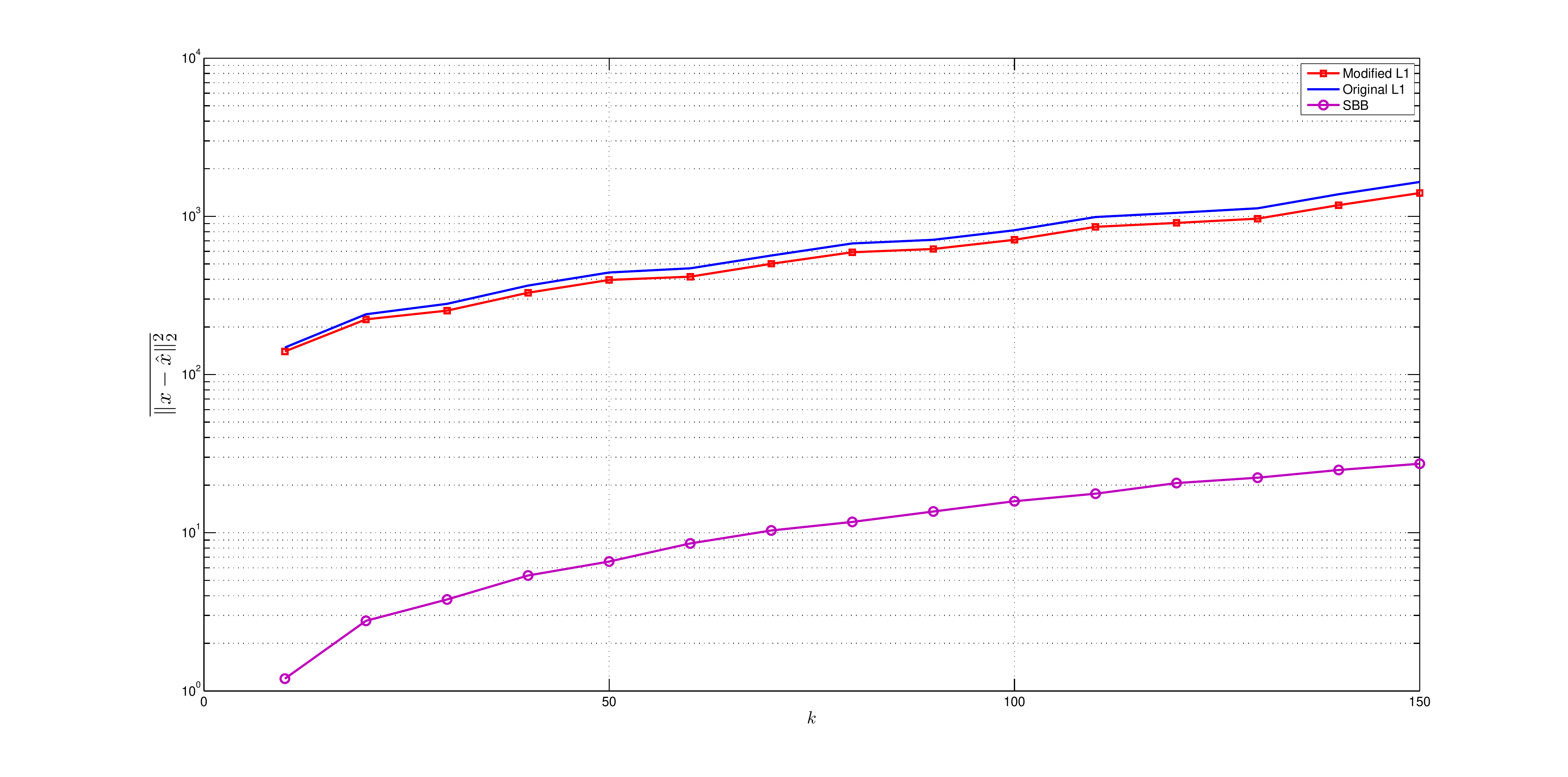}
\caption{Comparison between SBB, $\ell_1$ minimization, and modified $\ell_1$ minimization in terms of MS reconstruction error in the presence of Gaussian measurement noise with zero mean and variance $\sigma^2 = 0.25$ (n = 1000, m=500).}
\label{Comparison1}
\end{figure}

As the measure of performance, we consider the mean square error (MSE) between the original and the recovered signals. For each value of $k$, we perform simulations until we obtain $100$ ``successful'' recovery instances. The results for the 3 algorithms are reported in Fig. \ref{Comparison1}, where for each algorithm, MSE averaged over all simulated cases for a given value of $k$ is shown.

As can be seen, the SBB recovery algorithm significantly (by about two orders of magnitude) outperforms both $\ell_1$-based algorithms.
%%%%%%%%%%%%%%%%%%%%%%%%%%%%%%%%%%%%%%%%%%%%%%%%%%%%%%%%%%%%%%%%%%%%%%%%%%%%%%%%%%%%%%%%%%%%%%%%%%%%%%%%%%%%%
%%%%%%%%%%%%%%%%%%%%%%%%%%%%%%%%%%%%%%%%%%%%%%%%%%%%%%%%%%%%%%%%%%%%%%%%%%%%%%%%%%%%%%%%%%%%%%%%%%%%%%%%%%%%%
\section{Simulation Results}
\label{simulation}
In this section, we present simulation results obtained by running the recovery algorithms over random biregular graphs to recover sparse signals of finite length $n$. We also present analytical results obtained through the mathematical analysis described in Section \ref{analysis} for the asymptotic regime when $n \rightarrow\infty$. This includes the success threshold of different VB algorithms over different biregular graphs. The comparison of asymptotic and finite-length results shows that there is a good agreement between the two for moderately large block lengths ($n \geq 10^5$).

In all simulations, a signal element belongs to the support set with probability $\alpha^{(0)}$, unless otherwise specified. Also, each nonzero signal element (variable) is drawn according to a standard Gaussian distribution. The biregular graphs are constructed randomly with no parallel edges and all the edge weights are equal to one. In each set of simulations, the sensing graph is fixed and each simulation point is generated by averaging over 1000 random instances of the input signal, unless specified otherwise. We repeated each simulation with different randomly generated graphs (with the same variable and check node degrees), and observed that the results were almost identical for every graph. Each simulation is run until the algorithm makes no further progress. In this case, if the signal is recovered perfectly, the recovery is called successful, otherwise a failure is declared.

For the analytical results, based on the fact that $\alpha^{(\ell)}$ is a non-increasing function of iteration number $\ell$, we consider the following stopping criteria:
\begin{enumerate}
	\item $\alpha^{(\ell)} \leq 10^{-7}$,
	\item $\alpha^{(\ell)} > 10^{-7}$ and $|\alpha^{(\ell)} - \alpha^{(\ell-1)}| < 10^{-8}$.
\end{enumerate}
If the analysis stops based on the first stopping criterion, the algorithm is considered successful. If, on the other hand, it stops based on the second criterion, the algorithm is considered unsuccessful and a failure is reported. To calculate the success threshold, a binary search is performed until the separation between the start and the end of the search region is less than $10^{-5}$.

To motivate the use of recovery algorithms over sparse graphs, as the first simulation result, we present the comparison between the SBB algorithm and two benchmark $\ell_1$-based algorithms, $\ell_1$ minimization \cite{CRTFeb06} and iterative weighted $\ell_1$ minimization \cite{CWB04}. The setup is as follows. For SBB, we choose a random $(3,6)$ biregular sensing graph with 1000 variable nodes and 500 check nodes. The sensing matrix used for the two $\ell_1$-based algorithms consists of 500 rows and 1000 columns. The elements are initially i.i.d. standard Gaussian random variables. Then the rows are made orthonormal.

The cost functions used in $\ell_1$ and weighted $\ell_1$ minimization algorithms are $\|\bs{v}\|_1:=\sum_i |\bs{v}_i|$ and $\|\bs{W} \bs{v}\|_{1}:=\sum_i w_i|\bs{v}_i|$, respectively, where $\bs{v}$ is the original signal of interest with elements $\bs{v}_i$, and $\bs{W}$ is a diagonal matrix with positive diagonal elements $w_i$ representing the weights. Weighted $\ell_1$ minimization is an iterative algorithm in which the weights at iteration $t$ ($w^{(t)}_i$) are updated according to $w^{(t)}_i = 1/(|\bs{v}^{(t-1)}_i| + \epsilon)$, where $\bs{v}^{(t-1)}_i$ is the estimate of the signal element $\bs{v}_i$ at iteration $t-1$. The weighted $\ell_1$ is not very sensitive to the parameter $\epsilon$ as noted in \cite{CWB04}. We found $\epsilon = 0.1$ is a good choice based on our simulations. Regarding the maximum number of iterations for the weighted $\ell_1$ minimization algorithm, authors in \cite{CWB04} show that as this parameter increases, better results are achieved, with the cost of longer running time. The improvement gained by increasing the number of iterations beyond $6$ however, is negligible \cite{CWB04}. Therefore, in our simulations, we choose a conservative maximum number of iterations equal to $10$. As $\ell_1$ and weighted $\ell_1$ minimization algorithms output an estimate which is very close to the original signal, but not exactly the same, we declare a success for these two algorithms if the difference between every original signal element and its corresponding estimate is less than $10^{-2}$. Lastly, we use the L1MAGIC package in \cite{L1} as the optimization engine for simulating $\ell_1$ and weighted $\ell_1$ minimization algorithms.

\begin{figure}[!h]
\vspace{10pt}
\centering
\includegraphics[height=250 pt]{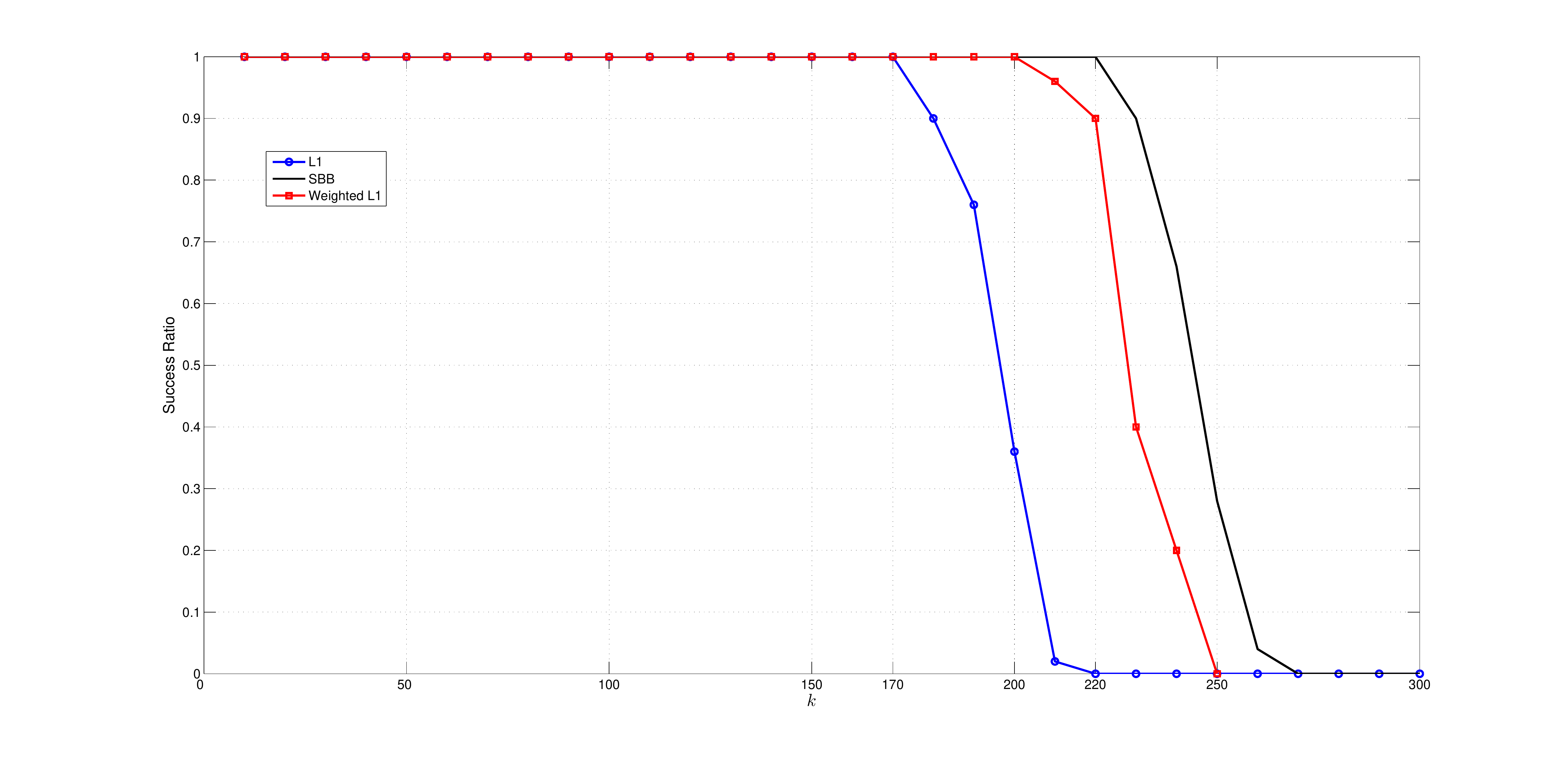}
\caption{Comparison between success ratios of $\ell_1$, weighted $\ell_1$ and SBB algorithms for $n=1000, m=500$.}
\label{L1_WL1_SBB_Success}
\end{figure}

\begin{figure}[!h]
\centering
\includegraphics[height=250 pt]{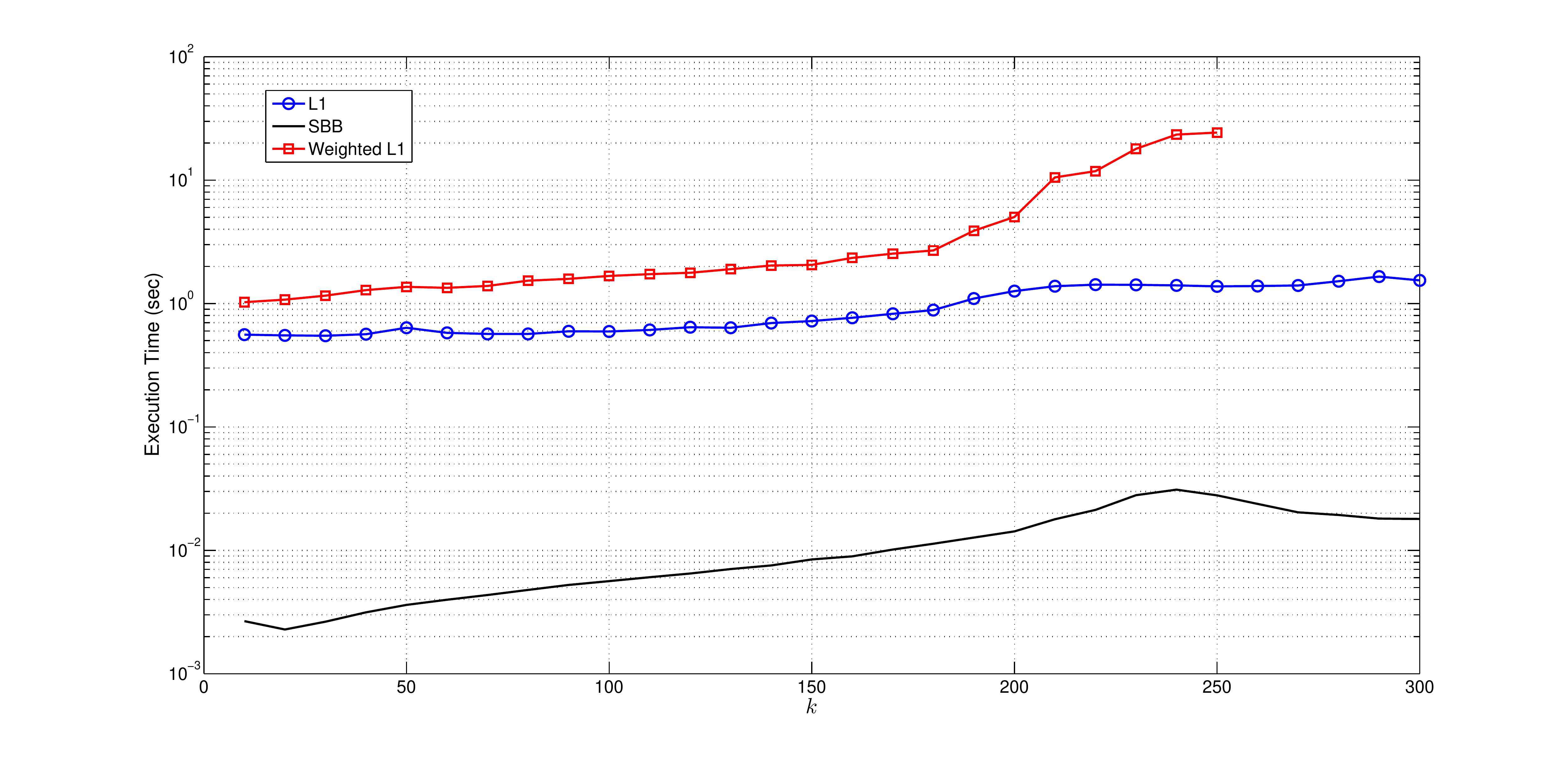}
\caption{Comparison between the average running times of $\ell_1$, weighted $\ell_1$ and SBB algorithms for $n=1000, m=500$.}
\label{Total_Exe_time}
\end{figure}

To have a fair comparison, the same signal vectors are used for all the algorithms. We also choose the size of the support set deterministically, and let the size range from 10 to 300. For each support size, 100 instances of the signal vector are generated. Each signal vector is then measured according to the corresponding sensing mechanism for each class of algorithms. The success or failure of the recovery algorithms over the resulting measurements are then averaged over the 100 instances, and plotted in Figure \ref{L1_WL1_SBB_Success}. In Figure \ref{Total_Exe_time} the average running time, in seconds, is plotted for the three algorithms. The algorithms were implemented in MATLAB and were run on a computer with an AMD Phenom 9650 Quad-Core Processor 2.3 GHz, 3 GB RAM and a Windows 7 operating system. As can be seen, the SBB algorithm recovers signals with more nonzero elements at a speed which is about 2 orders of magnitude faster compared to that of the $\ell_1$ algorithms.

\begin{figure}[!h]
\centering
\includegraphics[height=250 pt]{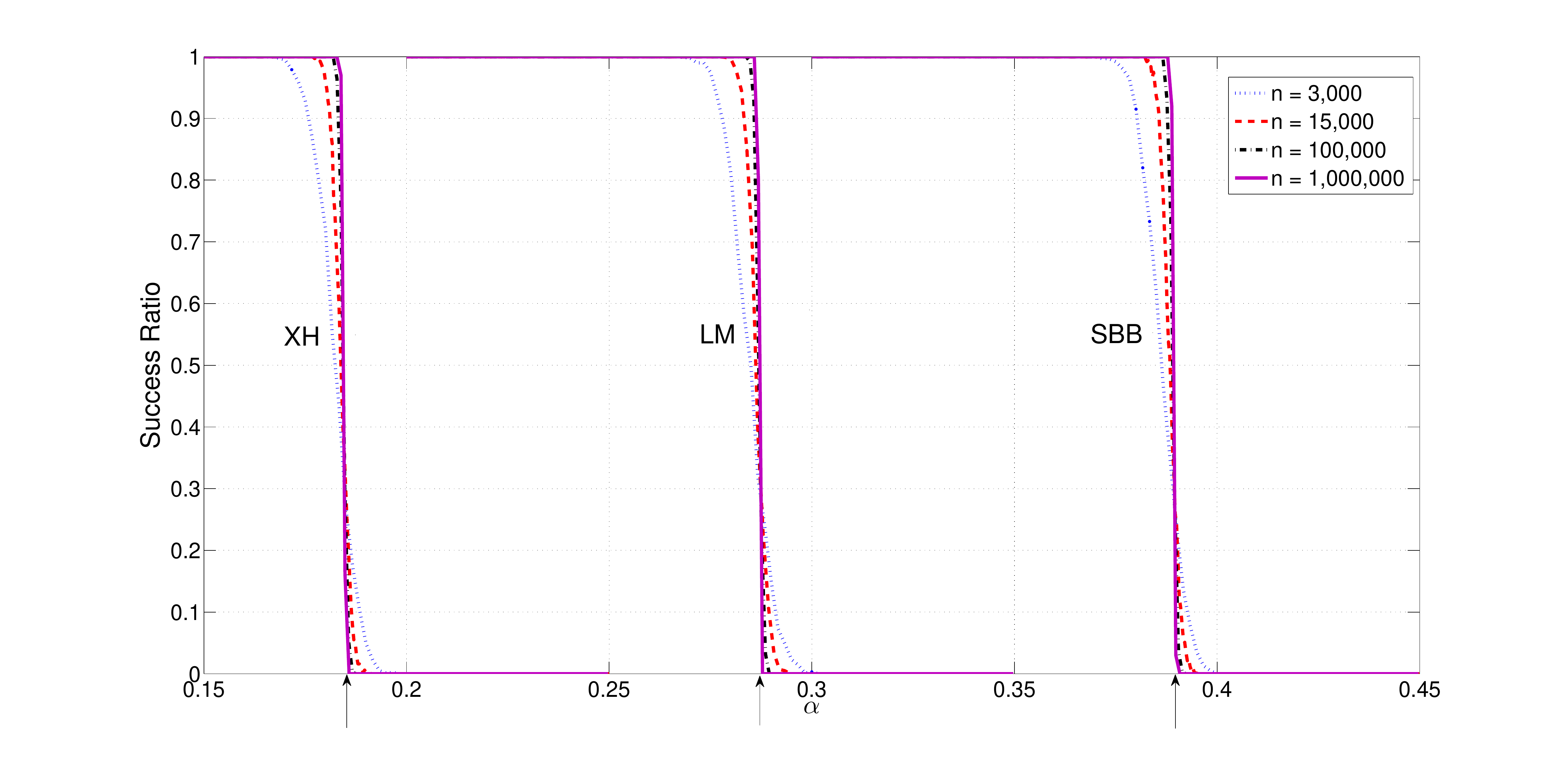}
\caption{Success ratio of XH, LM and SBB algorithms vs. $\alpha = \alpha^{(0)}$ for $(5,6)$ graphs with $n=\left\{3, 15, 100 \text{ and } 1000\right\}\times 10^3$. Analytical thresholds are shown by arrows.}
\label{VarLength}
\end{figure}

For the next experiment, we apply XH, SBB and LM algorithms to four randomly constructed $(5,6)$ regular graphs with $n=\{3, 15, 100, 1000\}\times 10^3$. The success ratio of the algorithms vs. the initial density factor $\alpha = \alpha^{(0)}$ are shown in Figure \ref{VarLength}. From the figure, we can see that, for all algorithms, by increasing $n$, the transition part of the curves becomes sharper such that the curves for $n=10^6$ practically look like a step function. In the figure, we have also shown the success threshold of the algorithms for $(5,6)$ graphs, obtained based on the proposed analysis, by arrows. As can be seen, the thresholds match very well with the waterfall region of the simulation curves.

\vspace{1cm}

In Table \ref{success_threshold_1}, we have listed the analytical success thresholds of the iterative recovery algorithms for graphs with different $d_v$ and $d_c$ values. The result for XH algorithm on $(3,4)$ graphs, and more generally for graphs with $d_v = 3$, is missing as the algorithm performs poorly on such graphs.\footnote{The reason is that for $d_v = 3$, a variable node is verified with the common value of $\lceil d_v/2 \rceil = 2$ check nodes. However, if two nonzero variable nodes share the same two check nodes (a cycle of length 4 exists in the graph), then a false verification may occur.} For every graph, the Genie algorithm has the best performance. This is followed by SBB, LM and XH algorithms, respectively. Careful inspection of the results in Table \ref{success_threshold_1} indicates that the oversampling ratio $r_o = d_v/\alpha d_c$, improves consistently by decreasing both $d_v$ and $d_c$ values. In fact, among the results presented in Table \ref{success_threshold_1}, the application of the Genie and SBB to $(3,4)$ graphs results in the lowest oversampling ratio of $\approx 1.16$ and $\approx 1.67$, respectively.

In Table \ref{success_threshold_2}, we have listed the analytical success thresholds of the iterative recovery algorithms for graphs with compression ratio $d_v/d_c = 0.5$ and different $d_v$ and $d_c$ values. In general, as we decrease $d_v$, algorithms perform better in terms of recovery capability.\footnote{These results are consistent with the results observed for the Belief Propagation (BP) decoding of binary LDPC codes based on biregular graphs.} This also implies that for a fixed compression ratio, the oversampling ratio improves by decreasing $d_v$ and $d_c$. 

In Tables \ref{iter_threshold_1} and \ref{iter_threshold_2}, we have listed the number of iterations required for different recovery algorithms to recover signals with density factor equal to the success thresholds reported in Tables \ref{success_threshold_1} and \ref{success_threshold_2} minus $0.0001$, respectively. These results, which are obtained by the asymptotic analysis are in close agreement with finite-length simulation results at block lengths of about $10^5$. These results indicate that with a few exceptions the better performance comes at the expense of a larger number of iterations. In particular, among the practical recovery algorithms, SBB requires the largest number of iterations for convergence. 

\begin{table}[!h]
	\caption{Success Thresholds for different graphs and algorithms}
	\centering	
%	{\footnotesize{
	\begin{tabular}{|l|c|c|c|c|c|}
		\hline
		$(d_v,d_c)$ & $(3,4)$ & $(5,6)$ & $(5,7)$ & $(5,8)$ & $(7,8)$\\
		\hline
		\hline
		 Genie& 0.6474 & 0.5509 & 0.4786 & 0.4224 & 0.4708\\
		\hline
		 SBB& 0.4488 & 0.3892 & 0.3266 & 0.2806 & 0.3335\\
		\hline
		 LM & 0.3440 & 0.2871 & 0.2305 & 0.1907 & 0.2385\\
		\hline
		 XH& - & 0.1846 & 0.1552 & 0.1339 & 0.1435\\
		\hline
	\end{tabular}
%	}}
	\label{success_threshold_1}
\end{table}
\begin{table}[!h]
	\caption{Success Thresholds for different graphs and algorithms for fixed compression ratio $r_c = 0.5$}
	\begin{center}	
	\begin{tabular}{|l|c|c|c|c|c|c|}
		\hline
		Graphs: $(d_v,d_c)$ & $(3,6)$ & $(4,8)$ & $(5,10)$ & $(6,12)$ & $(7,14)$ & $(8,16)$\\
		\hline
		\hline
		 Genie& 0.4294 & 0.3834 & 0.3415 & 0.3074 & 0.2797 & 0.2568\\
		\hline
		 SBB& 0.2574 & 0.2394 & 0.2179 & 0.1992 & 0.1835 & 0.1703\\
		\hline
		 LM & 0.1702 & 0.1555 & 0.1391 & 0.1253 & 0.1140 & 0.1048\\
		\hline
		 XH& - & 0.1875 & 0.1050 & 0.1170 & 0.0791 & 0.0834\\
		\hline		
	\end{tabular}
	\end{center}
	\label{success_threshold_2}
\end{table}
\begin{table}[!h]
	\caption{Number of iterations required for different recovery algorithms over different graphs to recover a signal with density ratio equal to the success threshold minus $0.0001$}
	\begin{center}	
	\begin{tabular}{|l|c|c|c|c|c|}
		\hline
		Graphs: $(d_v,d_c)$ & $(3,4)$ & $(5,6)$ & $(5,7)$ & $(5,8)$ & $(7,8)$\\
		\hline
		\hline
		 Genie& 106 & 66 & 66 & 62 & 55 \\
		\hline
		 SBB& 655 & 178 & 165 & 200 & 344 \\
		\hline
		 LM & 258 & 139 & 103 & 126 & 108 \\
		\hline
		 XH& - & 63 & 58 & 54 & 41 \\
		\hline		
	\end{tabular}
	\end{center}
	\label{iter_threshold_1}
\end{table}
\begin{table}[!h]
	\caption{Number of iterations required for different recovery algorithms over different graphs with fixed compression ratio $r_c = 0.5$, to recover a signal with density ratio equal to the success threshold minus $0.0001$}
	\begin{center}	
	\begin{tabular}{|l|c|c|c|c|c|c|}
		\hline
		Graphs: $(d_v,d_c)$ & $(3,6)$ & $(4,8)$ & $(5,10)$ & $(6,12)$ & $(7,14)$ & $(8,16)$\\
		\hline
		\hline
		 Genie& 93 & 69 & 57 & 50 & 46 & 41\\
		\hline
		 SBB& 247 & 167 & 172 & 163 & 127 & 108\\
		\hline
		 LM & 142 & 94 & 136 & 97 & 55 & 67\\
		\hline
		 XH& - & 64 & 48 & 38 & 32 & 28\\
		\hline		
	\end{tabular}
	\end{center}
	\label{iter_threshold_2}
\end{table}

\begin{figure}[!h]
\centering
\includegraphics[height=250 pt]{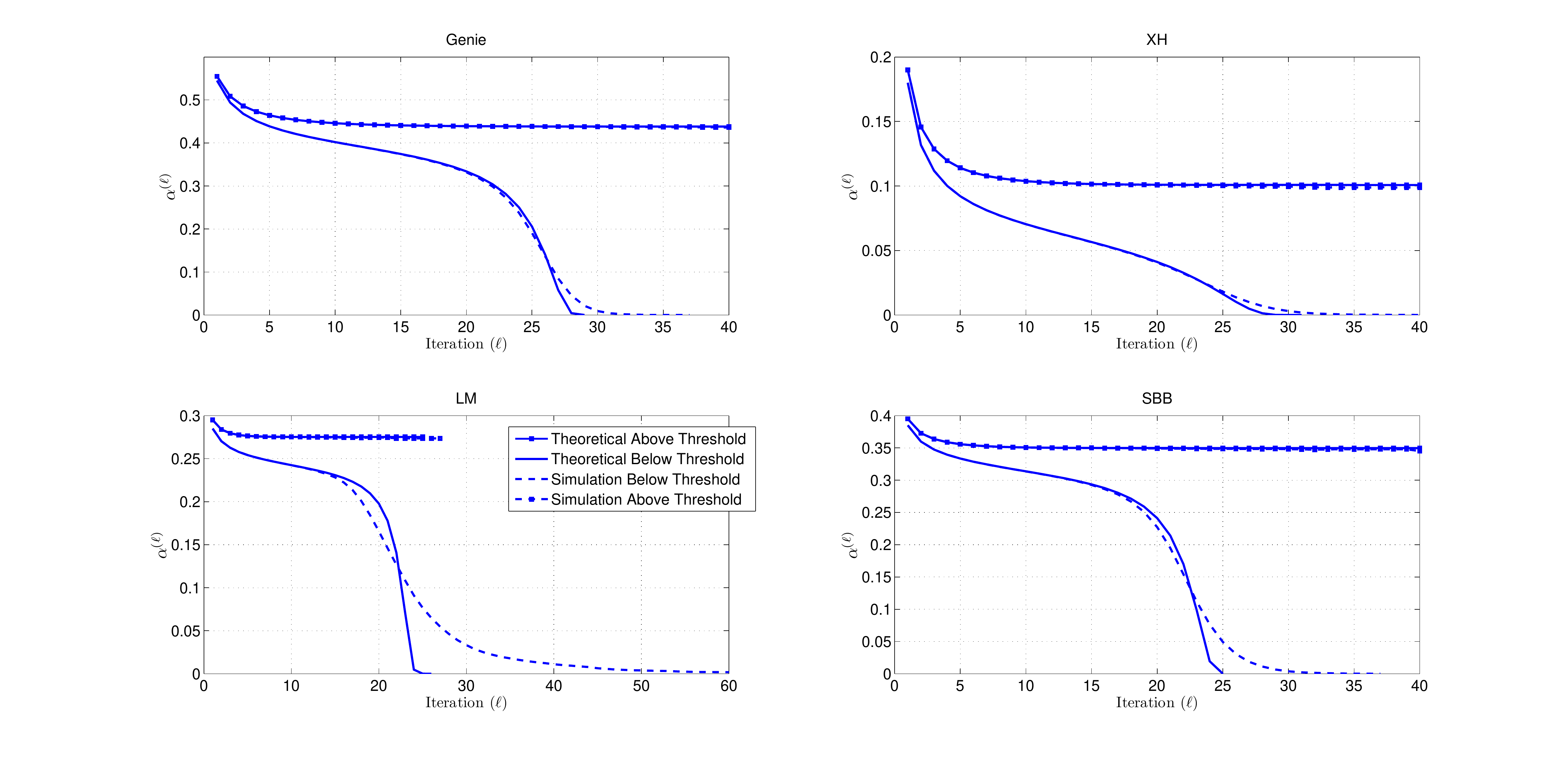}
\caption{Evolution of $\alpha^{(\ell)}$ vs. iteration number $\ell$ for the four recovery algorithms over a $(5,6)$ graph (finite-length simulations are for $n=10^5$).}
\label{G4G6Evolution100k}
\end{figure}

To further investigate the degree of agreement between our theoretical asymptotic analysis and finite-length simulation results, we have presented in Fig. \ref{G4G6Evolution100k} the evolution of $\alpha^{(\ell)}$ with iterations $\ell$ for the four algorithms Genie, LM, SBB, and XH over a $(5,6)$ graph. For each algorithm, two values of $\alpha^{(0)}$ are selected: one above and one below the success threshold presented in Table \ref{success_threshold_1}. The theoretical results are shown by solid lines while simulations for $n = 10^5$ are presented with dotted lines. As one can see, the two sets of results are in close agreement particularly for the cases where $\alpha^{(0)}$ is above the threshold and for smaller values of $\ell$.

\begin{figure}[!ht]
\centering
\includegraphics[height=250 pt]{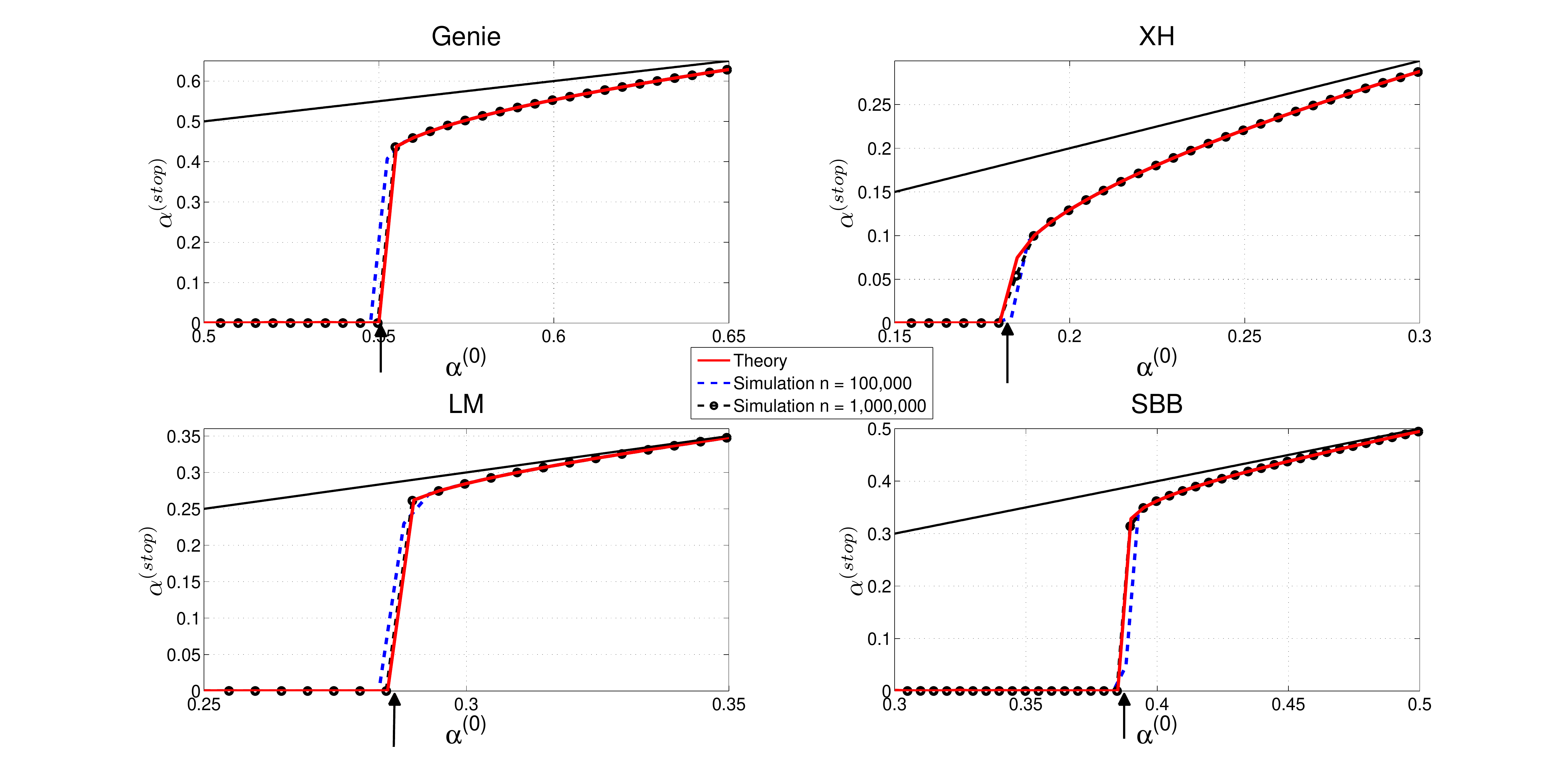}
\caption{Fraction of unrecoverable variable nodes for different recovery algorithms and different starting density factors over random (5,6) regular bipartite graphs. Each arrow represents the theoretical success threshold.}
\label{RecoveredVarNodes}
\end{figure}

Next, for different values of $\alpha^{(0)}$, we estimate the average fraction of unverified nonzero variable nodes $\alpha^{(\ell)}$ using the analysis, and denote the value of $\alpha^{(\ell)}$ at the time that the analysis stops (because one of the stopping criteria is met) as $\alpha^{(stop)}$. These values are plotted vs. the corresponding values of $\alpha^{(0)}$ in Fig. \ref{RecoveredVarNodes} for the four VB recovery algorithms over the $(5,6)$ sensing graphs. In the same figure, we have also given the corresponding simulation results for two randomly selected $(5,6)$ sensing graphs with $n = 10^{5}$ and $10^{6}$. The simulation results for both lengths closely match the analytical results, with those of $n = 10^{6}$ being practically identical to the analytical results. We have indicated the success threshold of the algorithms by arrows. From the figure, it can also be seen that as $\alpha^{(0)}$ increases and tends to one, the curves tend to the asymptote  $\alpha^{(stop)} = \alpha^{(0)}$.

%\setcounter{figure}{0}
%\vspace{5cm}

As the last experiment, we compare the running time and the accuracy of the proposed asymptotic analysis against those of the differential equation approach presented in \cite{ZP07J}. For comparison, a biregular $(3,6)$ graph and the SBB algorithm were chosen. The binary search for the success threshold starts with the interval $[0.2,0.3]$ and ends when the separation between the start and the end of the search region in less than $10^{-5}$. The analysis is implemented in MATLAB and executed on the same computer described before. Using the proposed analysis, we obtain the success threshold of 0.2574 in 23.1 seconds. Table \ref{pcompare} summarizes the results of running the analysis of~\cite{ZP07J} on the same machine for different values of $n$. The reported thresholds increase with the increase in $n$. For $n=100,000$, the running time is roughly 100 times that of our proposed method. Moreover, the obtained threshold of $0.2591$ is only in agreement with the threshold of $0.2574$, obtained by the proposed method, up to two decimal points. In fact, experiments similar to those reported in Fig.~\ref{G4G6Evolution100k} reveal that the accuracy of the threshold obtained by the method of~\cite{ZP07J} is lower than our results. In particular, our simulations show that the SBB algorithm over $(3,6)$ graphs with $n=10^5$ fails for $\alpha^{(0)} = 0.259$, which would imply that the threshold $0.2591$ is only accurate up to two decimal points. 

\begin{table}[!h]
	\caption{Success threshold and running time of the analysis of \cite{ZP07J} for SBB over a random $(3,6)$ regular graph.}
	\begin{center}	
	\begin{tabular}{|l|c|c|c|c|c|c|}
		\hline
		$n$ & $100$ & $1,000$ & $10,000$ & $20,000$ & $50,000$ & $100,000$\\
		\hline
		 Success Threshold & 0.2465 & 0.2577 & 0.2589 & 0.2590 & 0.2590 & 0.2591 \\
		\hline
		 Running Time (seconds) & 1.1 & 9.9 & 103.9 & 220.6 & 647.4 & 2044.1 \\
		\hline
	\end{tabular}
	\end{center}
	\label{pcompare}
\end{table}
%%%%%%%%%%%%%%%%%%%%%%%%%%%%%%%%%%%%%%%%%%%%%%%%%%%%%%%%%%%%%%%%%%%%%%%%%%%%%%%%%%%%%%%%%%%%%%%%%%%%%%%%%%%%%
%%%%%%%%%%%%%%%%%%%%%%%%%%%%%%%%%%%%%%%%%%%%%%%%%%%%%%%%%%%%%%%%%%%%%%%%%%%%%%%%%%%%%%%%%%%%%%%%%%%%%%%%%%%%%
\section{Acknowledgment}
The authors wish to thank H. D. Pfister for providing them with the latest version of \cite{ZP07J}.
%%%%%%%%%%%%%%%%%%%%%%%%%%%%%%%%%%%%%%%%%%%%%%%%%%%%%%%%%%%%%%%%%%%%%%%%%%%%%%%%%%%%%%%%%%%%%%%%%%%%%%%%%%%%%
%%%%%%%%%%%%%%%%%%%%%%%%%%%%%%%%%%%%%%%%%%%%%%%%%%%%%%%%%%%%%%%%%%%%%%%%%%%%%%%%%%%%%%%%%%%%%%%%%%%%%%%%%%%%%
\newpage
\bibliographystyle{IEEEtran}
\bibliography{Journal_biblio}

\begin{thebibliography}{10}
\providecommand{\url}[1]{#1}
\csname url@samestyle\endcsname
\providecommand{\newblock}{\relax}
\providecommand{\bibinfo}[2]{#2}
\providecommand{\BIBentrySTDinterwordspacing}{\spaceskip=0pt\relax}
\providecommand{\BIBentryALTinterwordstretchfactor}{4}
\providecommand{\BIBentryALTinterwordspacing}{\spaceskip=\fontdimen2\font plus
\BIBentryALTinterwordstretchfactor\fontdimen3\font minus
  \fontdimen4\font\relax}
\providecommand{\BIBforeignlanguage}[2]{{%
\expandafter\ifx\csname l@#1\endcsname\relax
\typeout{** WARNING: IEEEtran.bst: No hyphenation pattern has been}%
\typeout{** loaded for the language `#1'. Using the pattern for}%
\typeout{** the default language instead.}%
\else
\language=\csname l@#1\endcsname
\fi
#2}}
\providecommand{\BIBdecl}{\relax}
\BIBdecl

\bibitem{D06}
D.~Donoho, ``Compressed sensing,'' \emph{IEEE Trans. Inform. Theory}, vol. 52
  (4), pp. 1289--1306, April 2006.

\bibitem{CRTFeb06}
E.~Cand\`{e}s, J.~Romberg, and T.~Tao, ``Robust uncertainty principles: Exact
  signal reconstruction from highly incomplete frequency information,''
  \emph{IEEE Trans. Inform. Theory}, pp. 489--509, February 2006.

\bibitem{XH07}
W.~Xu and B.~Hassibi, ``Efficient compressive sensing with deterministic
  guarantees using expander graphs,'' in \emph{Proc. Information Theory
  Workshop (ITW)}, September 2007, pp. 414--419.

\bibitem{WV09}
Y.~Wu and S.~Verd\'{u}, ``Fundamental limits of almost lossless analog
  compression,'' in \emph{Proc. IEEE Int. Symp. Information Theory (ISIT)},
  2009, pp. 359 -- 363.

\bibitem{B07}
R.~G. Baraniuk, ``Compressive sensing,'' \emph{IEEE Signal Processing
  Magazine}, vol.~24, pp. 118--124, July 2007.

\bibitem{T04}
J.~Tropp, ``Topics in sparse approximation,'' Ph.D. dissertation, University of
  Texas at Austin, 2004.

\bibitem{BGIKS08}
R.~Berinde, A.~Gilbert, P.~Indyk, and K.~Strauss, ``Combining geometry and
  combinatorics: A unified approach to sparse signal recovery,'' in \emph{46th
  Annual Allerton Conference on Communication, Control, and Computing},
  September 2008, pp. 798--805.

\bibitem{TG07}
J.~Tropp and A.~Gilbert, ``Signal recovery from random measurements via
  orthogonal matching pursuit,'' \emph{IEEE Trans. Inform. Theory}, vol. 53
  (12), pp. 4655--4666, December 2007.

\bibitem{NV09}
D.~Needell and R.~Vershynin, ``Uniform uncertainty principle and signal
  recovery via regularized orthogonal matching pursuit,'' \emph{Foundations of
  Computational Mathematics}, vol. 9 (3), pp. 317--334, June 2009.

\bibitem{MPTJ06}
S.~Mendelson, A.~Pajor, and N.~Tomczak-Jaegermann, ``Uniform uncertainty
  principle for bernoulli and sub-gaussian ensembles,'' \emph{Constructive
  Approximation}, vol. 28 (3), pp. 277--289, December 2008.

\bibitem{BDDW08}
R.~Baraniuk, M.~Davenport, R.~DeVore, and M.~Wakin, ``A simple proof of the
  restricted isometry property for random matrices,'' \emph{Constructive
  Approximation, Springer New York}, vol.~28, no.~3, pp. 253--263, December
  2008.

\bibitem{CT05}
E.~Cand\`{e}s and T.~Tao, ``Decoding by linear programming,'' \emph{IEEE Trans.
  Inform. Theory}, vol.~51, no.~12, pp. 4203--4215, December 2005.

\bibitem{C08}
V.~Chandar, ``A negative result concerning explicit matrices with the
  restricted isometry property,'' \emph{Preprint}, March 2008.

\bibitem{CM06}
G.~Cormode and M.~Muthukrishnan, ``Combinatorial algorithms for compressed
  sensing,'' in \emph{Proc. Structural Information and Communication Complexity
  (SIROCCO)}, 2006, pp. 280--294.

\bibitem{I08}
P.~Indyk, ``Explicit constructions for compressed sensing of sparse signals,''
  in \emph{Proc. Symp. on Discrete Algorithms (SODA)}, 2008.

\bibitem{GSTV06}
A.~C. Gilbert, M.~J. Strauss, J.~A. Tropp, and R.~Vershynin, ``Algorithmic
  linear dimension reduction in the $\ell_1$ norm for sparse vectors,'' in
  \emph{44th Annual Allerton Conference on Communication, Control, and
  Computing}, 2006.

\bibitem{GSTV07}
------, ``One sketch for all: Fast algorithms for compressed sensing,'' in
  \emph{Proc. 39th ACM Symposium on Theory of Computing (STOC)}, 2007, pp.
  237--246.

\bibitem{SBB206}
S.~Sarvotham, D.~Baron, and R.~Baraniuk, ``Sudocodes - fast measurement and
  reconstruction of sparse signals,'' in \emph{Proc. IEEE Int. Symp.
  Information Theory (ISIT)}, July 2006, pp. 2804--2808.

\bibitem{CSW10}
V.~Chandar, D.~Shah, and G.~W. Wornell, ``A simple message-passing algorithm
  for compressed sensing,'' in \emph{Proc. IEEE Int. Symp. Information Theory
  (ISIT)}, June 2010, pp. 1968--1972.

\bibitem{ZP09}
\BIBentryALTinterwordspacing
F.~Zhang and H.~D. Pfister, ``On the iterative decoding of high rate ldpc codes
  with applications in compressed sensing.'' [Online]. Available:
  \url{http://arxiv.org/abs/0903.2232}
\BIBentrySTDinterwordspacing

\bibitem{ZP08}
------, ``Compressed sensing and linear codes over real numbers,'' in
  \emph{Proc. Information Theory and Applications Workshop}, February 2008, pp.
  558--561.

\bibitem{ZP07}
------, ``List-message passing achieves capacity on the q-ary symmetric channel
  for large q,'' in \emph{IEEE Global Telecommunications Conference
  (GLOBECOM)}, November 2007, pp. 283--287.

\bibitem{ZP07J}
------, ``List-message passing achieves capacity on the q-ary symmetric channel
  for large q,'' \emph{Preprint Submitted to IEEE Trans. Inform. Theory}.

\bibitem{LMPDK08}
Y.~Lu, A.~Montanari, B.~Prabhakar, S.~Dharmapurikar, and A.~Kabbani, ``Counter
  braids: A novel counter architecture for per-flow measurement,'' in
  \emph{Proc. International Conference on Measurement and Modeling of Computer
  Systems ACM SIGMETRICS}, June 2008, pp. 121--132.

\bibitem{APT10}
M.~Akcakaya, J.~Park, and V.~Tarokh, ``Low density frames for compressive
  sensing,'' in \emph{Proc. IEEE Int. Conf. Acoustics, Speech, and Signal
  Processing (ICASSP)}, March 2010, pp. 3642--3645.

\bibitem{BSB10}
D.~Baron, S.~Savotham, and R.~G. Baraniuk, ``Bayesian compressive sensing via
  belief propagation,'' \emph{IEEE Transactions on Signal Processing}, vol. 58
  (1), pp. 269--280, January 2010.

\bibitem{R94}
C.~P. Robert, \emph{The Bayesian Choise: A Decision Theoretic
  Motivation}.\hskip 1em plus 0.5em minus 0.4em\relax Springer-Verlag, 1994.

\bibitem{PS82}
C.~C. Paige and M.~A. Saunders, ``Lsqr: Sparse linear equations and least
  squares problems,'' \emph{ACM Transactions on Mathematical Software (TOMS)},
  vol.~8, pp. 195--209, June 1982.

\bibitem{LM05}
M.~Luby and M.~Mitzenmacher, ``Verification-based decoding for packet-based
  low-density parity-check codes,'' \emph{IEEE Trans. Inform. Theory}, vol. 51
  (1), pp. 120--127, January 2005.

\bibitem{LMP08}
Y.~Lu, A.~Montanari, and B.~Prabhakar, ``Counter braids: Asymptotic optimality
  of the message passing decoding algorithm,'' in \emph{46th Annual Allerton
  Conference on Communication, Control, and Computing}, September 2008, pp. 209
  -- 216.

\bibitem{LMSS01}
M.~G. Luby, M.~Mitzenmacher, M.~A. Shokrollahi, and D.~A. Spielman, ``Efficient
  erasure correcting codes,'' \emph{IEEE Trans. Inform. Theory}, vol.~47, pp.
  569--584, February 2001.

\bibitem{LMSSF01}
M.~Luby, M.~Mitzenmacher, M.~A. Shokrollahi, and D.~Spielman, ``Improved
  low-density parity-check codes using irregular graphs,'' \emph{IEEE Trans.
  Infor. Theory}, vol.~47, pp. 585--598, February 2001.

\bibitem{M}
D.~J. MacKay, \emph{Information Theory, Inference, and Learning
  Algorithms}.\hskip 1em plus 0.5em minus 0.4em\relax Cambridge University
  Press, 2003.

\bibitem{BG96}
C.~Berrou and A.~Glavieux, ``Near optimum error correcting coding and decoding:
  turbo-codes,'' \emph{IEEE Trans. Comm.}, vol. 44 (10), pp. 1261--1271,
  October 1996.

\bibitem{RU01}
T.~J. Richardson and R.~L. Urbanke, ``The capacity of low-density parity-check
  codes under message-passing decoding,'' \emph{IEEE Trans. Inform. Theory},
  vol. 47 (2), pp. 599--618, February 2001.

\bibitem{AS:2008}
N.~Alon and J.~H. Spencer, \emph{The Probabilistic Method}.\hskip 1em plus
  0.5em minus 0.4em\relax Wiley Series in Discrete Mathematics and
  Optimization, 2008.

\bibitem{CRTAug06}
E.~Cand\`{e}s, J.~Romberg, and T.~Tao, ``Stable signal recovery from incomplete
  and inaccurate measurements,'' \emph{Communications on Pure Mathematics},
  vol.~59, pp. 1207--1223, August 2006.

\bibitem{T01}
H.~L.~V. Trees, \emph{Detection, Estimation, and Modulation Theory, Part
  I}.\hskip 1em plus 0.5em minus 0.4em\relax John Wiley \& Sons, 2001.

\bibitem{L1}
\BIBentryALTinterwordspacing
 [Online]. Available: \url{http://www.acm.caltech.edu/l1magic/}
\BIBentrySTDinterwordspacing

\bibitem{CWB04}
E.~Cand\`{e}s, M.~Wakin, and S.~Boyd, ``Enhancing sparsity by reweighted l1
  minimization,'' \emph{The Journal of Fourier Analysis and Applications},
  vol.~14, no. 5-6, pp. 877--905, December 2008.

\end{thebibliography}
%%%%%%%%%%%%%%%%%%%%%%%%%%%%%%%%%%%%%%%%%%%%%%%%%%%%%%%%%%%%%%%%%%%%%%%%%%%%%%%%%%%%%%%%%%%%%%%%%%%%%%%%%%%%%
%%%%%%%%%%%%%%%%%%%%%%%%%%%%%%%%%%%%%%%%%%%%%%%%%%%%%%%%%%%%%%%%%%%%%%%%%%%%%%%%%%%%%%%%%%%%%%%%%%%%%%%%%%%%%
\newpage
\appendices
\section{Detailed Description of the Analysis for Genie}
\label{app_original_Genie}
\subsection{General Setup}
To derive the update rules in the analysis of the Genie, assume that we are at the start of the first half-round of iteration $\ell$. We thus have the probabilities $\alpha^{(\ell)}$, $\ra{p}^{(\ell-1,1)}_{\ma{N}_i}$, and $\ra{p}^{(\ell-1,2)}_{\ma{K}_j}$, and are interested in deriving the same probabilities for iteration $\ell+1$. The update rules relate the two sets of probabilities; one at the beginning of iteration $\ell$ and the other at the end of iteration $\ell$. Hence, we first derive the update rules for iteration $\ell$, then we discuss the initial conditions for iteration 0.

In the following, we use the notation $s_e$ to refer to the status bit in the message transmitted over edge $e$ from variable node to check node.

\subsection{Derivation of Formulas}
When a variable node is recovered in the second half-round of iteration $\ell-1$, the $d_v$ edges adjacent to the variable node carry the recovery message to the neighboring nodes. Therefore, check nodes neighbor to the recovered variable node face a reduction in their degree. We denote by $\ra{p}^{(\ell)}_{\ma{N}_{j \downarrow i}}$ the probability that the degree of a check node in the induced subgraph is reduced from $i$ to $j\leq i$ after the first half-round of iteration $\ell$. This happens if out of $i$ edges emanating from the check node and incident to the set of unresolved variable nodes $\ma{K}^{(\ell)}$, $i-j$ of them carry a message from their variable nodes indicating that they have been recovered.

\begin{figure}[ht]
\centering
\includegraphics[bb = 0 0 400 200 pt,scale = .5]{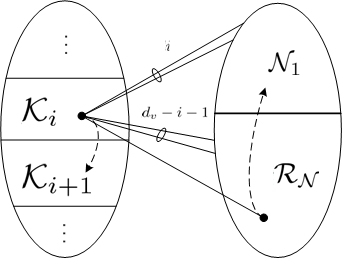}
\caption{A variable node in $\ma{K}_i$ turns to a variable node in $\ma{K}_{i+1}$.}
\label{appA_rec}
\end{figure}

On the other side of the graph, when a variable node in $\ma{K}^{(\ell-1,2)}_i$ ($1\leq i\leq d_v$) is recovered, by definition, out of $d_v$ check nodes receiving the recovery message, $i$ have degree $1$ and $d_v-i$ have degree $j$ ($2\leq j\leq d_c$). In the asymptotic case, as $n$ grows large, we may assume that for each recovered variable node, the set of $i$ check nodes of degree $1$ and the set of $d_v-i$ check nodes of degree more than $1$ are distributed independently and uniformly with respect to the set of all check nodes of degree $1$ and the set of all check nodes of degree more than $1$, respectively. As one set contains check nodes of only degree $1$ and the other contains check nodes with a variety of degrees, we shall differentiate between $\ra{p}^{(\ell)}_{\ma{N}_{j \downarrow i}}$ for $j=1$ and $j>1$. Once these probabilities are found, the new distribution of check node degrees $\ra{p}^{(\ell,1)}_{\ma{N}_i}$ can be derived using the total probability law:
\[
\ra{p}^{(\ell,1)}_{\ma{N}_i} = \sum_{j=i}^{d_c}{\ra{p}^{(\ell-1,1)}_{\ma{N}_{j}} \ra{p}^{(\ell)}_{\ma{N}_{j \downarrow i}}}, \hspace{20pt} 0\leq i\leq d_c.
\]
To find the probability $\ra{p}^{(\ell)}_{\ma{N}_{j \downarrow i}}$, $j\geq 2$, we need the conditional probability that an edge connecting a check node in the set $\ma{N}^{(\ell-1,1)}_j$ and an unverified variable node, carries a recovered message to the check node in the first half-round of iteration $\ell$. We denote this conditional probability by $\ra{p}^{(\ell)}_{d>1}$. Assuming this probability is known, we have:
\[
\ra{p}^{(\ell)}_{\ma{N}_{j \downarrow i}} = {j\choose{j-i}} \left(p^{(\ell)}_{d>1}\right)^{j-i}\left(1-p^{(\ell)}_{d>1}\right)^i,\hspace{20pt} j=2,\cdots,d_c,\hspace{20pt} i=0,\cdots,j.
\]
The probability $\ra{p}^{(\ell)}_{d>1}$ can be computed as follows.
\begin{align*}
\ra{p}^{(\ell)}_{d>1} &= \Pr[s_e = 1|c\in\{\ma{N}^{(\ell-1,1)}_2,\cdots,\ma{N}^{(\ell-1,1)}_{d_c}\},v\in\ma{K}^{(\ell-1)}],\\
&= \ds_{i=1}^{d_v} {\Pr[v\in \ma{K}^{(\ell-1,2)}_i|c\in\{\ma{N}^{(\ell-1,1)}_2,\cdots,\ma{N}^{(\ell-1,1)}_{d_c}\},v\in\ma{K}^{(\ell-1)}]},\\
&= \ds_{i=1}^{d_v} {\frac{\Pr[c\in\{\ma{N}^{(\ell-1,1)}_2,\cdots,\ma{N}^{(\ell-1,1)}_{d_c}\}|v\in \ma{K}^{(\ell-1,2)}_i,v\in\ma{K}^{(\ell-1)}] \Pr[v\in \ma{K}^{(\ell-1,2)}_i|v\in\ma{K}^{(\ell-1)}]}{\Pr[c\in\{\ma{N}^{(\ell-1,1)}_2,\cdots,\ma{N}^{(\ell-1,1)}_{d_c}\}|v\in\ma{K}^{(\ell-1)}]}},\\
&= \ds_{i=1}^{d_v} {\frac{\left(1-\Pr[c\in\ma{N}^{(\ell-1,1)}_1|v\in \ma{K}^{(\ell-1,2)}_i,v\in\ma{K}^{(\ell-1)}]\right) \Pr[v\in \ma{K}^{(\ell-1,2)}_i|v\in\ma{K}^{(\ell-1)}]}{\left(1-\Pr[c\in\ma{N}^{(\ell-1,1)}_1|v\in\ma{K}^{(\ell-1)}]\right)}},\\
&= \ds_{i=1}^{d_v} {\df{\left(1-\df{i}{d_v}\right)\ra{p}^{(\ell-1,2)}_{\ma{K}_i}}{1-p^{(\ell)}}}
= \df{\ds_{i=1}^{d_v} {\ra{p}^{(\ell-1,2)}_{\ma{K}_i}} - \ds_{i=1}^{d_v} {\df{i}{d_v}\ra{p}^{(\ell-1,2)}_{\ma{K}_i}}}{1 - p^{(\ell)}},
\end{align*}
where, $p^{(\ell)}$ is defined as the probability of the edge $e$ being adjacent to the check node $c \in \ma{N}^{(\ell-1,1)}_1$ conditioned on the fact that the variable node $v \in \ma{K}^{(\ell-1)}$. This probability can be calculated as:
\begin{align*}
p^{(\ell)} &= \Pr[c\in\ma{N}^{(\ell-1,1)}_1|v\in\ma{K}^{(\ell-1)}],\\
&= \ds_{i=0}^{d_v} \Pr[c\in\ma{N}^{(\ell-1,1)}_1, v\in\ma{K}^{(\ell-1,2)}_i|v\in\ma{K}^{(\ell-1)}],\\
&= \ds_{i=0}^{d_v} \Pr[c\in\ma{N}^{(\ell-1,1)}_1| v\in\ma{K}^{(\ell-1,2)}_i,v\in\ma{K}^{(\ell-1)}] \Pr[v\in\ma{K}^{(\ell-1,2)}_i|v\in\ma{K}^{(\ell-1)}],\\
&= \ds_{i=0}^{d_v} \df{i}{d_v} \ra{p}^{(\ell-1,2)}_{\ma{K}_i}.
\end{align*}
Therefore, the probability $\ra{p}^{(\ell)}_{d>1}$ can be simplified as follows.
\begin{align*}
\ra{p}^{(\ell)}_{d>1} &= \df{\ds_{i=1}^{d_v} {\ra{p}^{(\ell-1,2)}_{\ma{K}_i}} - \ds_{i=1}^{d_v} {\df{i}{d_v}\ra{p}^{(\ell-1,2)}_{\ma{K}_i}}}{1 - p^{(\ell)}},\\
&= \df{1 - \ra{p}^{(\ell-1,2)}_{\ma{K}_0} - p^{(\ell)}}{1 - p^{(\ell)}},\\
&= 1 - \df{\ra{p}^{(\ell-1,2)}_{\ma{K}_0}}{1 - p^{(\ell)}}.
\end{align*}
By using Bayes' rule, the probability $p^{(\ell)}$ can be calculated more efficiently as:
\begin{align}
p^{(\ell)} &= \Pr[c\in\ma{N}^{(\ell-1,1)}_1|v\in\ma{K}^{(\ell-1)}],\notag\\
&= \df{\Pr[v\in\ma{K}^{(\ell-1)}|c\in\ma{N}^{(\ell-1,1)}_1]\Pr[c\in\ma{N}^{(\ell-1,1)}_1]}{\Pr[v\in\ma{K}^{(\ell-1)}]},\notag\\
&= \df{\df{1}{d_c}\times \ra{p}^{(\ell-1,1)}_{\ma{N}_1}}{\alpha^{(\ell-1)}},\notag\\
&= \df{\ra{p}^{(\ell-1,1)}_{\ma{N}_1}}{\alpha^{(\ell-1)} d_c}.
	\label{eq:p}	
\end{align}
In the Genie algorithm, all variable nodes neighbor to at least one check node of degree $1$ are recovered. Therefore, all check nodes of degree $1$ must be grouped as degree $0$ in the first half-round of the next iteration. Hence, we have:
\[
\ra{p}^{(\ell)}_{\ma{N}_{1 \downarrow 0}} = 1,
\hspace{1cm}
\ra{p}^{(\ell)}_{\ma{N}_{1 \downarrow 1}} = 0.
\]
So far, we found the update rules for the first half-round of iteration $\ell$. In the second half-round, variable nodes receive messages from their neighboring check nodes. The degrees reflected in the messages may re-group some variable nodes. According to the verification rule in the Genie algorithm, the only unverified set of variable nodes is the set $\ma{K}^{(\ell-1,2)}_0$. Variable nodes in this set have no connection to check nodes in the set $\ma{N}^{(\ell-1,1)}_1$ but $d_v$ connections to the sets $\{\ma{N}^{(\ell-1,1)}_2,\cdots,\ma{N}^{(\ell-1,1)}_{d_c}\}$. Suppose $v\in\ma{K}^{(\ell-1,2)}_i$. In this case, if one of the adjacent check nodes of $v$ in $\{\ma{N}^{(\ell-1,1)}_2,\cdots,\ma{N}^{(\ell-1,1)}_{d_c}\}$ turns to a check node in $\ma{N}^{(\ell,1)}_1$, $v$ will move from $\ma{K}^{(\ell-1,2)}_i$ to $\ma{K}^{(\ell,2)}_{i+1}$. This is shown in Figure \ref{appA_rec}.

Due to the random structure of the graph assumed in the asymptotic case, the single edges connected to the set of check nodes $\ma{N}^{(\ell,1)}_1$ are uniformly and independently distributed with respect to the candidate edges adjacent to the set of check nodes $\{\ma{N}^{(\ell-1,1)}_2,\cdots,\ma{N}^{(\ell-1,1)}_{d_c}\}$.

In the Genie, the probability $\ra{p}^{(\ell)}_{\ma{K}_{0 \uparrow j}}$ is defined as the probability of a variable node $v\in \ma{K}^{(\ell-1,2)}_0$ turning to $v\in\ma{K}^{(\ell,2)}_j$. As the set $\ma{K}^{(\ell-1,2)}_0$ is also the unverified support set, we have:
\begin{equation}
\label{appA_pxij}
\ra{p}^{(\ell,2)}_{\ma{K}_{j}} = \ra{p}^{(\ell)}_{\ma{K}_{0 \uparrow j}} = \Pr[v\in\ma{K}^{(\ell,2)}_j|v\in\ma{K}^{(\ell-1,2)}_0,v\in\ma{K}^{(\ell)}] = {d_v\choose j}\left(\ra{p}^{(\ell)}_{x}\right)^{j}\left(1-\ra{p}^{(\ell)}_{x}\right)^{d_v-j}, \hspace{20pt} j=0,\cdots,d_v,
\end{equation}
where $\ra{p}^{(\ell)}_{x}$ is defined as the probability that an edge adjacent to a variable node $v \in \ma{K}^{(\ell-1,2)}_0$ carries a message indicating that the adjacent check node $c$ has a degree equal to $1$; i.e., $c \in \ma{N}^{(\ell,1)}_1$. The probability $\ra{p}^{(\ell)}_{x}$ is calculated as follows:
\begin{align}
\ra{p}^{(\ell)}_x &= \Pr[c \in \ma{N}^{(\ell,1)}_1|v\in\ma{K}^{(\ell-1,2)}_0],\notag\\
&= \df{\Pr[c \in \ma{N}^{(\ell,1)}_1] \Pr[v\in\ma{K}^{(\ell-1,2)}_0|c \in \ma{N}^{(\ell,1)}_1]}{\ds_{i=0}^{d_c} \Pr[c \in \ma{N}^{(\ell,1)}_i] \Pr[v\in\ma{K}^{(\ell-1,2)}_0|c \in \ma{N}^{(\ell,1)}_i]},\notag\\
&= \df{\ra{p}^{(\ell,1)}_{\ma{N}_1} \times \df{1}{d_c}}{\ds_{i=0}^{d_c}{\df{i}{d_c} \ra{p}^{(\ell,1)}_{\ma{N}_i}}}
= \df{\ra{p}^{(\ell,1)}_{\ma{N}_1}}{\ds_{i=1}^{d_c}{i \ra{p}^{(\ell,1)}_{\ma{N}_i}}} = \df{\ra{p}^{(\ell,1)}_{\ma{N}_1}}{\alpha^{(\ell)} d_c} = \ra{p}^{(\ell+1)}.
\label{appA_p_x}
\end{align}
In the Genie algorithm, a variable node is resolved if among all its neighboring check nodes, there exists at least one with degree equal to $1$. Hence, the probability of a variable node in the set $\ma{K}^{(\ell)}$ being recovered is calculated as $\sum_{i=1}^{d_v}{\ra{p}^{(\ell,2)}_{\ma{K}_i}}$.
%\begin{align}
%\ds_{i=1}^{d_v}{\ra{p}^{(\ell,2)}_{\ma{K}_i}}.
%	\label{eq:v}
%\end{align}
Therefore, according to the total probability theorem, the probability of a variable node $v$ remaining unverified, i.e., $v\in\ma{K}^{(\ell+1)}$, is:
\[
\alpha^{(\ell+1)} = \alpha^{(\ell)}\left(1 - \ds_{i=1}^{d_v}{\ra{p}^{(\ell,2)}_{\ma{K}_i}}\right) = \alpha^{(\ell)} \ra{p}^{(\ell,2)}_{\ma{K}_0}.
\]

\subsection{Initial Probabilities for the Genie Algorithm}
In the first half-round of iteration $0$, all check nodes have degree $d_c$ in the subgraph induced by the unverified variable nodes. This degree is reflected in their outgoing messages. These messages carry no information to the variable nodes, and no variable node can be resolved based on them. On the other hand, variable nodes not in the support set are resolved with a value equal to $0$. Assuming an initial density factor of $\alpha^{(0)}$, $1-\alpha^{(0)}$ fraction of the messages from variable nodes carry a recovery notice to check nodes. Since at iteration $0$ no variable node in the support set was verified we have $\alpha^{(1)} = \alpha^{(0)}$. Hence, $\ra{p}^{(1)}_{d>1} = 1 - \alpha^{(1)}$. Since $\ra{p}^{(0,1)}_{\ma{N}_{d_c}}=1$, the set of probabilities $\ra{p}^{(1,1)}_{\ma{N}_i}$, is given by the following.
\begin{align*}
\ra{p}^{(1,1)}_{\ma{N}_i} &= \ra{p}^{(0,1)}_{\ma{N}_{d_c}} \ra{p}^{(1)}_{\ma{N}_{d_c \downarrow i}} = \ra{p}^{(1)}_{\ma{N}_{d_c \downarrow i}}, \hspace{20pt} 0\leq i\leq d_c,\\
&= {d_c \choose{d_c-i}} \left(p^{(1)}_{d>1}\right)^{d_c-i}\left(1-p^{(1)}_{d>1}\right)^i,\hspace{20pt} 0\leq i\leq d_c,\\
&= {d_c \choose{i}} \left(\alpha^{(1)}\right)^i \left(1 - \alpha^{(1)}\right)^{d_c-i},\hspace{20pt} 0\leq i\leq d_c.
\end{align*}
To find the probability $\ra{p}^{(1,2)}_{\ma{K}_i}$, we first find the probability $\ra{p}^{(1)}_x$ from (\ref{appA_p_x}) and then replace it in the (\ref{appA_pxij}).
%%%%%%%%%%%%%%%%%%%%%%%%%%%%%%%%%%%%%%%%%%%%%%%%%%%%%%%%%%%%%%%%%%%%%%%%%%%%%%%%%%%%%%%%%%%%%%%%%%%%%%%%%%%%%
%%%%%%%%%%%%%%%%%%%%%%%%%%%%%%%%%%%%%%%%%%%%%%%%%%%%%%%%%%%%%%%%%%%%%%%%%%%%%%%%%%%%%%%%%%%%%%%%%%%%%%%%%%%%%
\newpage
\section{Detailed Description of the Analysis for LM}
\label{app_original_LM}
We first prove Theorem \ref{LMModel} in which the verified variable nodes in the support set are characterized.
\begin{IEEEproof}[Proof of Theorem \ref{LMModel}]
Suppose that we partition the set of unknown variable nodes in the support set $\ma{K}^{(\ell)}$ into subsets $\ma{K}^{(\ell)}_i$, where $i$ represents the number of messages received by the variable node indicating that the transmitting check node has degree $1$; i.e., $i$ neighboring check nodes are in the set $\ma{N}^{(\ell)}_{1,0}$. Since check nodes in $\ma{N}_{1,0}$ have both degree $1$ and a non-zero value, each variable node $v$ in the set $\bigcup_{i=1}^{d_v}\ma{K}^{(\ell)}_i$ has at least one non-zero check node of degree one as a neighbor. Recall that in each iteration of the LM algorithm a non-zero variable node is verified if and only if it is connected to a check node of degree one. Therefore, $v \in \bigcup_{i=1}^{d_v}\ma{K}^{(\ell)}_i$ means that $v$ is recovered at iteration $\ell$ of the LM algorithm.

To prove the converse, we assume that a variable node $v$ is resolved at iteration $\ell$ of the LM algorithm. We show that $v \in \bigcup_{i=1}^{d_v}\ma{K}^{(\ell)}_i$. The fact that $v$ is resolved at iteration $\ell$ of the LM algorithm implies that it was connected to at least one check node of degree one. As we dealt with all zero-valued check nodes in the previous round, the degree-one check nodes must have a non-zero value. Therefore, the variable node $v$ is, by definition, in the set $\bigcup_{i=1}^{d_v}\ma{K}^{(\ell)}_i$.
\end{IEEEproof}
%********************************************************************************************************************************************
%********************************************************************************************************************************************
\subsection{Recovering Variable Nodes}
\begin{itemize}
	\item A variable node in $\ma{K}$ is recovered if it is connected to a check node in $\ma{N}_{1,0}$.
	\item A variable node in $\Delta$ is recovered if it is connected to a check node in $\bigcup_{i=1}^{d_c} \ma{N}_{0,i}$; i.e., a zero-valued check node.
\end{itemize}
%********************************************************************************************************************************************
%********************************************************************************************************************************************
\subsection{Iteration Zero}
\label{APBIterZ}
The messages passed in the first round of iteration zero do not contribute to the verification of variable nodes and do not change the grouping of the check nodes. Iteration zero thus consists of only one round and hence, two half-rounds. In the first half-round, check nodes pass their values along with their degrees ($d_c$) to their neighboring variable nodes. In the second half-round, variable nodes process the incoming messages. A variable node is verified if it receives at least one message with a value equal to zero. In this case, the variable node is verified with a value equal to zero according to the ZCN rule. The set of all variable nodes verified in this half-round make the set $\ma{R}^{(1)}$. 

Let $\ma{N}^{(0)}_i$ denote the set of check nodes with $i$ neighboring variable nodes in the support set. The probability $\ra{p}^{(0)}_{\ma{N}_i}$ defined as the probability that a check node belongs to the set $\ma{N}^{(0)}_i$ is calculated as follows:
\[
\ra{p}^{(0)}_{\ma{N}_i} \triangleq \Pr(c \in \ma{N}^{(0)}_i) = {d_c\choose i} \left(\alpha^{(0)} \right)^i \left(1 - \alpha^{(0)} \right)^{d_c-i},\hspace{20pt}i=0,\cdots,d_c.
\]
Hence, the probability that a check node $c$ has a value equal to zero ($c = 0$) is:
\[
\Pr[c = 0] = \ra{p}^{(0)}_{\ma{N}_0} = \left(1 - \alpha^{(0)} \right)^{d_c}.
\]
Let $\Delta^{(0,R2,2)}_j$ denote the set of zero-valued variable nodes that receive $j$ zero-valued messages. The probability $\ra{p}^{(0,R2,2)}_{\Delta_j}$ defined as the probability that a zero-valued variable node belongs to the set $\Delta^{(0,R2,2)}_j$ is calculated as follows:
\[
\ra{p}^{(0,R2,2)}_{\Delta_j} = \Pr[v \in \Delta^{(0,R2,2)}_j| v \notin \ma{K}^{(0)}] = {d_v\choose j} \left(\ra{p}^{(0)}_\delta \right)^j \left(1 - \ra{p}^{(0)}_\delta \right)^{d_v-j},\hspace{20pt}j=0,\cdots,d_v,
\]
where $\ra{p}^{(0)}_\delta$ is the probability that an edge adjacent to a zero-valued variable node carries a message with value zero, and is calculated as follows:
\begin{equation}
\ra{p}^{(0)}_\delta = \left( 1 - \alpha^{(0)} \right)^{d_c - 1}.
\label{eq:p_delta_0}
\end{equation}
Let $\ra{p}^{(1)}_\Delta$ denote the probability that a variable node has a zero value and does not receive even one message with value equal to zero in the second half-round of iteration zero. We have:
\begin{equation}
\ra{p}^{(1)}_\Delta \triangleq \Pr[v \in \Delta^{(1)}] = \Pr[v \notin \ma{K}^{(0)}] \Pr[v \in \Delta^{(0,R2,2)}_0| v \notin \ma{K}^{(0)}] = (1-\alpha^{(0)})\left(1 - \left(1 - \alpha^{(0)} \right)^{d_c - 1} \right)^{d_v}.
\label{eq:p_Delta_1}
\end{equation}
Since no element of the support set is verified at iteration zero, we have $\ma{K}^{(0)} = \ma{K}^{(1)}$, and hence,
\[
\alpha^{(1)} = \Pr[v \in \ma{K}^{(1)}] = \alpha^{(0)}.
\]
At the end of iteration $0$, all check nodes have degree $d_c$ in the subgraph induced by the unverified variable nodes. Thus, the set of all check nodes can be partitioned into subsets $\ma{N}^{(0,R2,1)}_{i,d_c-i}$, where $i$ denotes the number of neighboring variable nodes in the support set ($0\leq i\leq d_c$). The edges adjacent to a check node are partitioned into two sets: $\ma{K}$-edges and $\Delta$-edges. $\ma{K}$-edges are connected to variable nodes in the support set, while $\Delta$-edges are connected to zero-valued variable nodes. Therefore, a check node in the set $\ma{N}^{(0,R2,1)}_{i,d_c-i}$ ($0\leq i\leq d_c$) has $i$, $\ma{K}$-edges and $d_c-i$, $\Delta$-edges.
%************************************************************************************************************
%************************************************************************************************************
\subsection{Iteration One and Beyond}
Here we present the analysis for iteration one. Since the analysis of the second iteration and beyond is similar to that of iteration one, they are omitted. The summary of the formulas can be found in Section \ref{originalLM}.

The verified messages sent from variable nodes to check nodes at the end of iteration zero, are processed at check nodes at iteration $1$, HR1-R1. Based on the recovery process at iteration zero, all verified messages are sent from variable nodes in the sets $\Delta_j$, $1\leq j\leq d_v$. We partition the set of edges adjacent to a variable node in the set $\Delta_j$, $0\leq j\leq d_v$, into $\ma{N}_{=0}$-edges and $\ma{N}_{\neq 0}$-edges. Edges in the set $\ma{N}_{=0}$-edges are connected to zero-valued check nodes (check nodes in the set $\ma{N}_0$), while edges in the set $\ma{N}_{\neq 0}$-edges are connected to non-zero check nodes.

Since check nodes in the set $\ma{N}^{(0,R2,1)}_0$ receive $d_c$ verified messages, we are interested in the set of check nodes that are regrouped from $\ma{N}^{(0,R2,1)}_{i,d_c-i}$ to $\ma{N}^{(1,R1,1)}_{i,j}$ in HR1-R1 of iteration $1$ for $1\leq i\leq d_c$. Such a set of check nodes is denoted by $\ma{N}^{(1,R1)}_{i,d_c-i \downarrow j}$. The messages responsible for such regrouping are carried over $\ma{N}_{\neq 0}$-edges. To analyze the regrouping, we need to find the probability $\ra{p}_{\ma{E}_R}$ of an edge in the set of $\Delta$-edges to carry a verified message. Such edges are not connected to the set $\Delta_0$. Before finding the probability $\ra{p}_{\ma{E}_R}$, we introduce two notations $v_e$ and $c_e$ to denote the variable node and the check node connected by means of the edge $e$. Now, we have:
\begin{align*}
\ra{p}^{(1,R1)}_{\ma{E}_R} &= 1 - \Pr[v_e \in \Delta^{(0,R2,2)}_0|v_e \notin \ma{K}^{(1)}, c_e \notin \ma{N}^{(1)}_0],\\
&= 1 - \df{\Pr[v_e \in \Delta^{(0,R2,2)}_0|v_e \notin \ma{K}^{(1)}] \Pr[c_e \notin \ma{N}^{(1)}_0|v_e \notin \ma{K}^{(1)}, v_e \in \Delta^{(0,R2,2)}_0]}{\Pr[c_e \notin \ma{N}^{(1)}_0|v_e \notin \ma{K}^{(1)}]},\\
&= 1 - \df{\Pr[v_e \in \Delta^{(0,R2,2)}_0|v_e \notin \ma{K}^{(1)}] \Pr[c_e \notin \ma{N}^{(1)}_0|v_e \notin \ma{K}^{(1)}, v_e \in \Delta^{(0,R2,2)}_0]}{\ds_{i=0}^{d_v} \Pr[c_e \notin \ma{N}^{(1)}_0, v_e \in \Delta^{(0,R2,2)}_i|v_e \notin \ma{K}^{(1)}]},\\
&= 1 - \df{\ra{p}^{(0,R2,2)}_{\Delta_0} \times 1}{\ds_{i=0}^{d_v} \ra{p}^{(0,R2,2)}_{\Delta_i} \left(\df{d_v - i}{d_v}\right)},\\
&= 1 - \df{\ra{p}^{(0,R2,2)}_{\Delta_0}}{1 - \ra{p}^{(0)}_\delta} = 1 - \df{\ra{p}^{(1)}_\Delta}{1 - \ra{p}^{(0)}_\delta},
\end{align*}
where $\ra{p}^{(1)}_\Delta$ and $\ra{p}^{(0)}_\delta$ are given in \eqref{eq:p_Delta_1} and \eqref{eq:p_delta_0}, respectively. We thus have the following regrouping of check nodes based on the second index:
\[
\ra{p}^{(1,R1)}_{\ma{N}_{i,d_c-i \downarrow j}} = {d_c-i \choose j} \left( 1 - \ra{p}^{(1,R1)}_{\ma{E}_R} \right)^j \left( \ra{p}^{(1,R1)}_{\ma{E}_R} \right)^{d_c - i-j},\hspace{20pt}i=1,\cdots,d_v,\hspace{20pt}j=0,\cdots,d_c-i.
\]
Hence,
\begin{equation}
\ra{p}^{(1,R1,1)}_{\ma{N}_{i,j}} = \ra{p}^{(0,R2,1)}_{\ma{N}_{i,d_c-i}} \ra{p}^{(1,R1)}_{\ma{N}_{i,d_c-i \downarrow j}},\hspace{20pt}i=1,\cdots,d_v,\hspace{20pt}j=0,\cdots,d_c-i.
\label{eq:p_N_1R11}
\end{equation}
In HR2-R1, messages are sent from check nodes to variable nodes. At this stage, check nodes in the set $\ma{N}^{(1,R1,1)}_{1,0}$ transmit a message with its first coordinate equal to $1$. Variable nodes in the support set that receive at least one such message, are verified with the value contained in that message in HR2-R1. After processing the received messages, we divide the set of all variable nodes in the support set $\ma{K}^{(1)}$ into subsets $\ma{K}^{(1,R1,2)}_i$, $0\leq i\leq d_v$, where $i$ denotes the number of degree-$1$ messages a variable node receives. We denote the set of such variable nodes by $\ma{K}^{(1,R1)}_{0 \uparrow i}$. Let $p^{(1,R1)}$ denote the probability that an edge adjacent to a variable node in the support set carries a message with the first coordinate equal to $1$. Using the same notations $v_e$ and $c_e$ defined above, we have:
\begin{align}
p^{(1,R1)} &= \Pr[c_e \in \ma{N}^{(1,R1,1)}_{1,0} | v_e \in \ma{K}^{(1)}],\\
&= \df{\Pr[c_e \in \ma{N}^{(1,R1,1)}_{1,0}] \Pr[v_e \in \ma{K}^{(1)} | c_e \in \ma{N}^{(1,R1,1)}_{1,0}]}{\Pr[v_e \in \ma{K}^{(1)}]},\\
&= \df{\ra{p}^{(1,R1,1)}_{\ma{N}_{1,0}} \times 1/d_c}{\alpha^{(1)}} = \df{\ra{p}^{(1,R1,1)}_{\ma{N}_{1,0}}}{\alpha^{(1)} d_c}.
\label{eq:p_1R1}
\end{align}
Hence, the probability $\ra{p}^{(1,R1)}_{\ma{K}_{0 \uparrow i}}$ that a variable node $v \in \ma{K}^{(1)}$ belongs to the set $\ma{K}^{(1,R1,2)}_i$ is calculated as follows:
\begin{align}
\ra{p}^{(1,R1,2)}_{\ma{K}_i} &= \ra{p}^{(1,R1)}_{\ma{K}_{0 \uparrow i}} \triangleq \Pr(v \in \ma{K}^{(1,R1,2)}_i|v \in \ma{K}^{(1)}),\\
&= {d_v\choose i}\left(p^{(1,R1)} \right)^i \left(1 - p^{(1,R1)} \right)^{d_v-i},\hspace{20pt}i=0,\cdots,d_v.
\label{eq:p_K_1R12}
\end{align}
Based on the D1CN rule in the LM algorithm, variable nodes in the set $\bigcup_{i=1}^{d_v} \ma{K}^{(1,R1,2)}_i$ are verified. Therefore, the probability that a variable node in the support set remains unverified for iteration $2$ is as follows:
\[
\alpha^{(2)} = \alpha^{(1)} \left(1 - \ds_{i=1}^{d_v} \ra{p}^{(1,R1,2)}_{\ma{K}_i} \right).
\]
With this, the first round of iteration $1$ is over. In HR1-R2 of iteration $1$, check nodes receive messages from variable nodes. At this point, check nodes should be regrouped based on their first index, as some variable nodes in the support set have been verified at HR2-R1. A check node in the set $\ma{N}^{(1,R1,1)}_{i,k}$ is regrouped into the set $\ma{N}^{(1,R2,1)}_{j,k}$ at HR1-R2, if from $i$ edges in the set of $\ma{K}$-edges adjacent to the check node, $i-j$ of them carry a verified message. The set of such check nodes are denoted by $\ma{N}^{(1,R2)}_{i\downarrow j,k}$. To analyze this regrouping, we need the probability $\ra{p}^{(1,R2)}_d$ that a $\ma{K}$-edge carries a verified message at HR1-R2. To find this probability, we proceed as follows:
\begin{align*}
\ra{p}^{(1,R2)}_{d} &= 1 - \Pr[v_e \in \ma{K}^{(1,R1,2)}_{0} | v_e \in \ma{K}^{(1)},c_e \notin \ma{N}^{(1,R1,1)}_{1,0}],\\
&= 1 - \df{\Pr[v_e \in \ma{K}^{(1,R1,2)}_{0} | v_e \in \ma{K}^{(1)}] \Pr[c_e \notin \ma{N}^{(1,R1,1)}_{1,0} | v_e \in \ma{K}^{(1)},v_e \in \ma{K}^{(1,R1,2)}_{0}]}{\Pr[c_e \notin \ma{N}^{(1,R1,1)}_{1,0} | v_e \in \ma{K}^{(1)}]},\\
&= 1 - \df{\Pr[v_e \in \ma{K}^{(1,R1,2)}_{0} | v_e \in \ma{K}^{(1)}] \Pr[c_e \notin \ma{N}^{(1,R1,1)}_{1,0} | v_e \in \ma{K}^{(1)},v_e \in \ma{K}^{(1,R1,2)}_{0}]}{\ds_{i=0}^{d_v} \Pr[v_e \in \ma{K}^{(1,R1,2)}_{i} | v_e \in \ma{K}^{(1)}] \Pr[c_e \notin \ma{N}^{(1,R1,1)}_{1,0} | v_e \in \ma{K}^{(1)},v_e \in \ma{K}^{(1,R1,2)}_{i}]},\\
&= 1 - \df{\ra{p}^{(1,R1,2)}_{\ma{K}_{0}} \times 1}{\ds_{i=0}^{d_v} \ra{p}^{(1,R1,2)}_{\ma{K}_{i}} \left(\df{d_v-i}{d_v}\right)},\\
&= 1 - \df{d_v\ra{p}^{(1,R1,2)}_{\ma{K}_{0}}}{\ds_{i=0}^{d_v} (d_v-i)\ra{p}^{(1,R1,2)}_{\ma{K}_{i}}},\\
&= 1 - \df{\ra{p}^{(1,R1,2)}_{\ma{K}_{0}}}{1 - p^{(1,R1)}},
\end{align*}
where $\ra{p}^{(1,R1,2)}_{\ma{K}_{0}}$ and $p^{(1,R1)}$ are given by \eqref{eq:p_K_1R12} and \eqref{eq:p_1R1}, respectively. Hence, the probability $\ra{p}^{(1,R2)}_{\ma{N}_{i\downarrow j,k}}$ that a check node belongs to the set $\ma{N}^{(1,R2)}_{i\downarrow j,k}$ is calculated as follows:
\[
\ra{p}^{(1,R2)}_{\ma{N}_{i\downarrow j,k}} = {i\choose{j}} \left(\ra{p}^{(1,R2)}_{d}\right)^{i-j}\left(1-\ra{p}^{(1,R2)}_{d}\right)^j,\hspace{20pt} i=1,\cdots,d_c,\hspace{20pt} j=0,\cdots,i, \hspace{20pt} k=0,\cdots,d_c - i.
\]
Note that by construction, we have $\ra{p}^{(1,R2)}_{\ma{N}_{1\downarrow 0,0}} = 1$ and $\ra{p}^{(1,R2)}_{\ma{N}_{1\downarrow 1,0}} = 0$.
%\[
%\ra{p}^{(1,R2)}_{\ma{N}_{1\downarrow 0,0}} = 1, \hspace{20pt} \ra{p}^{(1,R2)}_{\ma{N}_{1\downarrow 1,0}} = 0.
%\]
After the regrouping, the probability $\ra{p}^{(1,R2,1)}_{\ma{N}_{j,k}}$ that a check node belongs to the set $\ma{N}^{(1,R2,1)}_{j,k}$ is calculated by:
\[
\ra{p}^{(1,R2,1)}_{\ma{N}_{j,k}} = \ds_{i=j}^{d_c} \ra{p}^{(1,R1,1)}_{\ma{N}_{i,k}} \ra{p}^{(1,R2)}_{\ma{N}_{i\downarrow j,k}},\hspace{20pt}j=0,\cdots,d_c, \hspace{20pt} k=0,\cdots,d_c-i.
\]
The measurement corresponding to check nodes in the set $\ma{N}^{(1,R2,1)}_{0,k}$, $1\leq k\leq d_c-1$, changes to zero, as the check nodes are no longer connected to an unverified variable node in the support set. Hence, the messages transmitted by such check nodes have a value equal to zero, which in turn verifies some variable nodes (in the set $\Delta$) at HR2-R2 of iteration $1$. This is indeed the last step in iteration $1$. 

As explained before, the probability $\ra{p}^{(1,R2)}_\delta$ is needed to find the probability that a zero-valued variable node is verified at this stage. This probability is calculated as follows:
\begin{align*}
\ra{p}^{(1,R2)}_\delta &= \ds_{j=1}^{d_c-1} \Pr[c_e \in \ma{N}^{(1,R2,1)}_{0,j} | v_e \in \Delta^{(1)}],\\
&= \ds_{j=1}^{d_c-1} \df{\Pr[c_e \in \ma{N}^{(1,R2,1)}_{0,j}] \Pr[v_e \in \Delta^{(1)} | c_e \in \ma{N}^{(1,R2,1)}_{0,j}]}{\ds_{i=0}^{d_c} \ds_{j=1}^{d_c-1} \Pr[c_e \in \ma{N}^{(1,R2,1)}_{i,j}] \Pr[v_e \in \Delta^{(1)} | c_e \in \ma{N}^{(1,R2,1)}_{i,j}]},\\
&= \ds_{j=1}^{d_c-1} \df{\ra{p}^{(1,R2,1)}_{\ma{N}_{0,j}} \left(\df{j}{d_c}\right)}{\ds_{i=0}^{d_c} \ds_{j=1}^{d_c-1} \ra{p}^{(1,R2,1)}_{\ma{N}_{i,j}} \left(\df{j}{d_c}\right)},\\
&= \ds_{j=1}^{d_c-1} j\df{\ra{p}^{(1,R2,1)}_{\ma{N}_{0,j}}}{\ds_{i=0}^{d_c} \ds_{j=1}^{d_c-1} j \ra{p}^{(1,R2,1)}_{\ma{N}_{i,j}}}.
\end{align*}
Note that the denominator is indeed $\Pr[v_e \in \Delta^{(1)}] \triangleq \ra{p}^{(1)}_\Delta$. Hence, the probability $\ra{p}^{(1,R2,2)}_{\Delta_i}$, as defined before, is calculated as follows:
\[
\ra{p}^{(1,R2,2)}_{\Delta_i} = {d_v \choose i} \left( \ra{p}^{(1,R2)}_\delta \right)^i \left( 1 - \ra{p}^{(1,R2)}_\delta \right)^{d_v - i}, \hspace{20pt} i=0,\cdots,d_v.
\]
Lastly, the probability that a variable node is zero-valued and remains unverified for iteration $2$ is as follows:
\[
\ra{p}^{(2)}_\Delta = \ra{p}^{(1)}_\Delta \ra{p}^{(1,R2,2)}_{\Delta_0}.
\]
%%%%%%%%%%%%%%%%%%%%%%%%%%%%%%%%%%%%%%%%%%%%%%%%%%%%%%%%%%%%%%%%%%%%%%%%%%%%%%%%%%%%%%%%%%%%%%%%%%%%%%%%%%%%%
%%%%%%%%%%%%%%%%%%%%%%%%%%%%%%%%%%%%%%%%%%%%%%%%%%%%%%%%%%%%%%%%%%%%%%%%%%%%%%%%%%%%%%%%%%%%%%%%%%%%%%%%%%%%%
\newpage
\section{Detailed Description of the Analysis for SBB}
\label{app_original_SBB}
\begin{IEEEproof}[Proof of Theorem \ref{SBBModel}]
First we show that if a variable node $v \in \bigcup_{i=2}^{d_v}\ma{K}^{(\ell,R1,2)}_i \cup \hat{\ma{K}}^{(\ell,R1,2)}_1$, it will be recovered in the SBB algorithm at iteration $\ell$, HR2-R1. A check node in the set $\ma{N}_1$ is only connected to one element of the support set. So, a variable node $v$ in the support set with $i$ neighbors in the set $\ma{N}_1$ receives $i$ messages from the neighboring check nodes with the same value. Hence, variable nodes in the set $\bigcup_{i=2}^{d_v}\ma{K}^{(\ell,R1,2)}_i$ are recovered. Variable nodes in the set $\ma{K}^{(\ell,R1,2)}_1$, however, can not be verified according to ECN rule because they are not connected to at least two check nodes with the same value. So, the only verification rule applicable would be the D1CN. With the definition of the set $\hat{\ma{K}}^{(\ell,R1,2)}_1$, a variable node $v \in \hat{\ma{K}}^{(\ell,R1,2)}_1$ is verified based on D1CN at iteration $\ell$ of the SBB algorithm.

To prove the converse, we have to show that when a variable node $v$ is verified with a non-zero value at iteration $\ell$ of the SBB algorithm, then we have $v \in \bigcup_{i=2}^{d_v}\ma{K}^{(\ell,R1,2)}_i \cup \hat{\ma{K}}^{(\ell,R1,2)}_1$. The statement is true based on the definition of the sets and the fact that false verification happens with probability zero.
\end{IEEEproof}
%************************************************************************************************************
%************************************************************************************************************
\subsection{Verifying Variable Nodes}
\begin{itemize}
	\item A variable node in $\ma{K}_1$ is verified if it is connected to a check node in $\ma{N}_{1,0}$.
	\item A variable node in $\ma{K}_i$, $2\leq i\leq d_v$, is verified if it is connected to a check node in $\ma{N}_{1} \triangleq \bigcup_{j=0}^{d_c-1} \ma{N}_{1,j}$.
	\item A variable node in $\Delta$ is verified if it is connected to a check node in $\bigcup_{i=1}^{d_c} \ma{N}_{0,i}$; i.e., a zero-valued check node.
\end{itemize}
%************************************************************************************************************
%************************************************************************************************************
\subsection{Iteration Zero}
The process for iteration zero is the same as that of the LM algorithm discussed in \ref{APBIterZ}. The analysis is thus not repeated here.
%************************************************************************************************************
%************************************************************************************************************
\subsection{Iteration One}
In the first half-round of the first round (HR1-R1) of any iteration, check nodes process the received messages from variable nodes and generate the outgoing messages accordingly. Note that HR1-R1 in the SBB and the LM algorithm are the same.

In this section, we adopt the notation $\ma{N}_i$, with any superscript, to denote the set $\bigcup_{j=0}^{d_c - i}\ma{N}_{i,j}$.
The verified messages sent from variable nodes to check nodes at the end of iteration zero, are processed at check nodes at iteration $1$, HR1-R1. Based on the recovery process at iteration zero, all verified messages are sent from variable nodes in the sets $\Delta_j$, $1\leq j\leq d_v$. We partition the set of edges adjacent to a variable node in the set $\Delta_j$, $0\leq j\leq d_v$, into $\ma{N}_{=0}$-edges and $\ma{N}_{\neq 0}$-edges. $\ma{N}_{=0}$-edges are connected to zero-valued check nodes (check nodes in the set $\ma{N}_0$), while $\ma{N}_{\neq 0}$-edges are connected to non-zero check nodes.

We are interested in the set of check nodes that are regrouped from $\ma{N}^{(0,R2,1)}_{i,d_c-i}$ to $\ma{N}^{(1,R1,1)}_{i,j}$ for $1\leq i\leq d_c$, and eventually the probability $\ra{p}^{(1,R1,1)}_{\ma{N}_{i,j}}$ in HR1-R1 of iteration $1$. The calculation of this probability is exactly the same as the derivation of \eqref{eq:p_N_1R11} in the analysis of the LM algorithm.

In HR2-R1, the following 2 types of variable nodes in the support set are verified: 
\begin{enumerate}
	\item variable nodes neighbor to at least one check node of degree $1$; i.e., variable nodes that receive at least one message with the first coordinate equal to $1$. These variable nodes are verified with the value contained in that message based on D1CN.
	\item variable nodes neighbor to at least two check nodes in the set $\ma{N}^{(1,R1,1)}_1$; i.e., variable nodes that receive at least two messages with the same value. These variable nodes are verified with the common value of the messages based on the ECN rule.
\end{enumerate}
Therefore, after processing the received messages, we divide the set of all unverified variable nodes in the support set $\ma{K}^{(1)}$ into subsets $\ma{K}^{(1,R1)}_i$, $0\leq i\leq d_v$, where $i$ denotes the number of neighboring check nodes in the set $\ma{N}^{(1,R1,1)}_1$. We denote the set of such variable nodes by $\ma{K}^{(1,R1)}_{0 \uparrow i}$. It is worth mentioning that some variable nodes in the set $\ma{K}^{(1,R1)}_i$ are verified according to D1CN and ECN. The remaining unverified variable nodes, after the recovery, make the sets $\ma{K}^{(1,R1,2)}_j$, $0\leq j\leq d_v$. The sets $\ma{K}^{(1,R1)}_i$ are removed from the summarized formulas in Section \ref{originalSBB} to prevent any confusion. They appear here because they simplify the notations and explanations.

Let $p^{(1,R1)}$ denote the conditional probability that an edge is adjacent to a check node in the set $\ma{N}^{(1,R1,1)}_1$ given that it is adjacent to a variable node in the support set. Using the same notations $v_e$ and $c_e$ defined before, we have:
\begin{align}
p^{(1,R1)} &= \Pr[c_e \in \ma{N}^{(1,R1,1)}_1 | v_e \in \ma{K}^{(1)}],\notag \\
&= \df{\Pr[c_e \in \ma{N}^{(1,R1,1)}_1] \Pr[v_e \in \ma{K}^{(1)} | c_e \in \ma{N}^{(1,R1,1)}_1]}{\Pr[v_e \in \ma{K}^{(1)}]},\notag \\
&= \ds_{j=0}^{d_c-1} \df{\ra{p}^{(1,R1,1)}_{\ma{N}_{1,j}} \times 1/d_c}{\alpha^{(1)}} = \ds_{j=0}^{d_c-1} \df{\ra{p}^{(1,R1,1)}_{\ma{N}_{1,j}}}{\alpha^{(1)}d_c}.
\label{eq:APPC_p1R1}
\end{align}
Hence, the probability $\ra{p}^{(1,R1)}_{\ma{K}_{0 \uparrow i}}$ that a variable node $v \in \ma{K}^{(1)}$ belongs to the set $\ma{K}^{(1,R1)}_i$ is calculated as follows:
\begin{align*}
\ra{p}^{(1,R1)}_{\ma{K}_i} &= \ra{p}^{(1,R1)}_{\ma{K}_{0 \uparrow i}} \triangleq \Pr(v \in \ma{K}^{(1,R1)}_i|v \in \ma{K}^{(1)}),\\
&= {d_v\choose i}\left(p^{(1,R1)} \right)^i \left(1 - p^{(1,R1)} \right)^{d_v-i},\hspace{20pt}i=0,\cdots,d_v.
\end{align*}
Based on the ECN rule, variable nodes in the set $\bigcup_{i=2}^{d_v} \ma{K}^{(1,R1)}_i$ are verified. A fraction $f^{(1,R1)}$ of variable nodes in the set $\ma{K}^{(1,R1)}_1$ that receive a message with the first coordinate equal to one are also verified based on the D1CN rule. Using the Bayes' rule, this fraction is calculated as follows:
\[
f^{(1,R1)} = \Pr[c_e \in \ma{N}^{(1,R1,1)}_{1,0} | v_e \in \ma{K}^{(1,R1)}_1, c_e \in \ma{N}^{(0,R2,2)}_{1}].
\]
By omitting the superscripts, we obtain:
\begin{align*}
f^{(1,R1)} &= \df{\Pr[c_e \in \ma{N}_{1,0}] \Pr[v_e \in \ma{K}_1|c_e \in \ma{N}_{1,0}] \Pr[c_e \in \ma{N}_{1}|c_e \in \ma{N}_{1,0},v_e \in \ma{K}_1]}{\Pr[c_e \in \ma{N}_{1}] \Pr[v_e \in \ma{K}_1|c_e \in \ma{N}_{1}]},\\
&= \df{\ra{p}^{(1,R1,1)}_{\ma{N}_{1,0}} \times X \times 1}{\ra{p}^{(1,R1,1)}_{\ma{N}_{1}} \times X},\\
&= \df{\ra{p}^{(1,R1,1)}_{\ma{N}_{1,0}}}{\ra{p}^{(1,R1,1)}_{\ma{N}_{1}}},
\end{align*}
where we have used the fact that $\Pr[v_e \in \ma{K}_1|c_e \in \ma{N}_{1,0}] = \Pr[v_e \in \ma{K}_1|c_e \in \ma{N}_{1}] \triangleq X$. Therefore, the probability that a variable node in the support set remains unverified for iteration $2$ is as follows:
\begin{align*}
\alpha^{(2)} &= \alpha^{(1)} \left(1 - f^{(1,R1)} \ra{p}^{(1,R1)}_{\ma{K}_1} - \ds_{i=2}^{d_v} \ra{p}^{(1,R1)}_{\ma{K}_i} \right),\\
&= \alpha^{(1)} \left( \ra{p}^{(1,R1)}_{\ma{K}_0} + \ra{p}^{(1,R1)}_{\ma{K}_1} - f^{(1,R1)} \ra{p}^{(1,R1)}_{\ma{K}_1} \right),\\
&= \alpha^{(1)} \left( \ra{p}^{(1,R1)}_{\ma{K}_0} + \left(1 - f^{(1,R1)}\right) \ra{p}^{(1,R1)}_{\ma{K}_1} \right).
\end{align*}
The final regrouping of variable nodes into sets $\ma{K}^{(1,R1,2)}_i$, $0\leq i\leq d_v$, is performed by taking into account the verification of some sets of variable nodes. We have:
\begin{align}
\ra{p}^{(1,R1,2)}_{\ma{K}_0} &= \df{1}{N^{(1,R1)}} \ra{p}^{(1,R1)}_{\ma{K}_0}.\notag \\
\ra{p}^{(1,R1,2)}_{\ma{K}_1} &= \df{1}{N^{(1,R1)}} \left(1 - f^{(1,R1)}\right) \ra{p}^{(1,R1)}_{\ma{K}_1}.\notag \\
\ra{p}^{(1,R1,2)}_{\ma{K}_i} &= 0, \hspace{20pt} 2\leq i\leq d_v.
\label{eq:APPC_p_K_1R12}
\end{align}
The normalization factor $N^{(1,R1)}$ is used to make the set of parameters $\ra{p}^{(1,R1,2)}_{\ma{K}_i}$ a valid probability measure, and is calculated as follows:
\[
N^{(1,R1)} = \ra{p}^{(1,R1)}_{\ma{K}_0} +  \left(1 - f^{(1,R1)}\right) \ra{p}^{(1,R1)}_{\ma{K}_1} = \df{\alpha^{(2)}}{\alpha^{(1)}}.
\]
With this, the analysis of the first round of iteration $1$ is over. In HR1-R2 of iteration $1$, check nodes receive messages from variable nodes. At this point, check nodes should be regrouped based on their first index, as some variable nodes in the support set have been verified at HR2-R1. Since a fraction of variable nodes in the set $\ma{K}^{(1,R1,2)}_1$ are left unverified, unlike the LM algorithm, not all check nodes in the set $\ma{N}^{(1,R1,1)}_{1,j}$ ($1 \leq j\leq d_c-1$) are regrouped into the set $\ma{N}^{(1,R2,1)}_{0,j}$; some will stay in the same set $\ma{N}^{(1,R2,1)}_{1,j}$. Hence, in addition to analyzing the set of check nodes $\ma{N}^{(1,R2)}_{i\downarrow k,j}$ that are regrouped from $\ma{N}^{(1,R1,1)}_{i,j}$ to $\ma{N}^{(1,R2,1)}_{k,j}$, we also have to analyze the set of check nodes $\ma{N}^{(1,R2)}_{1\downarrow 0,j}$ that are regrouped from $\ma{N}^{(1,R1,1)}_{1,j}$ to $\ma{N}^{(1,R2,1)}_{0,j}$. 

Suppose that edges adjacent to a check node are partitioned into two sets: $\ma{K}$-edges and $\Delta$-edges. $\ma{K}$-edges are connected to variable nodes in the set $\ma{K}^{(1)}$, while $\Delta$-edges are connected to zero-valued variable nodes. To analyze the regrouping of check nodes in the sets $\ma{N}^{(1,R1,1)}_{1}$ and $\ma{N}^{(1,R1,1)}_{i}$, $2\leq i\leq d_c$, we need the probabilities $\ra{p}^{(1,R2)}_{d=1}$ and $\ra{p}^{(1,R2)}_{d \neq 1}$ defined as follows. The probability $\ra{p}^{(1,R2)}_{d=1}$ is defined as the conditional probability of an edge carrying a verified message given that it is a $\ma{K}$-edge adjacent to a check node in the set $\ma{N}^{(1,R1,1)}_{1}$. The probability $\ra{p}^{(1,R2)}_{d \neq 1}$ is defined similarly with respect to the set of check nodes in $\bigcup_{i=2}^{d_c} \ma{N}^{(1,R1,1)}_{i}$. To find the probability $\ra{p}^{(1,R2)}_{d=1}$, we shall consider only the variable nodes in the set $\bigcup_{i=2}^{d_v} \ma{K}^{(1,R1,2)}_i$. This is because, variable nodes in the set $\ma{K}^{(1,R1,2)}_1$ are connected to check nodes in the set $\ma{N}^{(1,R1,1)}_{1,0}$ only. Let $f_e$ denote the status flag ($f_e\in \{0,1\}$) of the message carried over the edge $e$. We proceed as follows:
\begin{align*}
\ra{p}^{(1,R2)}_{d=1} &= \Pr[f_e = 1 | v_e \in \ma{K}^{(1)}, c_e \in \ma{N}^{(1,R1,1)}_{1}],\\
&= \Pr[v_e \in \bigcup_{i=2}^{d_v} \ma{K}^{(1,R1,2)}_i | v_e \in \ma{K}^{(1)}, c_e \in \ma{N}^{(1,R1,1)}_{1}],\\
&= 1 - \Pr[v_e \in \ma{K}^{(1,R1,2)}_1 | v_e \in \ma{K}^{(1)}, c_e \in \ma{N}^{(1,R1,1)}_{1}],\\
&= 1 - \df{\Pr[v_e \in \ma{K}^{(1,R1,2)}_1|v_e \in \ma{K}^{(1)}] \Pr[c_e \in \ma{N}^{(1,R1,1)}_{1}|v_e \in \ma{K}^{(1,R1,2)}_1, v_e \in \ma{K}^{(1)}]}{\Pr[c_e \in \ma{N}^{(1,R1,1)}_{1}|v_e \in \ma{K}^{(1)}]},\\
&= 1 - \df{\ra{p}^{(1,R1,2)}_{\ma{K}_1} 1/d_v}{\ra{p}^{(1,R1)}},\\
&= 1 - \df{\ra{p}^{(1,R1,2)}_{\ma{K}_1}}{d_v \ra{p}^{(1,R1)}},
\end{align*}
where $\ra{p}^{(1,R1,2)}_{\ma{K}_1}$ and $\ra{p}^{(1,R1)}$ are given by \eqref{eq:APPC_p_K_1R12} and \eqref{eq:APPC_p1R1}, respectively. Using the same approach, the calculation of the probability $\ra{p}^{(1,R2)}_{d \neq 1}$ follows. In each step, we omit some of the superscripts to simplify the presentation.
\begin{align*}
\ra{p}^{(1,R2)}_{d \neq 1} &= \Pr[f_e = 1 | v_e \in \ma{K}^{(1)}, c_e \in \bigcup_{j=2}^{d_c}\ma{N}^{(1,R1,1)}_{j}],\\
&= 1 - \Pr[v_e \in \ma{K}^{(1,R1,2)}_0 | v_e \in \ma{K}^{(1)}, c_e \in \bigcup_{j=2}^{d_c}\ma{N}_{j}] - 
\Pr[v_e \in \ma{K}^{(1,R1,2)}_1 , f_e = 0 | v_e \in \ma{K}^{(1)}, c_e \in \bigcup_{j=2}^{d_c}\ma{N}_{j}],\\
&= 1 - \Pr[v_e \in \ma{K}_0 | v_e \in \ma{K}, c_e \in \bigcup_{j=2}^{d_c}\ma{N}_{j}]\\
 &- \Pr[v_e \in \ma{K}_1 | v_e \in \ma{K}, c_e \in \bigcup_{j=2}^{d_c}\ma{N}_{j}] 
\Pr[f_e = 0 | v_e \in \ma{K}_1 , v_e \in \ma{K}, c_e \in \bigcup_{j=2}^{d_c}\ma{N}_{j}],\\
&= 1 - \df{\Pr[v_e \in \ma{K}_0 | v_e \in \ma{K}] \Pr[c_e \in \bigcup_{j=2}^{d_c}\ma{N}_{j} | v_e \in \ma{K}_0 , v_e \in \ma{K}]}{\Pr[c_e \in \bigcup_{j=2}^{d_c}\ma{N}_{j} | v_e \in \ma{K}]}\\
&- \df{\Pr[v_e \in \ma{K}_1 | v_e \in \ma{K}] \Pr[c_e \in \bigcup_{j=2}^{d_c}\ma{N}_{j} | v_e \in \ma{K}_1 , v_e \in \ma{K}]}{\Pr[c_e \in \bigcup_{j=2}^{d_c}\ma{N}_{j} | v_e \in \ma{K}]} \left(1 - f^{(1,R1)} \right),\\
&= 1 - \df{\Pr[v_e \in \ma{K}_0 | v_e \in \ma{K}] \Pr[c_e \in \bigcup_{j=2}^{d_c}\ma{N}_{j} | v_e \in \ma{K}_0 , v_e \in \ma{K}]}{1 - \Pr[c_e \in \ma{N}_{1} | v_e \in \ma{K}]}\\
&- \df{\Pr[v_e \in \ma{K}_1 | v_e \in \ma{K}] \Pr[c_e \in \bigcup_{j=2}^{d_c}\ma{N}_{j} | v_e \in \ma{K}_1 , v_e \in \ma{K}]}{1 - \Pr[c_e \in \ma{N}_{1} | v_e \in \ma{K}]} \left(1 - f^{(1,R1)} \right),\\
&= 1 - \df{\ra{p}^{(1,R1,2)}_{\ma{K}_0} \times 1}{1 - \ra{p}^{(1,R1)}} - \df{\ra{p}^{(1,R1,2)}_{\ma{K}_1}\left(\df{d_v-1}{d_v}\right)}{1 - \ra{p}^{(1,R1)}} \left(1 - f^{(1,R1)} \right).
\end{align*}
Hence, the probabilities $\ra{p}^{(1,R2)}_{\ma{N}_{i\downarrow k,j}}$ and $\ra{p}^{(1,R2)}_{\ma{N}_{1\downarrow 0,j}}$ that a check node belongs respectively to the set of check nodes $\ma{N}^{(1,R2)}_{i\downarrow k,j}$ and $\ma{N}^{(1,R2)}_{1\downarrow 0,j}$ are calculated as follows: 
\begin{align*}
\ra{p}^{(1,R2)}_{\ma{N}_{i\downarrow k,j}} &= {i\choose k} \left( \ra{p}^{(1,R2)}_{d \neq 1} \right)^{i-k}\left( 1 - \ra{p}^{(1,R2)}_{d \neq 1} \right)^k,\hspace{20pt} i=2,\cdots,d_c,\hspace{20pt} k=0,\cdots,i, & j=0,\cdots,d_c - i.\\
\ra{p}^{(1,R2)}_{\ma{N}_{1\downarrow 0,j}} &= {1\choose 1} \left( \ra{p}^{(1,R2)}_{d=1} \right)^{1}\left( 1 - \ra{p}^{(1,R2)}_{d=1} \right)^0 = \ra{p}^{(1,R2)}_{d=1}, \hspace{20pt} \ra{p}^{(1,R2)}_{\ma{N}_{1\downarrow 1,j}} = 1 - \ra{p}^{(1,R2)}_{d=1}, & j=1,\cdots,d_c - i.\\
\ra{p}^{(1,R2)}_{\ma{N}_{1\downarrow 0,0}} &= 1, \hspace{20pt} \ra{p}^{(1,R2)}_{\ma{N}_{1\downarrow 1,0}} = 0, \\
\ra{p}^{(1,R2)}_{\ma{N}_{0\downarrow 0,j}} &= 1, & j=1,\cdots,d_c - i.
\end{align*}
After the regrouping, the probability $\ra{p}^{(1,R2,1)}_{\ma{N}_{k,j}}$ that a check node belongs to the set $\ma{N}^{(1,R2,1)}_{k,j}$ is calculated as follows:
\[
\ra{p}^{(1,R2,1)}_{\ma{N}_{k,j}} = \ds_{i=k}^{d_c} \ra{p}^{(1,R1,1)}_{\ma{N}_{i,j}} \ra{p}^{(1,R2)}_{\ma{N}_{i\downarrow k,j}}, \hspace{20pt}k=0,\cdots,d_c, \hspace{20pt} j=0,\cdots,d_c-i.
\]
The measurement corresponding to check nodes in the set $\ma{N}^{(1,R2,1)}_{0,k}$, $1\leq k\leq d_c-1$, changes to zero, as the check nodes are no longer connected to an unverified variable node in the support set. Hence, the messages transmitted by such check nodes have a value equal to zero, which in turn verifies some variable nodes (in the set $\Delta$) at HR2-R2 of iteration $1$. This is indeed the last step in iteration $1$. 

As defined before, the probability $\ra{p}^{(1,R2)}_\delta$, is needed to find the probability that a zero-valued variable node is verified at this stage. This probability is calculated as follows:
\begin{align*}
\ra{p}^{(1,R2)}_\delta &= \ds_{j=1}^{d_c-1} \Pr[c_e \in \ma{N}^{(1,R2,1)}_{0,j} | v_e \in \Delta^{(1)}],\\
&= \ds_{j=1}^{d_c-1} \df{\Pr[c_e \in \ma{N}^{(1,R2,1)}_{0,j}] \Pr[v_e \in \Delta^{(1)} | c_e \in \ma{N}^{(1,R2,1)}_{0,j}]}{\ds_{i=0}^{d_c} \ds_{j=1}^{d_c-1} \Pr[c_e \in \ma{N}^{(1,R2,1)}_{i,j}] \Pr[v_e \in \Delta^{(1)} | c_e \in \ma{N}^{(1,R2,1)}_{i,j}]},\\
&= \ds_{j=1}^{d_c-1} \df{\ra{p}^{(1,R2,1)}_{\ma{N}_{0,j}} \left(\df{j}{d_c}\right)}{\ds_{i=0}^{d_c} \ds_{j=1}^{d_c-1} \ra{p}^{(1,R2,1)}_{\ma{N}_{i,j}} \left(\df{j}{d_c}\right)},\\
&= \ds_{j=1}^{d_c-1} j\df{\ra{p}^{(1,R2,1)}_{\ma{N}_{0,j}}}{\ds_{i=0}^{d_c} \ds_{j=1}^{d_c-1} j \ra{p}^{(1,R2,1)}_{\ma{N}_{i,j}}}.
\end{align*}
Note that the denominator is in fact $\Pr[v_e \in \Delta^{(1)}] \triangleq \ra{p}^{(1)}_\Delta$. Hence, the probability $\ra{p}^{(1,R2,2)}_{\Delta_i}$, defined as the probability that an unverified zero-valued variable node belongs to the set $\Delta^{(1,R2,2)}_i$, is calculated as follows:
\[
\ra{p}^{(1,R2,2)}_{\Delta_i} = {d_v \choose i} \left( \ra{p}^{(1,R2)}_\delta \right)^i \left( 1 - \ra{p}^{(1,R2)}_\delta \right)^{d_v - i}, \hspace{20pt} i=0,\cdots,d_v.
\]
Lastly, the probability that a variable node is zero-valued and remains unverified for iteration $2$ is given by:
\[
\ra{p}^{(2)}_\Delta = \ra{p}^{(1)}_\Delta \ra{p}^{(1,R2,2)}_{\Delta_0}.
\]
%************************************************************************************************************
%************************************************************************************************************
\subsection{Iterations Two and Beyond}
At iteration $2$, HR1-R1, the regrouping of check nodes based on their second index is similar to the process in HR1-R1 at iteration $1$. We have:
\[
\ra{p}^{(2,R1)}_{\ma{E}_R} = 1 - \df{\ra{p}^{(2)}_\Delta}{1 - \ra{p}^{(1,R2)}_\delta}.
\]
For the regrouping of check nodes based on the second index we thus have:
\[
\ra{p}^{(2,R1)}_{\ma{N}_{i,j \downarrow k}} = {j \choose k} \left( 1 - \ra{p}^{(2,R1)}_{\ma{E}_R} \right)^j \left( \ra{p}^{(2,R1)}_{\ma{E}_R} \right)^{j-k},\hspace{20pt} i=1,\cdots,d_c, \hspace{20pt} j=0,\cdots,d_c-i, \hspace{20pt} k=0,\cdots,j.
\]
Hence,
\[
\ra{p}^{(2,R1,1)}_{\ma{N}_{i,k}} = \ds_{j=k}^{d_c-i} \ra{p}^{(1,R2,1)}_{\ma{N}_{i,j}} \ra{p}^{(2,R1)}_{\ma{N}_{i,j \downarrow k}},\hspace{20pt}i=1,\cdots,d_c,\hspace{20pt}k=0,\cdots,d_c-i.
\]
At iteration $2$, HR2-R1, variable nodes in the support set $\ma{K}^{(2)}$ should be regrouped. Note that the set $\ma{K}^{(2)}$ consists of two partitions: $\ma{K}^{(1,R1,2)}_0$ and $\ma{K}^{(1,R1,2)}_1$. Since variable nodes in the set $\ma{K}^{(1,R1,2)}_1$ are connected to check nodes in the set $\ma{N}^{(2,R1,1)}_1$, and since the regrouping of the variable nodes is based on the number of their neighbors in the set $\ma{N}^{(2,R1,1)}_1$, we shall partition the set of check nodes in $\ma{N}^{(2,R1,1)}_1$ into two sets: $\ma{N}^{(2,R1,+)}_1$ and $\ma{N}^{(2,R1,C)}_1$. Check nodes in the set $\ma{N}^{(2,R1,+)}_1$ were regrouped into the set $\ma{N}^{(1,R2,1)}_1$ from all the other sets $\ma{N}^{(1,R1,1)}_i$, $2 \leq i\leq d_c$, at iteration $1$, HR1-R2. Check nodes in the set $\ma{N}^{(2,R1,C)}_1$, however, were regrouped into the set $\ma{N}^{(1,R2,1)}_1$ from the set $\ma{N}^{(1,R1,1)}_1$. Edges in the set $\ma{N}^{(2,R1,+)}_1$ are responsible for the regrouping of variable nodes at iteration $2$, HR2-R1. We denote by $\ra{p}^{(2,R1)}_k$ the conditional probability that an edge is adjacent to a check node in $\ma{N}^{(2,R1,+)}_1$ given that 1) it emanates from an unverified variable node 2) it is not adjacent to a check node in the set $\ma{N}^{(2,R1,C)}_1$. This is indeed the probability that a variable node has an edge that increases its index. This probability is calculated as follows:
\begin{align*}
\ra{p}^{(2,R1)}_k &= \Pr[c_e \in \ma{N}^{(2,R1,+)}_1 | v_e \in \ma{K}^{(2)} , c_e \notin \ma{N}^{(2,R1,C)}_1],\\
&= \df{\Pr[c_e \in \ma{N}^{(2,R1,+)}_1] \Pr[v_e \in \ma{K}^{(2)} | c_e \in \ma{N}^{(2,R1,+)}_1] \Pr[c_e \notin \ma{N}^{(2,R1,C)}_1 | v_e \in \ma{K}^{(2)} , c_e \in \ma{N}^{(2,R1,+)}_1]}{\Pr[v_e \in \ma{K}^{(2)} , c_e \notin \ma{N}^{(2,R1,C)}_1]},\\
&= \df{\Pr[c_e \in \ma{N}^{(2,R1,+)}_1] \Pr[v_e \in \ma{K}^{(2)} | c_e \in \ma{N}^{(2,R1,+)}_1] \Pr[c_e \notin \ma{N}^{(2,R1,C)}_1 | v_e \in \ma{K}^{(2)} , c_e \in \ma{N}^{(2,R1,+)}_1]}{\Pr[v_e \in \ma{K}^{(2)} , c_e \in \bigcup_{i=2}^{d_c} \ma{N}^{(2,R1,1)}_i \cup \ma{N}^{(2,R1,+)}_1]},\\
&= \df{\Pr[c_e \in \ma{N}^{(2,R1,+)}_1] \Pr[v_e \in \ma{K}^{(2)} | c_e \in \ma{N}^{(2,R1,+)}_1] \Pr[c_e \notin \ma{N}^{(2,R1,C)}_1 | v_e \in \ma{K}^{(2)} , c_e \in \ma{N}^{(2,R1,+)}_1]}{\Pr[c_e \in \ma{N}^{(2,R1,+)}_1] \Pr[v_e \in \ma{K}^{(2)} | c_e \in \ma{N}^{(2,R1,+)}_1] + \ds_{i=2}^{d_c} \Pr[\ma{N}^{(2,R1,1)}_i] \Pr[v_e \in \ma{K}^{(2)}|\ma{N}^{(2,R1,1)}_i]},\\
&= \df{\ra{p}^{(2,R1,+)}_{\ma{N}_1} \times 1/d_c}{\ra{p}^{(2,R1,+)}_{\ma{N}_1} \times 1/d_c + \ds_{i=2}^{d_c} \ra{p}^{(2,R1,1)}_{\ma{N}_i} \left(\df{i}{d_c}\right)},\\
&= \df{\ra{p}^{(2,R1,+)}_{\ma{N}_1}}{\ra{p}^{(2,R1,+)}_{\ma{N}_1} + \ds_{i=2}^{d_c} i \ra{p}^{(2,R1,1)}_{\ma{N}_i}},
\end{align*}
where $\ra{p}^{(2,R1,+)}_{\ma{N}_1}$ denotes the probability that a check node belongs to the set $\ma{N}^{(2,R1,+)}_1$. Since the two sets $\ma{N}^{(2,R1,+)}_1$ and $\ma{N}^{(2,R1,C)}_1$ are disjoint and their union is the set $\ma{N}^{(2,R1,1)}_1$, we have:
\[
\ra{p}^{(2,R1,+)}_{\ma{N}_1} = \ds_{j=2}^{d_c} \ds_{i=0}^{d_c-j} \ra{p}^{(1,R1,1)}_{\ma{N}_{j,i}} \ra{p}^{(1,R2)}_{\ma{N}_{j \downarrow 1,i}}, \hspace{20pt} \ra{p}^{(2,R1,C)}_{\ma{N}_1} = \ra{p}^{(2,R1,1)}_{\ma{N}_1} - \ra{p}^{(2,R1,+)}_{\ma{N}_1}.
\]
Hence, the probability $\ra{p}^{(2,R1)}_{\ma{K}_i \uparrow j}$, $i \in \{0,1\}$, that a variable node from $\ma{K}^{(1,R1,2)}_i$ is regrouped into $\ma{K}^{(2,R1,2)}_j$, $i \leq j \leq d_v$, is calculated as follows:
\[
\ra{p}^{(2,R1)}_{\ma{K}_{i \uparrow j}} = {d_v - i \choose j-i} \left( \ra{p}^{(2,R1)}_k \right)^{j-i} \left( 1 - \ra{p}^{(2,R1)}_k \right)^{d_v - j}, \hspace{20pt} i = 0,1, \hspace{20pt} j = i,\cdots,d_v.
\]
Finally, the probability $\ra{p}^{(2,R1)}_{\ma{K}_j}$ that a variable node in the support set belongs to the set $\ma{K}^{(2,R1)}_j$ is calculated by:
\[
\ra{p}^{(2,R1)}_{\ma{K}_j} = \ds_{i=0}^1 \ra{p}^{(1,R1,2)}_{\ma{K}_i} \ra{p}^{(2,R1)}_{\ma{K}_{i \uparrow j}}, \hspace{20pt} j=0,\cdots,d_v.
\]
The probability $\ra{p}^{(2,R1,2)}_{\ma{K}_j}$ that a variable node in the support set belongs to the set $\ma{K}^{(2,R1,2)}_j$, is calculated based on the set of verified variable nodes at this stage. Variable nodes in the set $\ma{K}^{(2,R1)}_j$, $2\leq j\leq d_v$, are all verified. Variable nodes in the set $\ma{K}^{(2,R1)}_0$ are left intact, and a fraction of the variable nodes in the set $\ma{K}^{(2,R1)}_1$ are verified. The procedure to find this fraction is as follows.

The set $\ma{K}^{(2,R1)}_1$ consists of two sets of variable nodes: $\ma{K}^{(2,R1)}_{0 \uparrow 1}$ and $\ma{K}^{(2,R1)}_{1 \uparrow 1}$. Variable nodes in the set $\ma{K}^{(2,R1)}_{0 \uparrow 1}$ are neighbor to check nodes in the set $\ma{N}^{(2,R1,+)}_1$, while variable nodes in the set $\ma{K}^{(2,R1)}_{1 \uparrow 1}$ are neighbor to check nodes in the set $\ma{N}^{(2,R1,C)}_1$. Since the structure and evolution of the two sets $\ma{N}^{(2,R1,+)}_1$ and $\ma{N}^{(2,R1,C)}_1$ are different, the sets $\ma{K}^{(2,R1)}_{0 \uparrow 1}$ and $\ma{K}^{(2,R1)}_{1 \uparrow 1}$ also evolve differently. The sets $\ma{N}^{(2,R1,+)}_1$ and $\ma{N}^{(2,R1,C)}_1$ are formed at iteration $1$, HR1-R2. We shall partition the two sets further into subsets $\ma{N}^{(2,R1,+)}_{1,j}$ and $\ma{N}^{(2,R1,C)}_{1,j}$, $0\leq j\leq d_c-1$. At iteration $2$, HR1-R1, the subsets are to be regrouped based on their second index, just like any other set of check nodes. Hence, we have:
\begin{align*}
\ra{p}^{(2,R1,1,+)}_{\ma{N}_{1,k}} &= \ds_{j=k}^{d_c-1} \ra{p}^{(1,R2,1,+)}_{\ma{N}_{1,j}} \ra{p}^{(2,R1)}_{\ma{N}_{1,j \downarrow k}},\hspace{20pt}k=0,\cdots,d_c-1,\\
\ra{p}^{(2,R1,1,C)}_{\ma{N}_{1,k}} &= \ds_{j=k}^{d_c-1} \ra{p}^{(1,R2,1,C)}_{\ma{N}_{1,j}} \ra{p}^{(2,R1)}_{\ma{N}_{1,j \downarrow k}},\hspace{20pt}k=0,\cdots,d_c-1,
\end{align*}
where,
\[
\ra{p}^{(2,R1)}_{\ma{N}_{1,j \downarrow k}} = {j \choose k} \left( 1 - \ra{p}^{(2,R1)}_{\ma{E}_R} \right)^j \left( \ra{p}^{(2,R1)}_{\ma{E}_R} \right)^{j-k} , \hspace{20pt} \ra{p}^{(2,R1)}_{\ma{E}_R} = 1 - \df{\ra{p}^{(2)}_\Delta}{1 - \ra{p}^{(1,R2)}_\delta},
\]
and,
\begin{align*}
\ra{p}^{(1,R2,1,+)}_{\ma{N}_{1,j}} &= \ds_{k=2}^{d_c} \ra{p}^{(1,R1,1)}_{\ma{N}_{k,j}} \ra{p}^{(1,R2)}_{\ma{N}_{k \downarrow 1,j}},\\
\ra{p}^{(1,R2,1,C)}_{\ma{N}_{1,j}} &= \ra{p}^{(1,R1,1)}_{\ma{N}_{1,j}} \ra{p}^{(1,R2)}_{\ma{N}_{1 \downarrow 1,j}}.
\end{align*}
Variable nodes in the two sets $\ma{K}^{(2,R1)}_{0 \uparrow 1}$ and $\ma{K}^{(2,R1)}_{1 \uparrow 1}$ are verified at iteration $2$, HR2-R1, if and only if, they are neighbor to check nodes in the sets $\ma{N}^{(2,R1,1,+)}_{1,0}$ and $\ma{N}^{(2,R1,1,C)}_{1,0}$, respectively. Therefore, the parameters $f^{(2,R1,+)}$ and $f^{(2,R1,C)}$, defined as the respective fraction of variable nodes in $\ma{K}^{(2,R1)}_{0 \uparrow 1}$ and $\ma{K}^{(2,R1)}_{1 \uparrow 1}$ that are verified at iteration $2$, HR2-R1, are calculated as follows:
\begin{align*}
f^{(2,R1,+)} &= \df{\ra{p}^{(2,R1,1,+)}_{\ma{N}_{1,0}}}{\ds_{k=0}^{d_c - 1} \ra{p}^{(2,R1,1,+)}_{\ma{N}_{1,k}}}.\\
f^{(2,R1,C)} &= \df{\ra{p}^{(2,R1,1,C)}_{\ma{N}_{1,0}}}{\ds_{k=0}^{d_c - 1} \ra{p}^{(2,R1,1,C)}_{\ma{N}_{1,k}}}.
\end{align*}
Finally, for the set of probabilities $\ra{p}^{(2,R1,2)}_{\ma{K}_j}$, we have:
\begin{align*}
\ra{p}^{(2,R1,2)}_{\ma{K}_0} &= \df{1}{N^{(2,R1)}} \ra{p}^{(2,R1)}_{\ma{K}_0} = \df{1}{N^{(2,R1)}} \ra{p}^{(1,R1,2)}_{\ma{K}_0} \ra{p}^{(2,R1)}_{\ma{K}_{0 \uparrow 0}},\\
\ra{p}^{(2,R1,2,+)}_{\ma{K}_1} &= \df{1}{N^{(2,R1)}} \ra{p}^{(1,R1,2)}_{\ma{K}_0} \ra{p}^{(2,R1)}_{\ma{K}_{0 \uparrow 1}} \left(1 - f^{(2,R1,+)} \right),\\
\ra{p}^{(2,R1,2,C)}_{\ma{K}_1} &= \df{1}{N^{(2,R1)}} \ra{p}^{(1,R1,2)}_{\ma{K}_1} \ra{p}^{(2,R1)}_{\ma{K}_{1 \uparrow 1}} \left(1 - f^{(2,R1,C)} \right),\\
\ra{p}^{(2,R1,2)}_{\ma{K}_j} &= 0, \hspace{20pt} j=2,\cdots,d_v.
\end{align*}
The normalization factor $N^{(2,R1)}$ is used to make the set of parameters $\ra{p}^{(2,R1,2)}_{\ma{K}_i}$ a valid probability measure, and is calculated by:
\[
N^{(2,R1)} = \ra{p}^{(1,R1,2)}_{\ma{K}_0} \ra{p}^{(2,R1)}_{\ma{K}_{0 \uparrow 0}} + \ra{p}^{(1,R1,2)}_{\ma{K}_0} \ra{p}^{(2,R1)}_{\ma{K}_{0 \uparrow 1}} \left(1 - f^{(2,R1,+)} \right) + \ra{p}^{(1,R1,2)}_{\ma{K}_1} \ra{p}^{(2,R1)}_{\ma{K}_{1 \uparrow 1}} \left(1 - f^{(2,R1,C)} \right).
\]
The probability that a variable node belongs to the support set and remains unverified after iteration $2$, $\alpha^{(3)}$, is calculated as follows:
\begin{align*}
\alpha^{(3)} &= \alpha^{(2)} \left( 
\ra{p}^{(1,R1,2)}_{\ma{K}_0} \ra{p}^{(2,R1)}_{\ma{K}_{0 \uparrow 0}} + \ra{p}^{(1,R1,2)}_{\ma{K}_0} \ra{p}^{(2,R1)}_{\ma{K}_{0 \uparrow 1}} \left(1 - f^{(2,R1,+)} \right) + \ra{p}^{(1,R1,2)}_{\ma{K}_1} \ra{p}^{(2,R1)}_{\ma{K}_{1 \uparrow 1}} \left(1 - f^{(2,R1,C)} \right)
\right),\\
&= \alpha^{(2)} N^{(2,R1)}.
\end{align*}
For iteration $2$, HR1-R2, we find the probability $\ra{p}^{(1,R2)}_{d \neq 1}$ in the following. (For simplicity, some superscripts are omitted. They appear when there is a risk of ambiguity.)
\begin{align*}
\ra{p}^{(2,R2)}_{d \neq 1} &= \Pr[f_e = 1 | v_e \in \ma{K}^{(2)}, c_e \in \bigcup_{j=2}^{d_c}\ma{N}^{(2,R1,1)}_{j}],\\
&= 1 - \Pr[v_e \in \ma{K}^{(2,R1)}_0 , f_e = 0 | v_e \in \ma{K}^{(2)}, c_e \in \bigcup_{j=2}^{d_c}\ma{N}_{j}]\\
&- \Pr[v_e \in \ma{K}^{(2,R1)}_1 , f_e = 0 | v_e \in \ma{K}^{(2)}, c_e \in \bigcup_{j=2}^{d_c}\ma{N}_{j}],\\
&= 1 - \Pr[v_e \in \ma{K}_0 | v_e \in \ma{K}, c_e \in \bigcup_{j=2}^{d_c}\ma{N}_{j}]\\
&- \Pr[v_e \in \ma{K}^+_1 | v_e \in \ma{K}, c_e \in \bigcup_{j=2}^{d_c}\ma{N}_{j}] 
\Pr[f_e = 0 | v_e \in \ma{K}^+_1 , v_e \in \ma{K}, c_e \in \bigcup_{j=2}^{d_c}\ma{N}_{j}]\\
&- \Pr[v_e \in \ma{K}^C_1 | v_e \in \ma{K}, c_e \in \bigcup_{j=2}^{d_c}\ma{N}_{j}] 
\Pr[f_e = 0 | v_e \in \ma{K}^C_1 , v_e \in \ma{K}, c_e \in \bigcup_{j=2}^{d_c}\ma{N}_{j}],\\
&= 1 - \df{\Pr[v_e \in \ma{K}_0 | v_e \in \ma{K}] \Pr[c_e \in \bigcup_{j=2}^{d_c}\ma{N}_{j} | v_e \in \ma{K}_0 , v_e \in \ma{K}]}{\Pr[c_e \in \bigcup_{j=2}^{d_c}\ma{N}_{j} | v_e \in \ma{K}]}\\
&- \df{\Pr[v_e \in \ma{K}^+_1 | v_e \in \ma{K}] \Pr[c_e \in \bigcup_{j=2}^{d_c}\ma{N}_{j} | v_e \in \ma{K}^+_1 , v_e \in \ma{K}]}{\Pr[c_e \in \bigcup_{j=2}^{d_c}\ma{N}_{j} | v_e \in \ma{K}]} \left(1 - f^{(2,R1,+)} \right)\\
&- \df{\Pr[v_e \in \ma{K}^C_1 | v_e \in \ma{K}] \Pr[c_e \in \bigcup_{j=2}^{d_c}\ma{N}_{j} | v_e \in \ma{K}^C_1 , v_e \in \ma{K}]}{\Pr[c_e \in \bigcup_{j=2}^{d_c}\ma{N}_{j} | v_e \in \ma{K}]} \left(1 - f^{(2,R1,C)} \right),\\
&= 1 - \df{\Pr[v_e \in \ma{K}_0 | v_e \in \ma{K}] \Pr[c_e \in \bigcup_{j=2}^{d_c}\ma{N}_{j} | v_e \in \ma{K}_0 , v_e \in \ma{K}]}{1 - \Pr[c_e \in \ma{N}_{1} | v_e \in \ma{K}]}\\
&- \df{\Pr[v_e \in \ma{K}^+_1 | v_e \in \ma{K}] \Pr[c_e \in \bigcup_{j=2}^{d_c}\ma{N}_{j} | v_e \in \ma{K}^+_1 , v_e \in \ma{K}]}{1 - \Pr[c_e \in \ma{N}_{1} | v_e \in \ma{K}]} \left(1 - f^{(2,R1,+)} \right)\\
&- \df{\Pr[v_e \in \ma{K}^C_1 | v_e \in \ma{K}] \Pr[c_e \in \bigcup_{j=2}^{d_c}\ma{N}_{j} | v_e \in \ma{K}^C_1 , v_e \in \ma{K}]}{1 - \Pr[c_e \in \ma{N}_{1} | v_e \in \ma{K}]} \left(1 - f^{(2,R1,C)} \right),\\
&= 1 - \df{\ra{p}^{(1,R1,2)}_{\ma{K}_0} \ra{p}^{(2,R1)}_{\ma{K}_{0 \uparrow 0}} \times 1}{1 - \ra{p}^{(2,R1)}}\\
&- \df{\ra{p}^{(1,R1,2)}_{\ma{K}_0} \ra{p}^{(2,R1)}_{\ma{K}_{0 \uparrow 1}} \left(\df{d_v-1}{d_v}\right)}{1 - \ra{p}^{(2,R1)}} \left(1 - f^{(2,R1,+)} \right)\\
&- \df{\ra{p}^{(1,R1,2)}_{\ma{K}_1} \ra{p}^{(2,R1)}_{\ma{K}_{1 \uparrow 1}} \left(\df{d_v-1}{d_v}\right)}{1 - \ra{p}^{(2,R1)}} \left(1 - f^{(2,R1,C)} \right).
\end{align*}
Hence, the probability $\ra{p}^{(2,R2)}_{\ma{N}_{i\downarrow k,j}}$ that a check node belongs to the set of check nodes $\ma{N}^{(2,R2)}_{i\downarrow k,j}$ is calculated as follows: 
\[
\ra{p}^{(2,R2)}_{\ma{N}_{i\downarrow k,j}} = {i\choose k} \left( \ra{p}^{(2,R2)}_{d \neq 1} \right)^{i-k}\left( 1 - \ra{p}^{(2,R2)}_{d \neq 1} \right)^k,\hspace{20pt} i=2,\cdots,d_c,\hspace{20pt} k=0,\cdots,i,\hspace{20pt} j=0,\cdots,d_c - i.
\]
Note that the probability $\ra{p}^{(2,R1)}$ is calculated by:
\[
\ra{p}^{(2,R1)} = \df{\ra{p}^{(2,R1,1,+)}_{\ma{N}_1} + \ra{p}^{(2,R1,1,C)}_{\ma{N}_1}}{\alpha^{(2)} d_c}.
\]
As we explain in the following, the evolution of the sets $\ma{N}^{(2,R1,1,+)}_1$ and $\ma{N}^{(2,R1,1,C)}_1$ is a bit more involved. A variable node in the set $\ma{K}^{(2,R1)}_{0 \uparrow i}$, $1\leq i\leq d_v$, has $i$ neighboring check nodes in the set $\ma{N}^{(2,R1,1,+)}_1$. On the other hand, a variable node in the set $\ma{K}^{(2,R1)}_{1 \uparrow i}$, $1\leq i\leq d_v$, has $1$ neighboring check node in the set $\ma{N}^{(2,R1,1,C)}_1$ and $i-1$ neighboring check nodes in the set $\ma{N}^{(2,R1,1,+)}_1$. Now let us consider a check node $c \in \ma{N}^{(2,R1,1,+)}_1$. Suppose, $c$ is neighbor to a variable node $v \in \ma{K}^{(2,R1)}_{0 \uparrow 1}$. Variable node $v$ is verified if and only if $c$ belongs to the subset $\ma{N}^{(2,R1,1,+)}_{1,0}$. Hence, $c$ is regrouped as a zero-valued check node if it belongs to the set $\ma{N}^{(2,R1,1,+)}_{1,0}$. Now, suppose $c$ is neighbor to a variable node $v' \in \bigcup_{i=2}^{d_v} \ma{K}^{(2,R1)}_{0 \uparrow i}$, or $v' \in \bigcup_{i=2}^{d_v} \ma{K}^{(2,R1)}_{1 \uparrow i}$. Since the variable node $v'$ is verified with probability $1$, check node $c$ is regrouped into the set of zero-valued check nodes with probability $1$ as well. A similar argument holds true for the set of check nodes in $\ma{N}^{(2,R1,1,C)}_1$ and variable nodes in $\ma{K}^{(2,R1)}_{1 \uparrow 1}$. Therefore, to regroup check nodes in the sets $\ma{N}^{(2,R1,1,+)}_1$ and $\ma{N}^{(2,R1,1,C)}_1$, we need to divide them further based on their neighbors; i.e., whether or not they are neighbor to variable nodes in the sets $\ma{K}^{(2,R1)}_{0 \uparrow 1}$ and $\ma{K}^{(2,R1)}_{1 \uparrow 1}$, respectively.

We partition the set $\ma{N}^{(2,R1,1,+)}_1$ into subsets $\ma{N}^{(2,R2,+,O)}_1$ and $\ma{N}^{(2,R2,+,F)}_1$. We also partition the set $\ma{N}^{(2,R1,1,C)}_1$ into subsets $\ma{N}^{(2,R2,C,O)}_1$ and $\ma{N}^{(2,R2,C,F)}_1$. Check nodes in sets $\ma{N}^{(2,R2,+,F)}_1$ and $\ma{N}^{(2,R2,C,F)}_1$ are neighbor to variable nodes in sets $\ma{K}^{(2,R1)}_{0 \uparrow 1}$ and $\ma{K}^{(2,R1)}_{1 \uparrow 1}$, respectively. Any other check node in the set $\ma{N}^{(2,R1,1,+)}_1$, not being part of the set $\ma{N}^{(2,R2,+,F)}_1$ is grouped into the set $\ma{N}^{(2,R2,+,O)}_1$. Similarly, any other check node in the set $\ma{N}^{(2,R1,1,C)}_1$, not being part of the set $\ma{N}^{(2,R2,C,F)}_1$ is grouped into the set $\ma{N}^{(2,R2,C,O)}_1$. 

Let $\ra{p}^{(2,R2,+,O)}_{\ma{N}_1}$ and $\ra{p}^{(2,R2,+,F)}_{\ma{N}_1}$ denote the probabilities that a check node belongs to sets $\ma{N}^{(2,R2,+,O)}_1$ and $\ma{N}^{(2,R2,+,F)}_1$, respectively. Probabilities $\ra{p}^{(2,R2,C,O)}_{\ma{N}_1}$ and $\ra{p}^{(2,R2,C,F)}_{\ma{N}_1}$ are defined similarly. The calculation of the probability $\ra{p}^{(2,R2,+,F)}_{\ma{N}_{1,i}}$, $0\leq i\leq d_c-1$, follows:
\begin{align*}
\ra{p}^{(2,R2,+,F)}_{\ma{N}_{1,i}} &= \Pr[c \in \ma{N}^{(2,R2,+,F)}_{1,i}],\\
&= \Pr[c \in \ma{N}^{(2,R1,1,+)}_{1,i}] \Pr[c \in \ma{N}^{(2,R2,+,F)}_{1,i} | c \in \ma{N}^{(2,R1,1,+)}_{1,i}],\\
&= \Pr[c \in \ma{N}^{(2,R1,1,+)}_{1,i}] \Pr[v_e \in \ma{K}^{(2,R2)}_{0 \uparrow 1} | c_e \in \ma{N}^{(2,R1,1,+)}_{1,i}, v_e \in \ma{K}^{(2)}],\\
&= \Pr[c \in \ma{N}^{(2,R1,1,+)}_{1,i}] \df{\Pr[v_e \in \ma{K}^{(2,R2)}_{0 \uparrow 1} | v_e \in \ma{K}^{(2)}] \Pr[c_e \in \ma{N}^{(2,R1,1,+)}_{1,i} | v_e \in \ma{K}^{(2,R2)}_{0 \uparrow 1} , v_e \in \ma{K}^{(2)}]}{\Pr[c_e \in \ma{N}^{(2,R1,1,+)}_{1,i} | v_e \in \ma{K}^{(2)}]},\\
&= \Pr[c \in \ma{N}^{(2,R1,1,+)}_{1,i}] \df{\Pr[v_e \in \ma{K}^{(2,R2)}_{0 \uparrow 1} | v_e \in \ma{K}^{(2)}] \Pr[c_e \in \ma{N}^{(2,R1,1,+)}_{1,i} | v_e \in \ma{K}^{(2,R2)}_{0 \uparrow 1} , v_e \in \ma{K}^{(2)}]}{A + B},\\
\end{align*}
where,
\begin{align*}
A &= \ds_{j=1}^{d_v} \Pr[v_e \in \ma{K}^{(2,R2)}_{0 \uparrow j} , c_e \in \ma{N}^{(2,R1,1,+)}_{1,i} | v_e \in \ma{K}^{(2)}],\\
&= \ds_{j=1}^{d_v} j \ra{p}^{(1,R1,2)}_{\ma{K}_0} \ra{p}^{(2,R1)}_{\ma{K}_{0 \uparrow j}},\\
B &= \ds_{j=2}^{d_v} \Pr[v_e \in \ma{K}^{(2,R2)}_{1 \uparrow j} , c_e \in \ma{N}^{(2,R1,1,+)}_{1,i} | v_e \in \ma{K}^{(2)}],\\
&= \ds_{j=2}^{d_v} (j-1) \ra{p}^{(1,R1,2)}_{\ma{K}_1} \ra{p}^{(2,R1)}_{\ma{K}_{1 \uparrow j}}.
\end{align*}
Hence,
\begin{align*}
\ra{p}^{(2,R2,+,F)}_{\ma{N}_{1,i}} &= \ra{p}^{(2,R1,1,+)}_{\ma{N}_{1,i}} \df{\ra{p}^{(1,R1,2)}_{\ma{K}_0} \ra{p}^{(2,R1)}_{\ma{K}_{0 \uparrow 1}}}{\ds_{j=1}^{d_v} j \ra{p}^{(1,R1,2)}_{\ma{K}_0} \ra{p}^{(2,R1)}_{\ma{K}_{0 \uparrow j}} + \ds_{j=2}^{d_v} (j-1) \ra{p}^{(1,R1,2)}_{\ma{K}_1} \ra{p}^{(2,R1)}_{\ma{K}_{1 \uparrow j}}},\\
\ra{p}^{(2,R2,+,O)}_{\ma{N}_{1,i}} &= \ra{p}^{(2,R1,1,+)}_{\ma{N}_{1,i}} - \ra{p}^{(2,R2,+,F)}_{\ma{N}_{1,i}}.\\
\end{align*}
Following a similar approach, we have:
\begin{align*}
\ra{p}^{(2,R2,C,F)}_{\ma{N}_{1,i}} &= \ra{p}^{(2,R1,1,C)}_{\ma{N}_{1,i}} \df{\ra{p}^{(2,R1)}_{\ma{K}_{1 \uparrow 1}}}{\ds_{j=1}^{d_v} \ra{p}^{(2,R1)}_{\ma{K}_{1 \uparrow j}}},\\
\ra{p}^{(2,R2,C,O)}_{\ma{N}_{1,i}} &= \ra{p}^{(2,R1,1,C)}_{\ma{N}_{1,i}} - \ra{p}^{(2,R2,C,F)}_{\ma{N}_{1,i}}.
\end{align*}
Since check nodes in the sets $\ma{N}^{(2,R2,+,O)}_1$, $\ma{N}^{(2,R2,C,O)}_1$, $\ma{N}^{(2,R2,+,F)}_{1,0}$, and $\ma{N}^{(2,R2,C,F)}_{1,0}$ receive $d_c$ verified messages at iteration $2$, HR1-R2, all such check nodes are grouped into the set $\ma{N}^{(2,R2,1)}_0$. On the other hand, check nodes in the sets $\ma{N}^{(2,R2,+,F)}_{1,i}$, and $\ma{N}^{(2,R2,C,F)}_{1,i}$, $1\leq i\leq d_c-1$, are grouped into the sets $\ma{N}^{(2,R2,1,C)}_{1,i}$. 
After the regrouping, the probability $\ra{p}^{(2,R2,1)}_{\ma{N}_{k,j}}$ that a check node belongs to the set $\ma{N}^{(2,R2,1)}_{k,j}$ is calculated as follows:
\begin{align*}
\ra{p}^{(2,R2,1)}_{\ma{N}_{k,j}} &= \ds_{i=k}^{d_c} \ra{p}^{(2,R1,1)}_{\ma{N}_{i,j}} \ra{p}^{(2,R2)}_{\ma{N}_{i\downarrow k,j}}, \hspace{20pt} k=2,\cdots,d_c, & j=0,\cdots,d_c-i,\\
\ra{p}^{(2,R2,1,+)}_{\ma{N}_{1,j}} &= \ds_{i=2}^{d_c} \ra{p}^{(2,R1,1)}_{\ma{N}_{i,j}} \ra{p}^{(2,R2)}_{\ma{N}_{i\downarrow 2,j}},  & j=0,\cdots,d_c-1, \\
\ra{p}^{(2,R2,1,C)}_{\ma{N}_{1,j}} &= \ra{p}^{(2,R2,C,F)}_{\ma{N}_{1,j}} + \ra{p}^{(2,R2,+,F)}_{\ma{N}_{1,j}},  & j=1,\cdots,d_c-1,\\
\ra{p}^{(2,R2,1)}_{\ma{N}_{0,j}} &= \ra{p}^{(2,R1,1)}_{\ma{N}_{0,j}} + \ra{p}^{(2,R2,C,O)}_{\ma{N}_{1,j}} + \ra{p}^{(2,R2,+,O)}_{\ma{N}_{1,j}} + \ds_{i=2}^{d_c} \ra{p}^{(2,R1,1)}_{\ma{N}_{i,j}} \ra{p}^{(2,R2)}_{\ma{N}_{i\downarrow 0,j}},  & j=0,\cdots,d_c-1.
\end{align*}
As previously defined, the probability $\ra{p}^{(2,R2)}_\delta$, is needed to find the probability that a zero-valued variable node is verified at this stage. This probability is calculated as follows:
\begin{align*}
\ra{p}^{(2,R2)}_\delta &= \ds_{j=1}^{d_c-1} \Pr[c_e \in \ma{N}^{(2,R2,1)}_{0,j} | v_e \in \Delta^{(2)}],\\
&= \ds_{j=1}^{d_c-1} \df{\Pr[c_e \in \ma{N}^{(2,R2,1)}_{0,j}] \Pr[v_e \in \Delta^{(2)} | c_e \in \ma{N}^{(2,R2,1)}_{0,j}]}{\ds_{i=0}^{d_c} \ds_{j=1}^{d_c-1} \Pr[c_e \in \ma{N}^{(2,R2,1)}_{i,j}] \Pr[v_e \in \Delta^{(2)} | c_e \in \ma{N}^{(2,R2,1)}_{i,j}]},\\
&= \ds_{j=1}^{d_c-1} \df{\ra{p}^{(2,R2,1)}_{\ma{N}_{0,j}} \left(\df{j}{d_c}\right)}{\ds_{i=0}^{d_c} \ds_{j=1}^{d_c-1} \ra{p}^{(2,R2,1)}_{\ma{N}_{i,j}} \left(\df{j}{d_c}\right)},\\
&= \ds_{j=1}^{d_c-1} j\df{\ra{p}^{(2,R2,1)}_{\ma{N}_{0,j}}}{\ds_{i=0}^{d_c} \ds_{j=1}^{d_c-1} j \ra{p}^{(2,R2,1)}_{\ma{N}_{i,j}}}.
\end{align*}
Note that the denominator is indeed $\Pr[v_e \in \Delta^{(2)}] \triangleq \ra{p}^{(2)}_\Delta$. Hence, the probability $\ra{p}^{(2,R2,2)}_{\Delta_i}$, defined as the probability that an unverified zero-valued variable node belongs to the set $\Delta^{(2,R2,2)}_i$, is calculated as follows:
\[
\ra{p}^{(2,R2,2)}_{\Delta_i} = {d_v \choose i} \left( \ra{p}^{(2,R2)}_\delta \right)^i \left( 1 - \ra{p}^{(2,R2)}_\delta \right)^{d_v - i}, \hspace{20pt} i=0,\cdots,d_v.
\]
Lastly, the probability that a variable node is zero-valued and remains unverified for iteration $3$ is given by:
\[
\ra{p}^{(3)}_\Delta = \ra{p}^{(2)}_\Delta \ra{p}^{(2,R2,2)}_{\Delta_0}.
\]
The analysis of an iteration $\ell$, $\ell \geq 3$, is similar to that of iteration 2. The update rules for a generic iteration $\ell$, $\ell \geq 2$ is given in subsection \ref{originalSBB}.
%%%%%%%%%%%%%%%%%%%%%%%%%%%%%%%%%%%%%%%%%%%%%%%%%%%%%%%%%%%%%%%%%%%%%%%%%%%%%%%%%%%%%%%%%%%%%%%%%%%%%%%%%%%%%
%%%%%%%%%%%%%%%%%%%%%%%%%%%%%%%%%%%%%%%%%%%%%%%%%%%%%%%%%%%%%%%%%%%%%%%%%%%%%%%%%%%%%%%%%%%%%%%%%%%%%%%%%%%%%
\newpage
\section{Details of Concentration Results}
\label{conc_bound}

\subsection{Probability of Tree-Like Neighborhood}\label{AppD1}
Consider a particular variable node $v$. We enumerate $M_{\ell}$ and $C_{\ell}$, respectively, the number of variable nodes and check nodes in $\mathcal{N}_v^{2\ell}$ under the assumption that $\mathcal{N}_v^{2\ell}$ is tree-like, as follows: \[M_{\ell}=1+d_v(d_c-1)\sum_{i=1}^{\ell}(d_c-1)^{i-1}(d_v-1)^{i-1},\] and \[C_{\ell}=d_v\sum_{i=1}^{\ell}(d_c-1)^{i-1}(d_v-1)^{i-1}.\]

Recall that $m$ and $n$ are the number of check nodes and the variable nodes, respectively. Fix ${\ell}^{*}$, and let $\ell<{\ell}^{*}$. Assuming that $\mathcal{N}_v^{2\ell}$ is tree-like, the probability that $\mathcal{N}_v^{2\ell+1}$ is tree-like is lower bounded by \[\left(1-\frac{C_{\ell^{*}}}{m}\right)^{C_{\ell+1}-C_{\ell}},\] and assuming that $\mathcal{N}_v^{2\ell+1}$ is tree-like, the probability that $\mathcal{N}_v^{2\ell+2}$ is tree-like is lower bounded by \[\left(1-\frac{M_{\ell^{*}}}{n}\right)^{M_{\ell+1}-M_{\ell}},\] for sufficiently large $n$ (see Appendix A in \cite{RU01}). Thus, the probability that $\mathcal{N}_v^{2\ell^{*}}$ is tree-like is bounded from below by (lower bounding is done by using the chain rule) \[\left(1-\frac{M_{\ell^{*}}}{n}\right)^{M_{\ell}^{*}}\left(1-\frac{C_{\ell^{*}}}{m}\right)^{C_{\ell}^{*}},\] for sufficiently large $n$, and since $\left(1-{M_{\ell^{*}}}/{n}\right)^{M_{\ell^{*}}}\geq \left(1-{M_{\ell^{*}}^2}/{n}\right)$, and $\left(1-{C_{\ell^{*}}}/{m}\right)^{C_{\ell^{*}}}\geq \left(1-{C_{\ell^{*}}^2}/{m}\right)$ for large $n$, then \begin{eqnarray*} \Pr[\mathcal{N}_v^{2\ell^{*}}\text{is not tree-like}]& \leq & 1-\left(1-\frac{M_{\ell^{*}}^2}{n}\right)\left(1-\frac{C_{\ell^{*}}^2}{m}\right) \\ & \leq & \frac{M_{\ell^{*}}^2}{n}+\frac{C_{\ell^{*}}^2}{m} \\ &=& \frac{M_{\ell^{*}}^2+C_{\ell^{*}}^2 (d_c/d_v)}{n}.\end{eqnarray*} Taking $\gamma=M_{\ell^{*}}^2+C_{\ell^{*}}^2(d_c/d_v)$, we see that $\gamma$ is a constant not depending on $n$, and thus, \[\Pr[\mathcal{N}_v^{2\ell^{*}}\text{is not tree-like}]\leq \frac{\gamma}{n}.\]

\subsection{Derivation of Bounds on the Difference between Two Consecutive Elements in Martingale Sequences}\label{AppD3}
Consider two realizations $T'$ and $T''$ from the ensemble of all graphs, inputs, and weights. We consider two cases:
\begin{itemize}
	\item Case I: Realizations $T'$ and $T''$ are the same except for the value of a variable node.
	\item Case II: Realizations $T'$ and $T''$ are the same except for the connection of two variable nodes $v_1$ and $v_2$ to two check nodes $c_1$ and $c_2$; i.e., $T': c_1 \in \ma{M}(v_1), c_2 \in \ma{M}(v_2)$ and $T'': c_1 \in \ma{M}(v_2), c_2 \in \ma{M}(v_1)$.
	\item Case III: Realizations $T'$ and $T''$ are the same except for one weight of an edge in the graph.
\end{itemize}
Suppose the difference between the two realizations results in the difference of $N(\ell)$ in verified variable nodes at iteration $\ell$. The goal in this appendix is to find an upper bound on $N(\ell)$ for Genie, LM and SBB algorithms. Without loss of generality, suppose that realization $T'$ verifies less number of variable nodes compared to $T''$. As we are seeking an upper bound, we assume the worst configuration for realization $T'$ and the best configuration for $T''$. The realization $T'$ being the worst configuration implies that the variable node/edge/weight that is the difference between the two realizations $T'$ and $T''$ results in no verification of variable nodes up to iteration $\ell$ under consideration for $T'$. On the other hand, we assume that the configuration in $T''$ is so that the verification of a variable node at an iteration, results in the maximum number of verifications in the next iteration. With these configurations for $T'$ and $T''$, in order to maximize the difference between the verified variable nodes between the two realizations at iteration $\ell$, we need to maximize the number of variable nodes that can be verified at iteration $0$.

Let the parameters $E(\ell), D(\ell), Z(\ell)$ denote the difference in the number of variable nodes verified at iteration $\ell$ between the two realizations due to the ECN, D1CN, and ZCN verification rules, respectively. Therefore, $N(\ell) = E(\ell)+ D(\ell)+ Z(\ell)$.

In what fallows we find an upper bound for $N(\ell)$ in the case of the Genie algorithm. A similar reasoning can be used for the other algorithms.

\subsubsection{Genie}
The only verification rule applied to the Genie is D1CN. We find the maximum $N(\ell)$ for each of the three cases discussed above. Focusing on case I, three possibilities exist:
\begin{enumerate}
	\item A variable node in $T'$ is non-zero, while the same variable node in $T''$ is zero.
	\item A variable node in $T'$ is zero, while the same variable node in $T''$ is non-zero.
	\item A variable node in both $T'$ and $T''$ is non-zero, but with two different values.
\end{enumerate}
In the first scenario, the worst configuration is such that the variable node under consideration remains unverified up to iteration $\ell$. As the corresponding variable node is zero in realization $T''$, this means that the $d_v$ neighboring check nodes have degrees smaller by one compared to their counterparts in realization $T'$. The best configuration $T''$ is then formed if each one of these $d_v$ check nodes has degree 1. Each such check node results in the verification of one variable node with the D1CN rule. So, $N(0) \leq d_v$.

A variable node verified based on D1CN at iteration $i-1$ ($0\leq i\leq \ell$) can reduce the degree of at most $d_v-1$ check nodes. In the best case, each such check node has degree $1$ which results in the verification of another variable node at iteration $i$. Therefore, we have:
\[
D(i) \leq D(i-1)(d_v-1),
\]
which results in
\[
N(\ell) \leq d_v(d_v-1)^{\ell}, \hspace{1cm} \ell\geq 0.
\]
In the second and third scenarios, considering the fact that $T'$ is the worst configuration and that the realizations $T'$ and $T''$ have the same weighted graph and the same input vector (except for the variable node under consideration), the realization $T''$ can not verify more variable nodes than $T'$. Therefore, in this case, $N(\ell) = 0$. This is assuming that no false verification happens in either realizations.

Now we consider Case II. Possible scenarios for the values of $v_1$ and $v_2$ are as follows:
\begin{enumerate}
	\item $v_1 = 0, v_2 \neq 0$ (due to the symmetry, this is the same as $v_1 \neq 0, v_2 = 0$),
	\item $v_1 = 0, v_2 = 0$,
	\item $v_1 \neq 0, v_2 \neq 0$.
\end{enumerate}
To make the worst realization, we assume that all the neighbors of $c_1$ are zero-valued variable nodes, and we let $c_2$ to be neighbor to only one other non-zero variable node, say $v_3$. In this realization, as $c_2$ has degree 2, then $v_2$ and $v_3$ can not be verified in iteration zero. However, when we switch the connections, $c_1$ and $c_2$ will both have degree 1. Therefore, both variable nodes $v_2$ and $v_3$ are verified based on the D1CN rule. So, in this scenario $N(0) = 2$. Again, to find the maximum number of variable nodes that can be verified in further iterations, we assume that the verification of each variable node at iteration $i-1$ results in the verification of $d_v-1$ other variable nodes at iteration $i$. We thus have:
\[
D(i) \leq D(i-1)(d_v-1),
\]
which results in
\[
N(\ell) \leq 2(d_v-1)^{\ell}, \hspace{1cm} \ell\geq 0.
\]
Due to the symmetry of the problem, the other two scenarios result in no difference in the number of verified variable nodes.

For Case III, the weights are, by definition, non-zero. Thus, the change in the weight of an edge in the graph, has no effect on the recovery of variable nodes.

Based on the above discussions, we have the following upper bound for the Genie algorithm:
\[
N(\ell) \leq d_v(d_v-1)^{(\ell)}, \hspace{1cm} \ell\geq 0.
\]

\subsubsection{LM}
This algorithm applies D1CN and ZCN verification rules in the first and second half-rounds of each iteration. Following the same steps as those for the Genie algorithm, one can show that the maximum $N(\ell)$ between all possible configurations for cases I, II, and III is achieved when a variable node changes its value from a non-zero value to a zero value. With the same logic as in the Genie, $d_v$ variable nodes can be verified with the D1CN rule at iteration $0$ in $T''$ that can not be verified in $T'$. We thus have
\[
N(0) = d_v.
\]

When a variable node is verified based on D1CN at the first half-round of iteration $i-1$, at most $d_v-1$ check nodes can have a value equal to zero, each of which results in the verification of $d_c - 1$ variable nodes with the ZCN rule in the second half-round of the same iteration.

On the other hand, when a variable node is verified based on ZCN at the second half-round of iteration $i-1$, at most $d_v-1$ check nodes can have a degree one for the first half-round of the next iteration. So, $d_v-1$ variable nodes can be verified using the D1CN rule at iteration $i$.

Putting these two steps together, we form the following recursive formulas:
\begin{align*}
D(i) &\leq Z(i-1) (d_v - 1),\\
Z(i) &\leq D(i) (d_v-1)(d_c-1).
\end{align*}

Solving the recursions with the initial condition $Z(0) = d_v$, we have:
\begin{align*}
Z(\ell) &\leq d_v(d_v-1)^{2\ell}(d_c-1)^{\ell},\\
D(\ell) &\leq d_v(d_v - 1)^{2\ell-1}(d_c-1)^{\ell-1},\\
N(\ell) = Z(\ell) + D(\ell) &\leq 2Z(\ell) = 2d_v(d_v-1)^{2\ell}(d_c-1)^{\ell}.
\end{align*}

\subsubsection{SBB}
This algorithm applies all ZCN, D1CN, and ECN verification rules. In order to find the upper bound on $N(\ell)$, we first find the number of variable nodes verified at a generic iteration $i$ due to the verification of a variable node at iteration $i-1$ based on each verification rule, separately. Then we find the maximum number of variable nodes that can possibly be verified at iteration zero based on each of the three Cases I, II, or III and use them as initial conditions to solve recursive formulas like the ones we saw for LM.

Assume two check nodes with the same value result in a variable node being verified according to the ECN rule. Therefore, the two check nodes result in the verification of $2d_c-2$ variable nodes in total with ZCN verification rule in the next round of the same iteration. Also, the other $d_v-2$ adjacent check nodes face a reduction in the degree as well as change in the value. Therefore, they can result in the verification of $(d_v-2)(d_c-1)$ variable nodes with ZCN in the next round of the same iteration, or the verification of $d_v-2$ variable nodes based on the D1CN or ECN in the next iteration. One may be able to find out the best combination of rules that results in the maximum number of verified variable nodes. However, as we are interested in finding an upper bound, we assume that all the events above can happen at the same time.

If a variable node is verified according to the ZCN rule, the $d_v-1$ adjacent check nodes do not face a change in their values, and thus can not contribute to the verification of other variable nodes based on ECN or ZCN. Therefore, each such check node verifies $1$ variable node based on the D1CN rule.

When a variable node is verified according to the D1CN rule, $d_v-1$ check nodes face a reduction in degree as well as change in value. With the same reasoning as that used in the ECN case, each such check node results in the verification of $d_c-1$, $1$, and $1$ variable nodes based on the ZCN, D1CN, and ECN rules, respectively.

Putting everything together, we have:
\begin{align*}
E(i) &\leq E(i-1)(d_v-2) + D(i-1)(d_v-1),\\
D(i) &\leq Z(i-1)(d_v-1) + D(i-1)(d_v-1) + E(i-1)(d_v-2),\\
Z(i) &\leq E(i)(d_v)(d_c-1) + D(i)(d_v-1)(d_c-1).
\end{align*}

Using the third inequality, we can rewrite the first two inequalities as follows:
\begin{align*}
E(i) &\leq E(i-1)(d_v-2) + D(i-1)(d_v-1) \leq d_v E(i-1) + d_v D(i-1),\\
D(i) &\leq E(i-1)\left(d_v(d_v-1)(d_c-1) + (d_v-2)\right) + D(i-1)(d_v-1)\left((d_v-1)(d_c-1)\right),\\
&\leq (d^2_vd_c+d_v) E(i-1) + d^2_vd_c D(i-1).\\
\end{align*}

Looking for an upper bound, we assume that the inequalities are satisfied with equalities. Therefore, we have:
\begin{align*}
E(i) &= d_v E(i-1) + d_v D(i-1),\\
D(i) &= (d^2_vd_c+d_v) E(i-1) + d^2_vd_c D(i-1) = d_vd_c E(i) + d_vE(i-1).\\
\end{align*}

Replacing $D(i-1)$ of the first equality by its value from the second one, we get:
\[
E(i) = \left(d_v + d^2_vd_c\right) E(i-1) + d^2_v E(i-2).
\]

This is a second order linear homogeneous recurrence relation. Using the characteristic polynomial technique, a closed form solution can be found. To find a more elegant solution, however, we derive the following upper bound on $E(i)$.
\[
E(i) = \left(d_v + d^2_vd_c\right) E(i-1) + d^2_v E(i-2) \leq \left(d_v + d^2_vd_c\right) E(i-1) + \df{\left(3d^2_v + 2d^3_vd_c\right)}{4} E(i-2).
\]
Considering the inequality as equality, and using the characteristic polynomial technique, we then obtain
\[
E(i) = K_1 \left(d^2_vd_c + \df{3}{2} d_v \right)^{i} + K_2 \left(- \df{d_v}{2} \right)^{i},
\]
for some $K_1$ and $K_2$ that depend on the initial conditions. We can simplify this even further as follows:
\[
E(i) \leq K_1 \left(d^2_vd_c + 2d_v \right)^{i}.
\]
By replacing this upper bound in $D(i)$ and $Z(i)$, we get:
\begin{align*}
D(i) &\leq d_vd_cK_1 \left(d^2_vd_c + 2d_v \right)^{i} + d_vK_1 \left(d^2_vd_c + 2d_v \right)^{i-1}
\leq 2d_vd_cK_1 \left(d^2_vd_c + 2d_v \right)^{i},\\
Z(i) &\leq d_vd_c (E(i) + D(i)) = d_vd_c K_1 \left(d^2_vd_c + 2d_v \right)^{i} (1+2d_vd_c),
\end{align*}
and thus
\begin{align*}
N(i) &= E(i) + D(i) + Z(i) \leq (d_vd_c+1) K_1 \left(d^2_vd_c + 2d_v \right)^{i} (1+2d_vd_c),\\
&\leq K_1 \left(d^2_vd_c + 2d_v \right)^{i+2},
\end{align*}
where $K_1$ is the maximum number of variable nodes that can be verified at iteration $0$ for the Cases I, II, and III. Once again, one can show that the maximum is achieved when a variable node changes value from a non-zero value to zero, and where $d_v$ variable nodes are verified using the ZCN rule. Therefore, we have $K_1 = d_v$. Hence,
\[
N(\ell) \leq d_v \left(d^2_vd_c + 2d_v \right)^{\ell+2}.
\]
%%%%%%%%%%%%%%%%%%%%%%%%%%%%%%%%%%%%%%%%%%%%%%%%%%%%%%%%%%%%%%%%%%%%%%%%%%%%%%%%%%%%%%%%%%%%%%%%%%%%%%%%%%%%%
%%%%%%%%%%%%%%%%%%%%%%%%%%%%%%%%%%%%%%%%%%%%%%%%%%%%%%%%%%%%%%%%%%%%%%%%%%%%%%%%%%%%%%%%%%%%%%%%%%%%%%%%%%%%%
\end{document}